\documentclass[final,5p,times,twocolumn,sort&compress]{elsarticle}

\usepackage{graphicx}
\usepackage{subfigure}
\usepackage[usenames]{color}
\usepackage{amssymb}
\usepackage{amsmath}
\usepackage{amsfonts}
\usepackage{placeins}
\usepackage{url}

\journal{Journal of Computational Science}

\begin{document}
\begin{frontmatter}
  
\title{An Evolutionary Algorithm Approach to Link Prediction in Dynamic Social Networks}
 
 \author{Catherine A. Bliss}
  \ead{Catherine.Bliss@uvm.edu}
     \author{Morgan R. Frank}
      \ead{Morgan.Frank@uvm.edu}
           \author{Christopher M. Danforth}
      \ead{Chris.Danforth@uvm.edu}
           \author{Peter Sheridan Dodds}
      \ead{pdodds@uvm.edu}
 \address{Computational Story Lab, Department of Mathematics and Statistics, Vermont Complex Systems Center\\ \& the Vermont Advanced Computing Core, 
 University of Vermont, Burlington, VT, 05405} 

\begin{abstract}
Many real world, complex phenomena have underlying structures of evolving networks where nodes and links are added and removed over time. A central scientific challenge is the description and explanation of network dynamics, with a key test being the prediction of short and long term changes. For the problem of short-term link prediction, existing methods attempt to determine neighborhood metrics that correlate with the appearance of a link in the next observation period. Recent work has suggested that the incorporation of topological features and node attributes can improve link prediction. We provide an approach to predicting future links by applying the Covariance Matrix Adaptation Evolution Strategy (CMA-ES) to optimize weights which are used in a linear combination of sixteen neighborhood and node similarity indices. We examine a large dynamic social network with over $10^6$ nodes (Twitter reciprocal reply networks), both as a test of our general method and as a problem of scientific interest in itself. Our method exhibits fast convergence and high levels of precision for the top twenty predicted links.  Based on our findings, we suggest possible factors which may be driving the evolution of Twitter reciprocal reply networks. \end{abstract}

\begin{keyword}
algorithms \sep data mining \sep link prediction \sep social networks \sep Twitter \sep complex networks \sep complex systems
\end{keyword}
\end{frontmatter}

\section{Introduction}
Time varying social networks can be used to model groups whose dynamics change over time. Individuals, represented by nodes, may enter or exit the network, while interactions, represented by links, may strengthen or weaken. Most network growth models capture global properties, but do not capture specific localized dynamics such as who will be connected to whom in the future. And yet, it is precisely this type of information that would be most valuable in applications such as national security, online social networking sites (people you may know), and organizational studies (predicting potential collaborators). 

In this paper, we focus primarily on the link prediction problem: given a snapshot of a network $G_t=(V,E_t)$, with nodes $V$ (nodes present across all time steps) and links $E_t$, at time $t$, we seek to predict the most likely links to newly occur in the next timestep, $t+1$~\cite{Liben-Nowell2007}. 

Link prediction strategies may be broadly categorized into three groups: similarity based strategies, maximum likelihood algorithms, and probabilistic models. As noted by Lu et al.~\cite{Lu2010}, the latter two approaches can be prohibitively time consuming for a large network over $10,000$ nodes. Given our interest in large, sparse networks with $N \gtrsim 10^6$, we focus primarily on local information and use similarity indices to characterize the likelihood of future interactions. We consider the two major classes of similarity indices: topological-based and node attribute (Table~\ref{tab:indices}).

There does not appear to be one best similarity index that is superior in all settings. Depending on the network under analysis, various measures have shown to be particularly promising~\cite{zhou2009predicting, Liben-Nowell2007, Wang2011, esslimani2011, backstrom2011, leroy2010, Yin2011}. These findings suggest that the predictors which work ``best'' for a given network may be related to the inherent structure within the individual network rather than a universal best set of predictors. Further, it is also plausible that the best link predictor may change as the network responds to endogenous and exogenous factors driving its evolution. 

Topological similarity indices encode information about the relative overlap between nodes' neighborhoods. We expect that the more ``similar'' two nodes' topological neighborhoods are (e.g., the more overlap in their shared friends), the more likely they may be to exhibit a future link. The common neighbors index, a building block of many other topological similarity indices, has been shown to correlate with the occurrence of future links~\cite{Newman2001}. Several variants of this index have been proposed and have been shown to be useful for link prediction in a variety of settings~\cite{barabasi2002evolution, ravasz2002hierarchical,wang2013similarity, zhou2009predicting, lu2011link, lin1998information, Salton:1986:IMI:576628, sorensen1948method, lichtenwalter2010new, Yang2012}. See~\cite{Lu2010} for a review. In their seminal paper on link prediction, Liben-Nowell and Kleinberg~\cite{Liben-Nowell2007} examined author collaboration networks derived from arXiv submissions in four subfields of Physics. They found that neighborhood similarity measures, such as the Jaccard~\cite{Salton:1986:IMI:576628}, Adamic-Adar~\cite{Adamic2003}, and the Katz coefficients~\cite{Katz1953} provided a large factor improvement over randomly predicted links.    
\begin{figure*}[ht!]
\includegraphics[width=\textwidth]{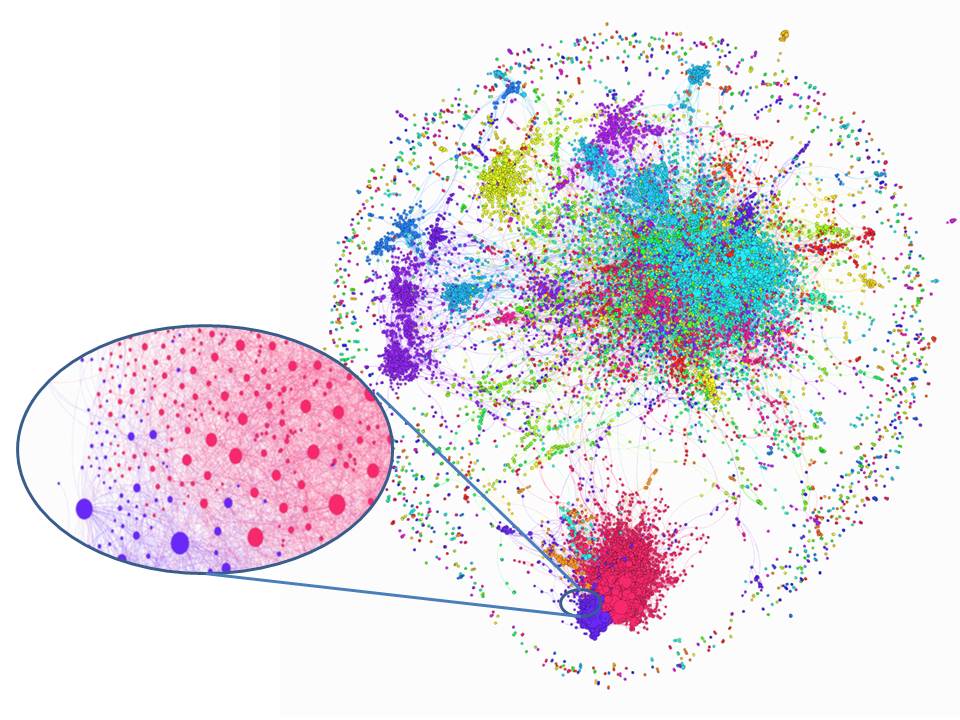}
\caption{A visualization of a one week Twitter reciprocal reply network exhibiting interactions between a core of 25,936 users who were active in each of networks in the period from September 9, 2008 to October 20, 2008. Note the large degree observed in one community (inset). The colors indicate modularity, a proxy for community structure, as detected by Gephi's implementation of Blondel's ``Fast unfolding of communities in large networks''~\cite{blondel2008}. }
\label{fig:twitter_plot}
\end{figure*}

As a complement for topological similarity indices, node-specific similarity indices examine node attributes, such as language, topical similarity, and behavior, in the case of social networks. Several studies have suggested that incorporating these measures can enhance link prediction~\cite{Aiello2012, Lu2010, Rowe2012, hutto2013longitudinal, Wang2011, romero2013, yin2010linkrec}.  In training algorithms for link prediction, researchers have used supervised learning including support vector machine~\cite{hasan2006}, decision trees~\cite{Wang2011}, bagged random forests~\cite{lichtenwalter2010new},  supervised random walks~\cite{backstrom2011}, multi-layer perceptrons, and others. Notably, Al Hasan et al.~\cite{hasan2006} use both topological and node-specific features to compare several supervised learning algorithms. They found that support vector machine (SVM) performed the best for the prediction of future links. While SVM is often considered a state of the art supervised learning model, one of its major drawbacks relates to kernel selection~\cite{burges1998}. Furthermore, Litchenwalter et al.~\cite{lichtenwalter2010new}, who use Weka's implementation of bagged random forests to produce ensembles of models and reduce variance, note the need to undersample due to the computational complexity of their method on large datasets. Of particular interest, Wang et al.~\cite{Wang2011} study a network of individuals constructed from mobile phone call data. They compare similarity indices used in isolation to a link predictor combining several indices (binary decision tree determined from supervised learning). These researchers found that the combination of node-specific and topological similarity indices outperform topological indices in isolation. While their results are promising, they acknowledge that the cost comes from looking at only a subset (e.g., 300 potential links which have Adamic-Adar scores $>$ 0.5 and Spatial Co-location rate $>$ 0.7) from the large potential set of user-user pairs two-links away (e.g., 266,750). 

Motivated by the above, we aim here to provide a link predictor encompassing both topological and user-specific information, which exhibits fast convergence and which does not require parametric thresholds nor undersampling due to computational complexity. 

In this paper, we fix a linear model for combining neighborhood similarity measures and node specific data and use an evolutionary algorithm to find the coefficients which optimize the proportion of correctly predicted links. Rather than pre-supposing that all similarity indices are of equal importance, we allow the weights of this linear combination to adjust using Covariance Matrix Adaptation Evolution Strategy (CMA-ES)~\cite{hansen2001completely}). Clearly, the optimal model combining similarity indices may not be linear and our assumption of this model structure is a limitation of our work. With that said, our work has several advantages over other methods for link prediction and our work reveals that a simple, linear model produces comparable results (if not better), with the added advantage of suggesting possible mechanisms driving the network's evolution over time. 

In many supervised learning approaches, link prediction efforts fit both a model structure and parameters. To surmount the challenge of large feature sets and large networks, researchers limit which features to include or perform undersampling due to computational complexity of these algorithms. Our approach of using CMA-ES for link prediction liberates researchers to include several indices in the link predictor, irrespective of their assumed performance. This is a strength of our method in that no assumption of network class nor prior knowledge about the system under analysis is required. 

Although we focus on the link prediction problem for a large, dynamic social network, our methods are independent of network type and may be applied to various biological, infrastructure, social and virtual networks. We demonstrate sixteen commonly used similarity indices here, but we emphasize that any other similarity indices may be interchanged for or added to the ones included in this study. The choice of which similarity measures to include will largely depend on available data (e.g., metadata for nodes and appropriate topological indices one has available in the context of the network one is studying) and the size of the network under consideration. 

Another limitation of several supervised learning approaches for link prediction is that the interpretation of the model may yield little information about the the network's evolutionary processes. Our methods provide transparency and the detection of indices which function as good predictors for future links which can help to elucidate possible mechanisms which may be driving the evolution of the network over time. 

In recent years, there has been a surge of interest in viewing Twitter activity through the lens of social network analysis. In many studies, nodes represent individuals and links represent following behavior~\cite{cha2010measuring, kwak2010twitter, huberman2008social}, reciprocated following~\cite{bollen2011twitter}, replies~\cite{romero2013} or reciprocated replies~\cite{Bliss2012}. 

Our application will be link prediction in Twitter reciprocal reply networks (RRNs), a construction first proposed by Bliss et al.~\cite{Bliss2012}.
We examine the evolution of these networks constructed at the time scale of weeks, where nodes represent users and links represent evidence of reciprocated replies during the time period of analysis. While many other studies have examined following and reciprocated following, we use reciprocated replies as evidence of social interaction and active engagement of individuals.\footnote{Following is a relatively passive activity and the establishment of a link between such users may misrepresent current attention to information in the network. Furthermore, follower networks typically do not account for the ``unfriending'' problem and the accumulation of dead links in a network can distort the representation of the true state of the system and spam.}

Due to the large size of networks that we seek to study and the hypothesis that friends of friends are more likely to become friends than individuals who have no friends in common~\cite{rapoport1963mathematical, Granovetter1973}, we restrict out attention to the prediction of new links at time $t+1$ which occur between individuals who were separated by a path length of 2 at time $t$ (i.e., triadic closure). Empirical evidence suggests that a preponderance of new links form between such 2-link neighbors in email reply networks~\cite{kossinets2006empirical}, Twitter follower networks~\cite{romero2010directed}, and Twitter RRNs.\footnote{We observe approximately 35\% of new links occurring between individuals connected by a path of length 2.}    

Previous link prediction efforts related to Twitter have largely focused on predicting follower relationships. Rowe, Stankovic and Alani~\cite{Rowe2012} use supervised learning to combine topological and node specific features (e.g., topics of tweets, tweet counts, re-tweets, etc.) to predict following behavior. Romero and Kleinberg also examined link prediction in follower networks and suggest that directed closure plays an important role in the formation of new links~\cite{romero2010directed}. Hutto, Yardi, and Gilbert~\cite{hutto2013longitudinal} examine 507 individuals and their followers to find that user-specific characteristics, such as message content and behavior should be given equal weight as topological characteristics for link prediction. Yin, Hong, and Davison examine 979 individuals and their neighbors (in Twitter follower networks) to predict following behavior over a six week time-scale~\cite{Yin2011}. Golder et al. examine Twitter users' desire to follow another user connected by a path length of two. They examine the correlation between shared interests and reciprocated following on users' expressed interest to make a new link (i.e., follow) and suggest that mutuality (reciprocated attention) is correlated with increased desire to follow~\cite{Golder2010}.  

We organize our paper as follows: In Section 2, we describe our data, the sixteen similarity indices, and the evolutionary algorithm used for evolving the weights on these indices. In Section 3 we present our results and in Section 4 discuss the significance of these findings, as well as suggest future directions for further work in this area. 

\section{Methods}
\subsection{Data}
Our data set consists of over 51 million tweets collected via the Twitter gardenhose API service from September 9, 2008 to December 1, 2008. This collection represents roughly 40\% of all messages sent during this period (Table A1). Using the criteria defined by Bliss et al.~\cite{Bliss2012}, we construct reciprocal reply networks\footnote{We also construct reply networks, whereby nodes represent users and directed, weighted links represent the number of replies sent from one individual to another during the week under analysis. Reply networks are used in the computation of the average path weight, one of our similarity indices.} as unweighted, undirected networks in which a link exists between nodes $u$ and $v$ if and only if these individuals exhibit reciprocal replies during the week under analysis (Fig.~\ref{fig:twitter_plot}). These networks range in size from $N=78296$ to $N=155753$ nodes (Table A2).

Since our task is to predict links, we do not wish to confound our task with the problem of node appearance or removal. To this end, we find a core of 25,936 users who were active in each of networks in the period from September 9, 2008 to October 20, 2008 and a core of 44,439 users who were active in each of the weeks in
the six weeks from October 21, 2008 and December 1, 2008. We train our link predictor on the new links that occur in a given Week $t$ (e.g., $e \in E_{t} \setminus E_{t-1}$) and validate on the new links that occur in week $t+1$ (e.g., $e \in E_{t+1} \setminus E_{t}$). We outline further details in the next two subsections. 
\begin{table*}
     \scriptsize
\begin{center}
\renewcommand{\arraystretch}{2}
		 \begin{tabular}{| l  | p{4cm} | p{9cm} |}
     \hline
\multicolumn{3}{ |l| }{\textbf{\textit{Topological similarity indices (abbreviation)}}}  \\ \hline  
Jaccard Index (J) & $       J(u,v)=\frac{|\Gamma(u)\cap\Gamma(v)|}{|\Gamma(u)\cup\Gamma(v)|}$ & Measures the probability that a neighbor of $u$ or $v$ is a neighbor of both $u$ and $v$. This measurement is a way of characterizing shared content and has been shown to be meaningful in information retrieval~\cite{Salton:1986:IMI:576628}.\\ 
Adamic-Adar Coefficient (A) & $        A(u,v)=\sum\limits_{z\in\Gamma(u)\cap\Gamma(v)}\frac{1}{log(|\Gamma(z)|)}$ & Quantifies features shared by nodes $u$ and $v$ and weights rarer features more heavily~\cite{Adamic2003}. Interpreting this in the context of neighborhoods, the Adamic-Adar Coefficient can be used to characterize neighborhood overlap between nodes $u$ and $v$, weighting the overlap of smaller such neighborhoods more heavily.\\ 
Common neighbors (C) & $C(u,v)=|\Gamma(u)\cap\Gamma(v)|$ & Measures the number of shared neighbors between $u$ and $v$. Despite the simplicity of this index, Newman~\cite{Newman2001} documented that the probability of future links occurring in a collaboration network was positively correlated with the number of common neighbors.          \\ 
Average Path Weight (P) &         $P(u,v)=\frac{\sum\limits_{p\in \mathcal{P}_2(u,v)\cup \mathcal{P}_3(u,v)}w_{p}}{|\mathcal{P}_2(u,v)|+| \mathcal{P}_3(u,v)|}$ & Computes the sum of the minimum weights on the directed paths between $u$ and $v$ divided by the number of paths between $u$ and $v$, where only paths of length 2 and 3 are considered due to the large size of this network. We take $w_p$ to be the minimum weight of the edges in the path, in the spirit that a path's strength is only as strong as its weakest edge.\\ 
Katz (K)  & $K=\sum\limits_{n=1}^\infty \beta^{n}A^n $ & Computed as such, the Katz is a global index~\cite{Katz1953}. This series converges to $\left( I- \beta A \right)^{-1} - I,$ when $\beta < \max(\lambda(A))$. When $\beta \ll 1$ then $K$ approximates the number of common neighbors. Due to the size of our network and computational expense of this index, we truncate to $n=3$. We set $\beta=1$ because we are not concerned with convergence \& to emphasize the number of paths of length greater than two. Previous observations suggest that individuals who appear to be connected by a path length of $n$ in Twitter RRNs may actually be connected by a path of shorter length due to role of missing data~\cite{Bliss2012}. \\ 
Preferential Attachment (Pr) & $		Pr(u,v)=k_{u}\times k_{v}$ & Gives higher scores to pairs of nodes for which one or both have high degree. This index arose from the observation that nodes in some networks acquire new links with a probability proportional to their degree~\cite{Newman2001} and preferential attachment random growth models~\cite{barabasi2002evolution}.\\ 
Resource Allocation (R) & $R(u,v)=\sum\limits_{z\in\Gamma(u)\cap\Gamma(v)}\frac{1}{|\Gamma(z)|}$ &  Considers the amount of a given resource one node has and assumes that each node will distribute its resource equally among all neighbors~\cite{zhou2009predicting}.\\ 
Hub promoted Index (Hp) & $		Hp(u,v)=\frac{|\Gamma(u)\cap\Gamma(v)|}{\min\{k_{u},k_{v}\}}$ & First proposed to measure the topological overlap of pairs of substrates in metabolic networks, this index assigns higher scores to links adjacent to hubs since the denominator depends on the minimum degree of the two users~\cite{ravasz2002hierarchical}.\\  
Hub depressed Index (Hd) & $		Hd(u,v)=\frac{|\Gamma(u)\cap\Gamma(v)|}{\max\{k_{u},k_{v}\}}$ &  When one of the nodes has large degree, the denominator will be larger and thus $Hd$ is smaller in the case where one of the users is a hub~\cite{lu2011link}.   \\   
Leicht-Holme-Newman Index (L) & $		L(u,v)=\frac{|\Gamma(u)\cap\Gamma(v)|}{k_{u} k_{v}}$ & Measures the number of common neighbors relative to the square of their  geometric mean. This index gives high similarities to pairs of nodes that have many common neighbors compared to the expected number of such neighbors~\cite{lin1998information}.\\  
Salton Index (Sa) & $Sa(u,v)=\frac{|\Gamma(u)\cap\Gamma(v)|}{\sqrt{k_{u} k_{v}}}$  & Measures the number of common neighbors relative to their geometric mean~\cite{Salton:1986:IMI:576628}.\\ 
Sorenson Index  (So) & $		So(u,v)=\frac{2|\Gamma(u)\cap\Gamma(v)|}{k_{u}+k_{v}}$ & Measures the number of common neighbors relative to their arithmetic mean. This index is similar to $J$, however $J$ counts the number of (unique) nodes in the shared neighborhood. This index was previously used to establish equal amplitude groups in plant sociology based on the similarity of species~\cite{sorensen1948method}.\\ \hline 
\multicolumn{3}{ |l| }{\textbf{\textit{Individual characteristics similarity indices}}} \\ \hline  
Id similarity (I) & $I(u,v)=1- \frac{\left| Id(u) - Id(v) \right|}{\max\left\{\left|Id(a)-Id(b)\right|\right\}_{a,b \in V}}$ & In 2008, user ids were numbered sequentially and a user's id served as a proxy for the relative length of time since opening a Twitter account. Id similarity characterizes the extent to which two individuals adopt Twitter simultaneously. \\ 
Tweet count similarity (T) & $T(u,v)=1- \frac{\left| T(u) - T(v) \right|}{\max\left\{\left|T(a)-T(b)\right|\right\}}_{a,b \in V}$ & Tweet count $T(u)$ measures the number of Tweets we have gathered for node $u$ in a given week. Tweet count similarity quantifies how similar two individuals' tweet counts are, with 1 representing identical tweet counts and 0 representing dissimilar tweet counts.\\
Happiness similarity (H) & 	$\text{H}(u,v)=1-\frac{\left|h(u)-h(v)\right|}{\max\left\{\left|h(a)-h(b)\right|\right\}_{a,b \in V}}$ & Building on previous work~\cite{Dodds2011}, happiness scores ($h(u)$ and $h(v)$) are computed as the average of happiness scores for words authored by users $u$ and $v$ during the week of analysis.  \\ 
Word similarity (W) & $W(u,v)=1-\frac{1}{2}\sum\limits_{n=1}^{50000} |f_{u,n}-f_{v,n}|$ & From a corpus consisting of the 50,000 most commonly occurring words used in Twitter from 2008 through 2011~\cite{Dodds2011}, the similarity of words used by $u$ and $v$ is computed by a modified Hamming distance, where $f_{u,n}$ represents the normalized frequency of word usage of the $n$th word by user $u$. The value of $W(u,v)$ ranges from 0 (dissimilar word usage) to 1 (similar word usage)~\cite{Bliss2012}.\\  \hline
           \end{tabular}
	 \end{center}
\caption{The sixteen similarity indices chosen for inclusion in the link predictor. We define the \emph{neighborhood of node $u$} to be $\Gamma(u)=\{v\in V | e_{u,v} \in E\},$ where $G=(V,E)$ is a network, consisting of vertices ($V$) and edges ($E$). The degree of node $u$ is represented by $k_u$, the adjacency matrix is denoted by $A$, and a path of length $n$ between $u,v \in V$ is denoted as  $\mathcal{P}_n(u,v)$. }
\label{tab:indices}
\end{table*}
\subsection{Similarity indices}
Similarity indices capture the shared characteristics or contexts of two nodes. We briefly describe 16 similarity indices chosen for inclusion in our link predictor, but wish to emphasize that any number of other similarity indices may be chosen for inclusion in the evolutionary algorithm. The choice of which similarity indices to include may largely depend on the metadata one has about the nodes and interactions, as well as the size of the network. 

\begin{figure*}[ht!]
\centering
\subfigure[Jaccard]{\includegraphics[width=.24\linewidth]{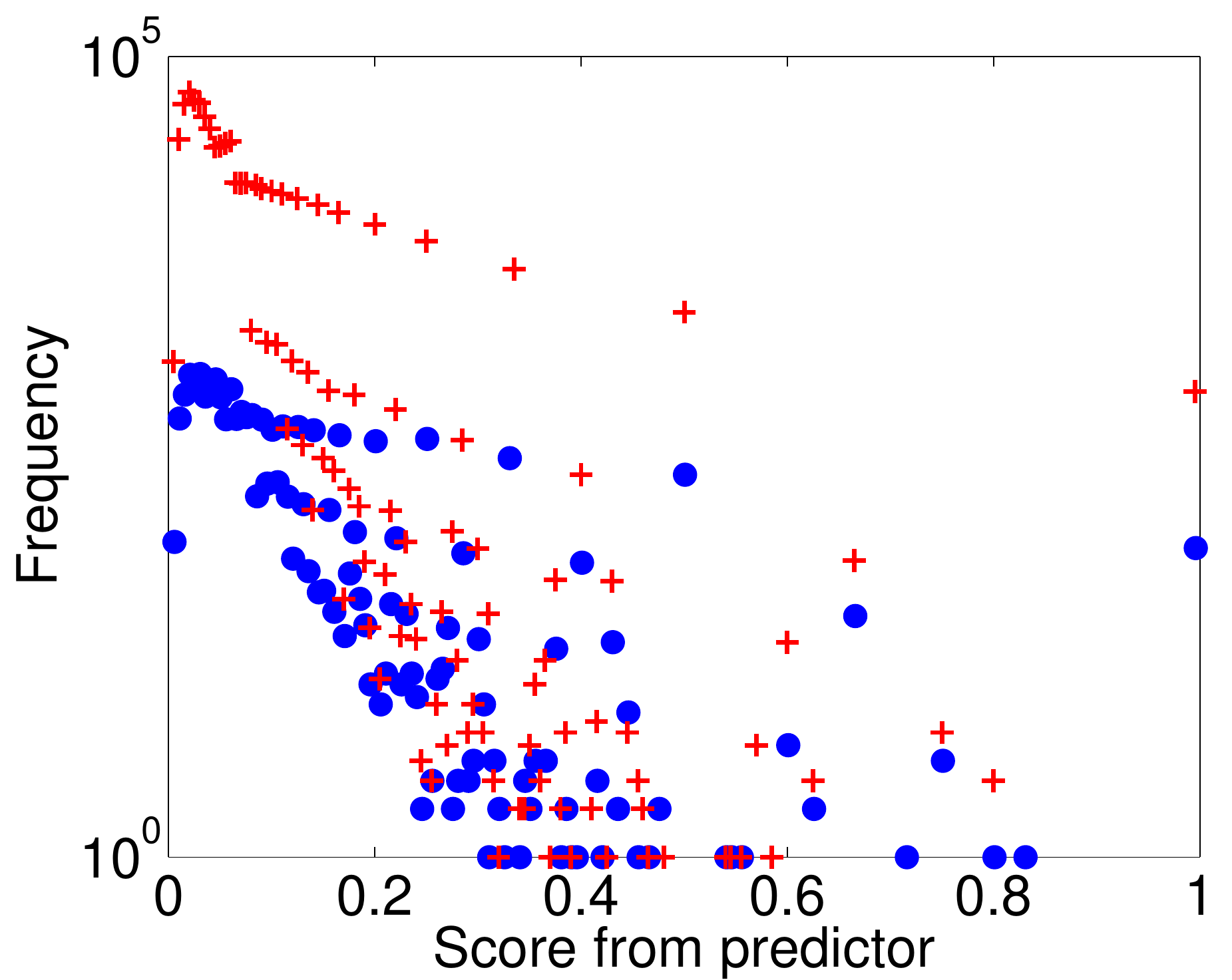}}
\subfigure[Adamic-Adar]{\includegraphics[width=.24\linewidth]{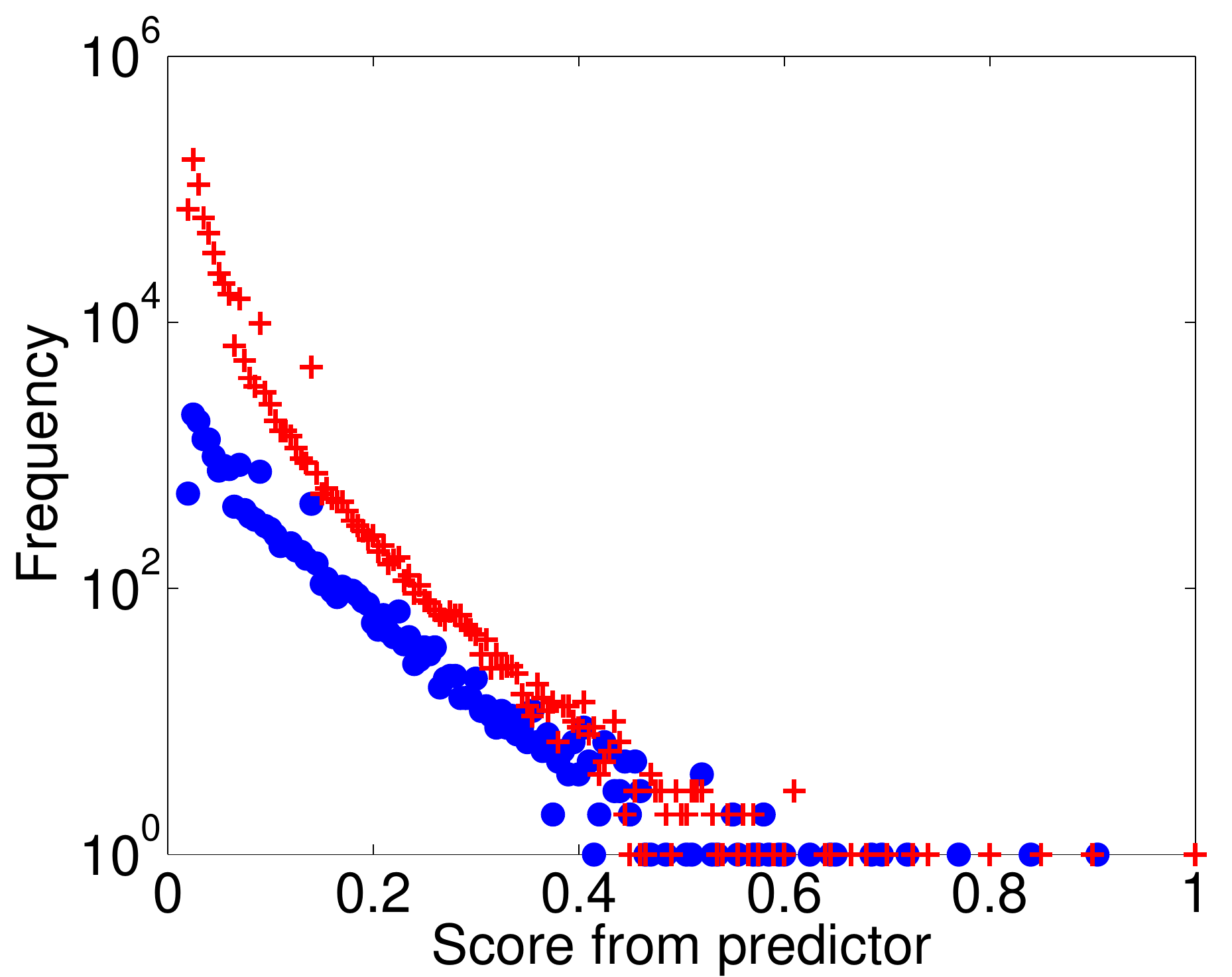}}
\subfigure[Common Neigh.]{\includegraphics[width=.24\linewidth]{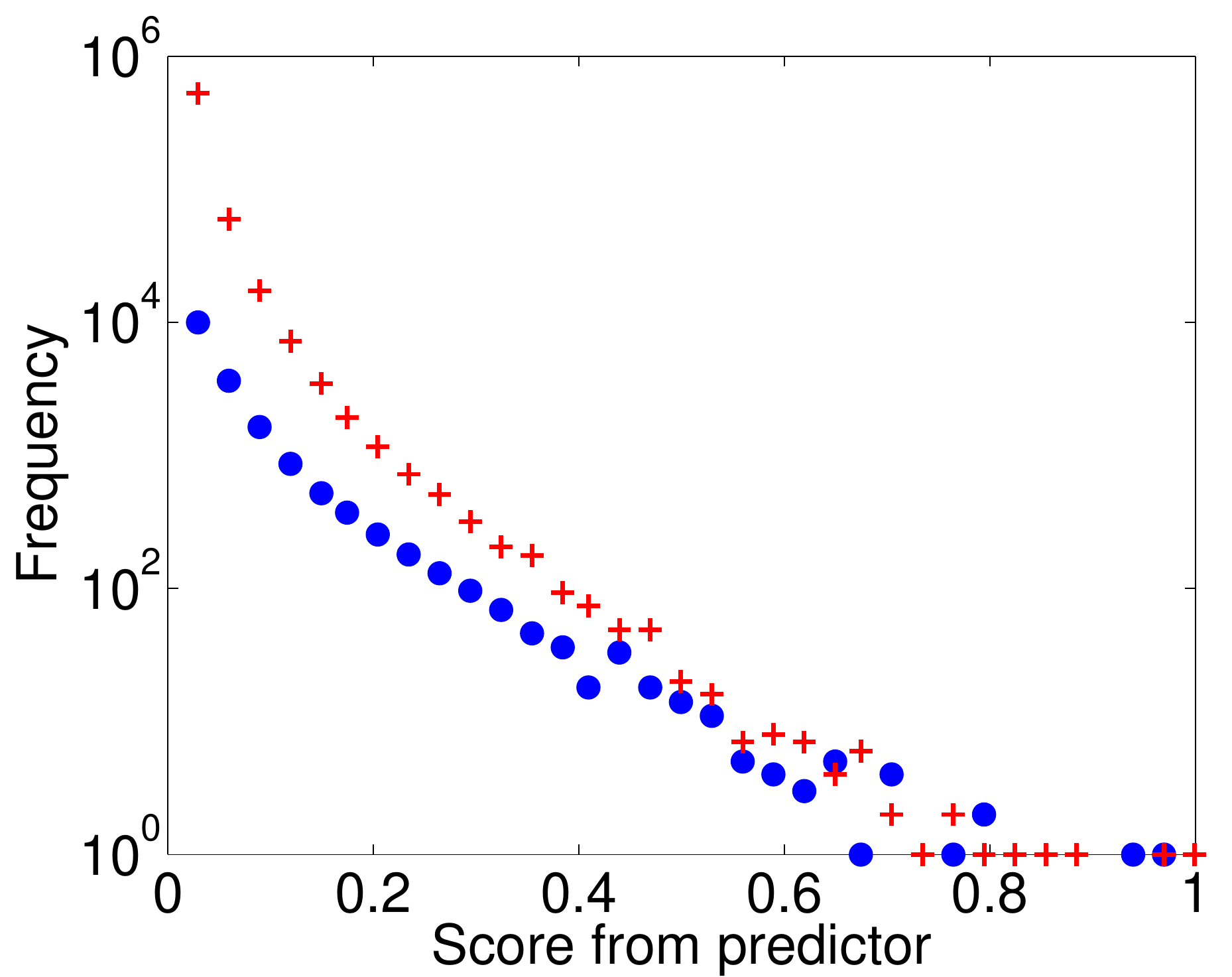}}
\subfigure[Paths]{\includegraphics[width=.24\linewidth]{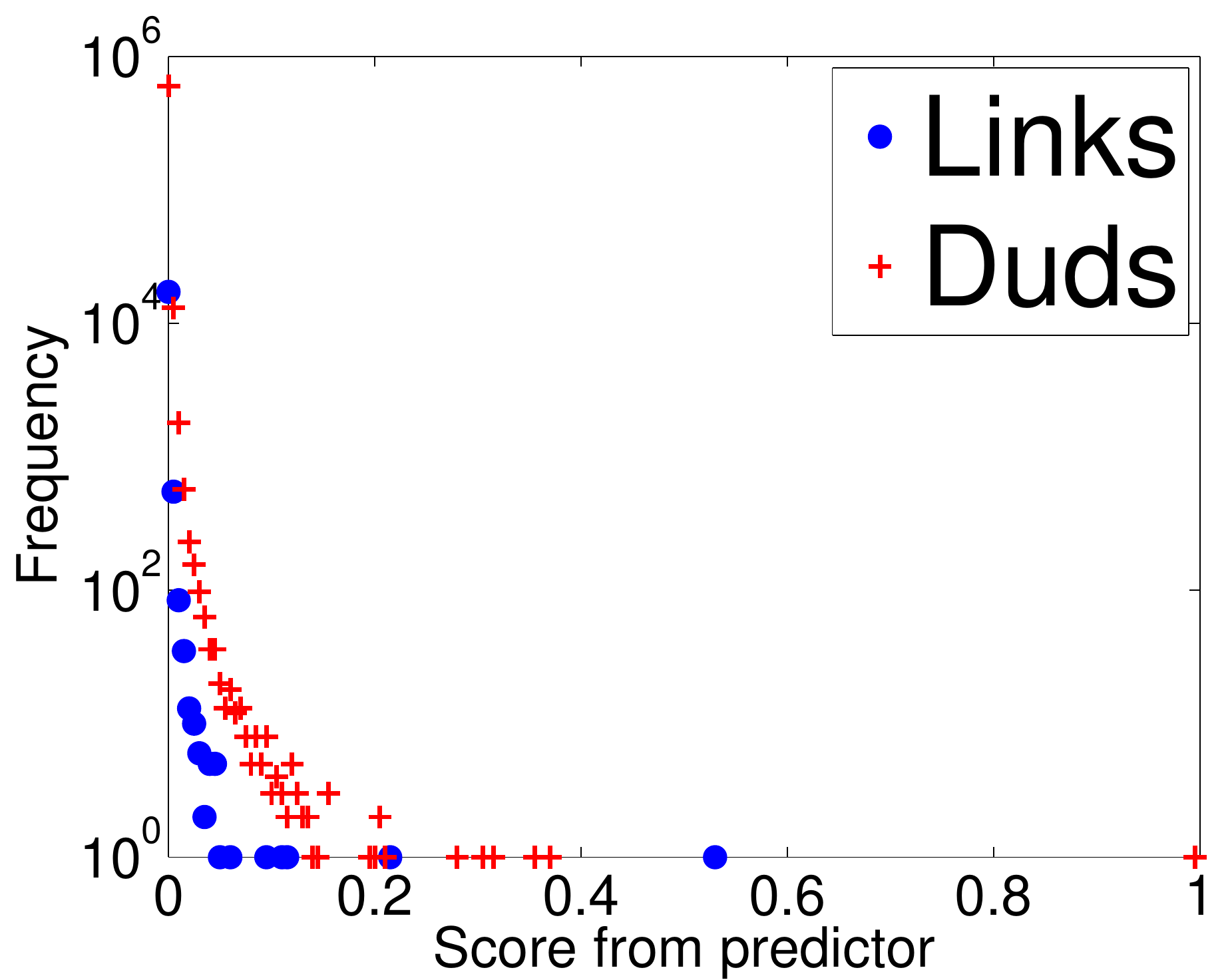}}\\
\subfigure[mod. Katz]{\includegraphics[width=.24\linewidth]{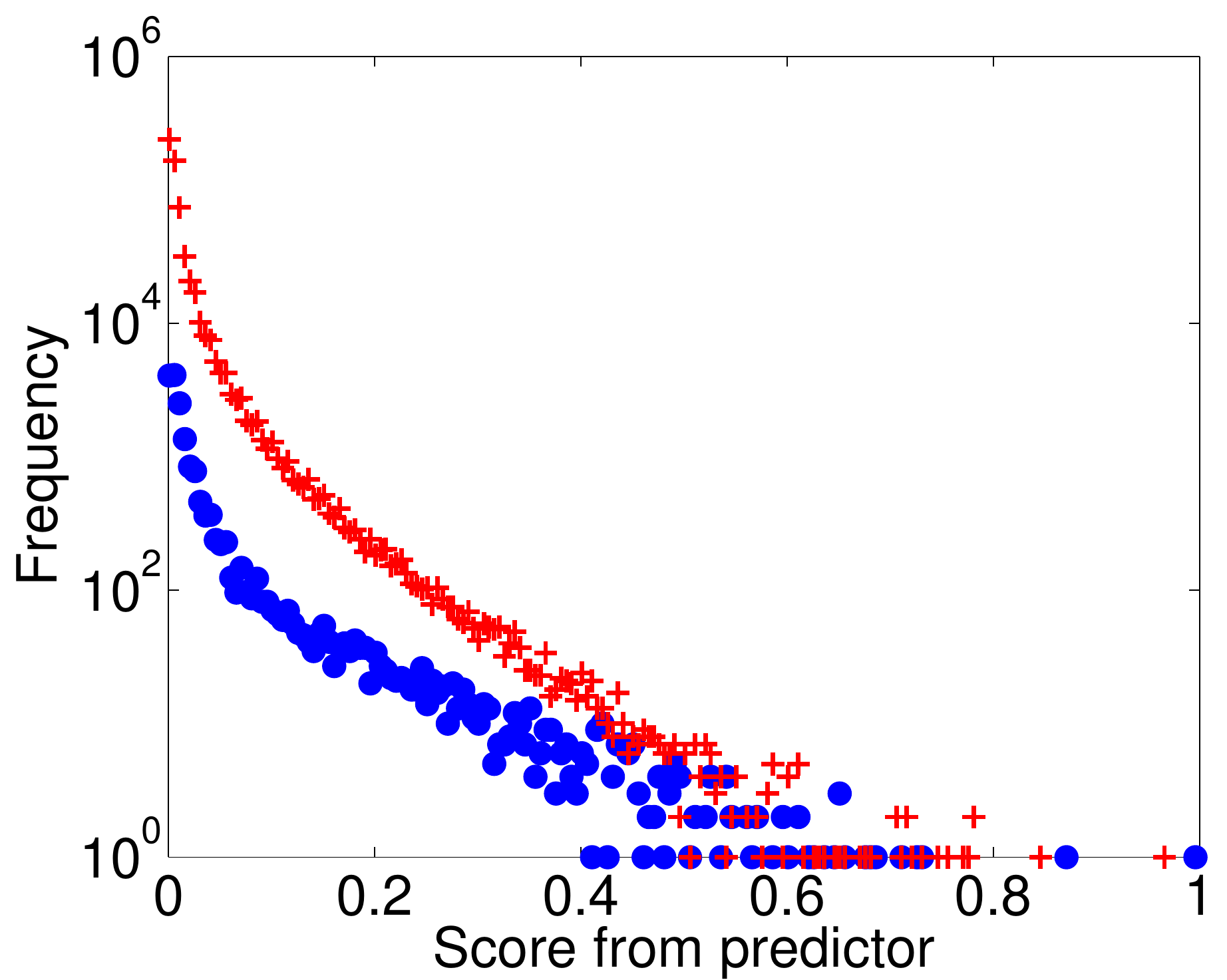}}
\subfigure[PrefAtt.]{\includegraphics[width=.24\linewidth]{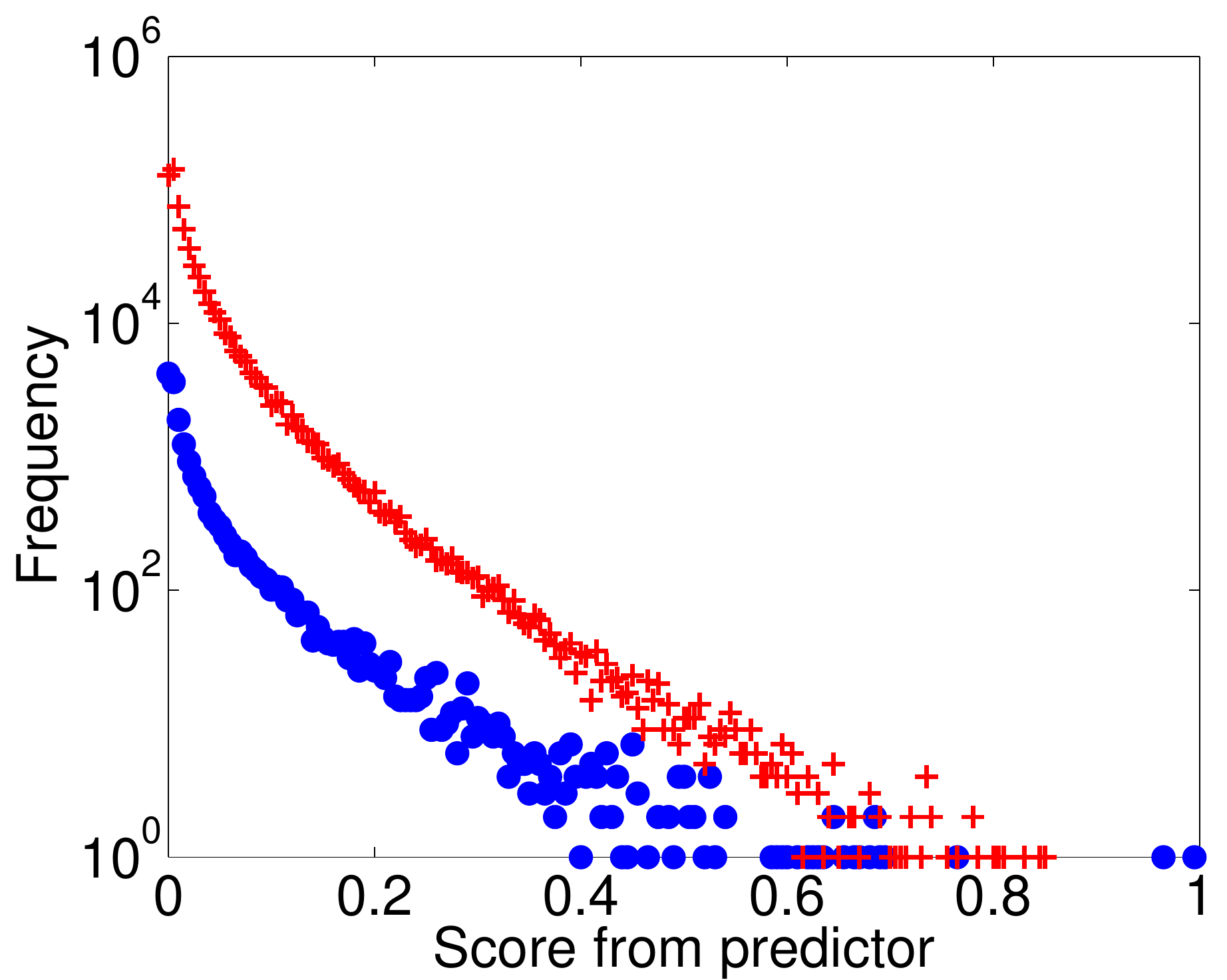}}
\subfigure[Resource All]{\includegraphics[width=.24\linewidth]{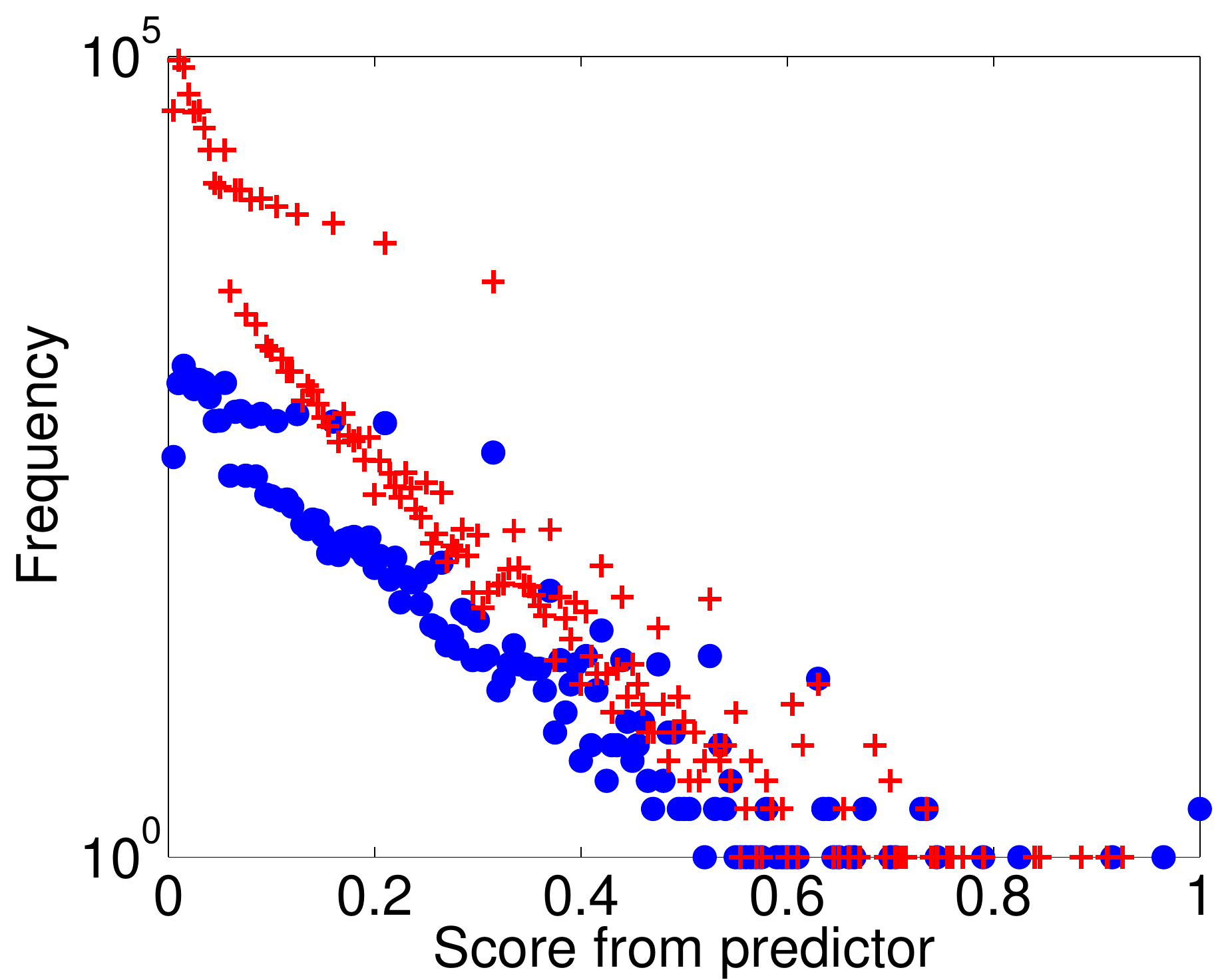}}
\subfigure[Hub depressed]{\includegraphics[width=.24\linewidth]{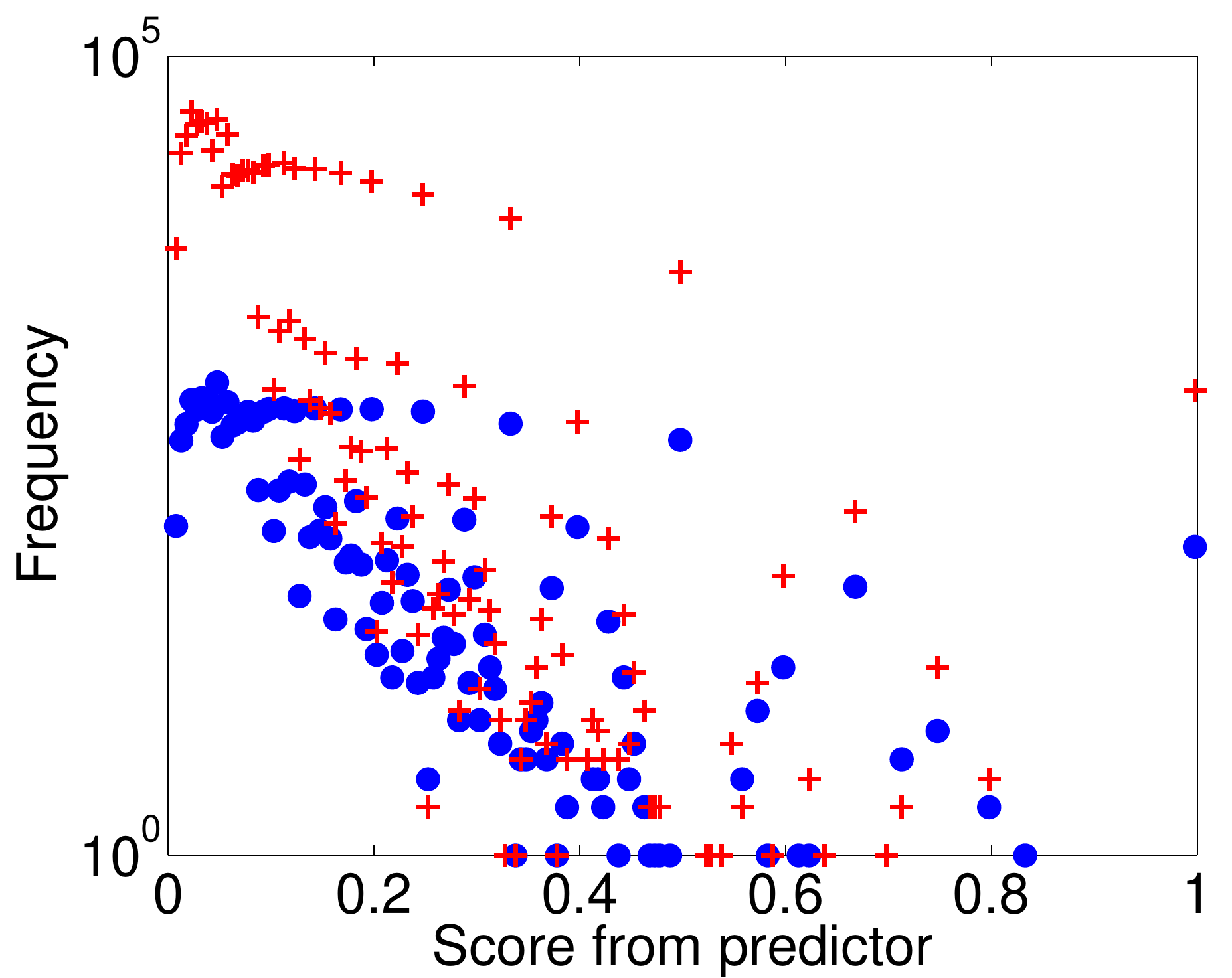}}\\
\subfigure[Hub promoted]{\includegraphics[width=.24\linewidth]{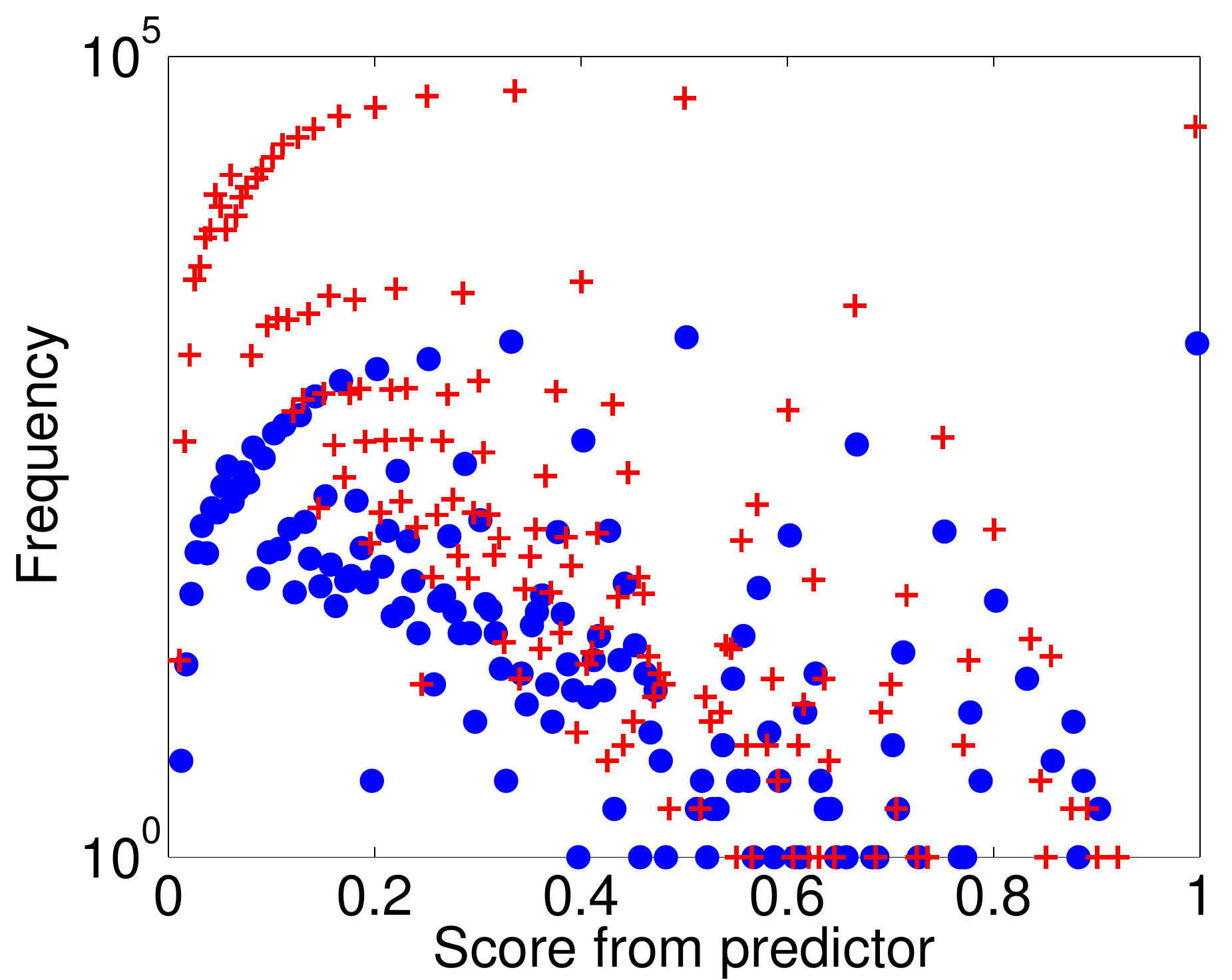}}
\subfigure[LHN]{\includegraphics[width=.24\linewidth]{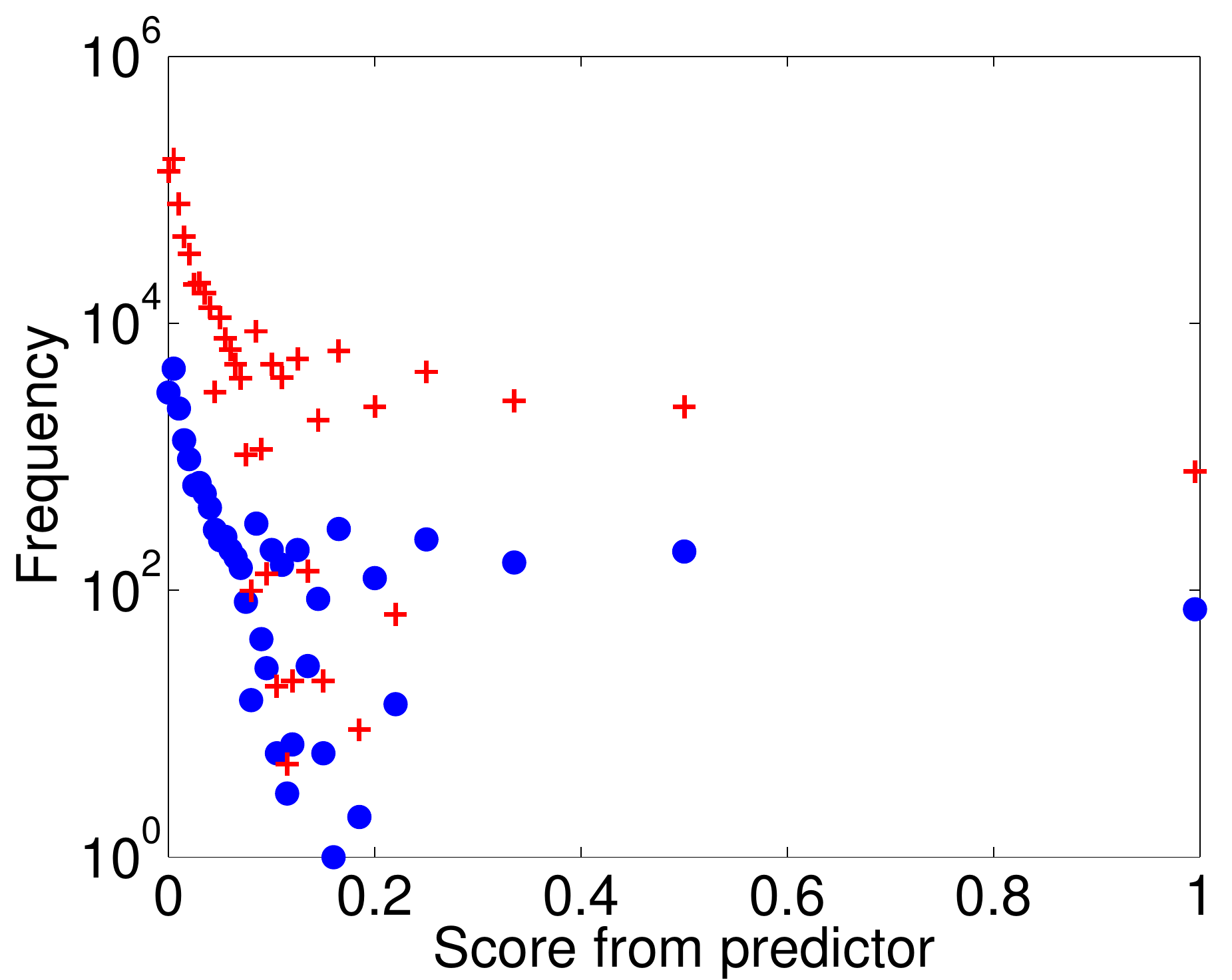}}
\subfigure[Salton]{\includegraphics[width=.24\linewidth]{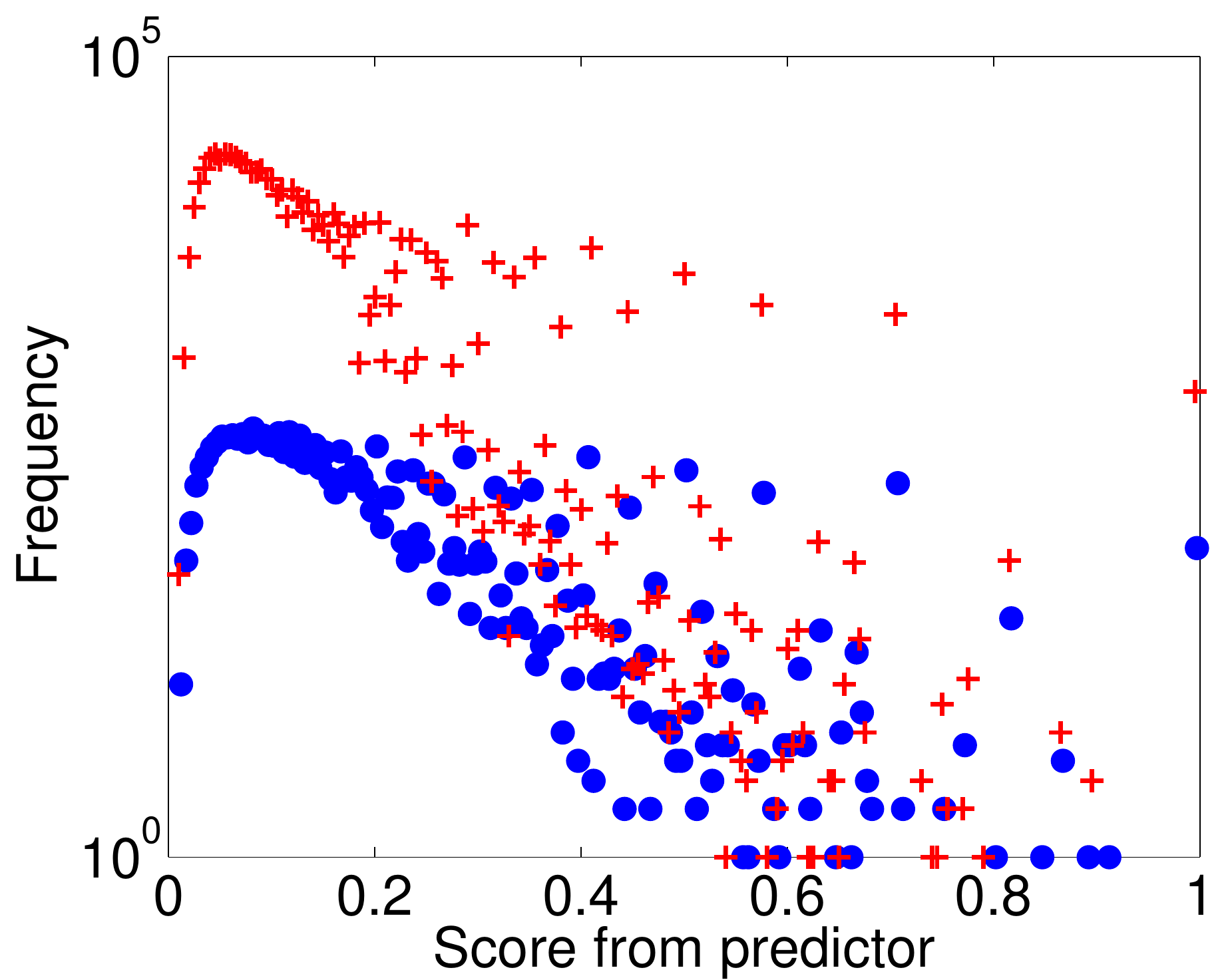}}
\subfigure[Sorenson]{\includegraphics[width=.24\linewidth]{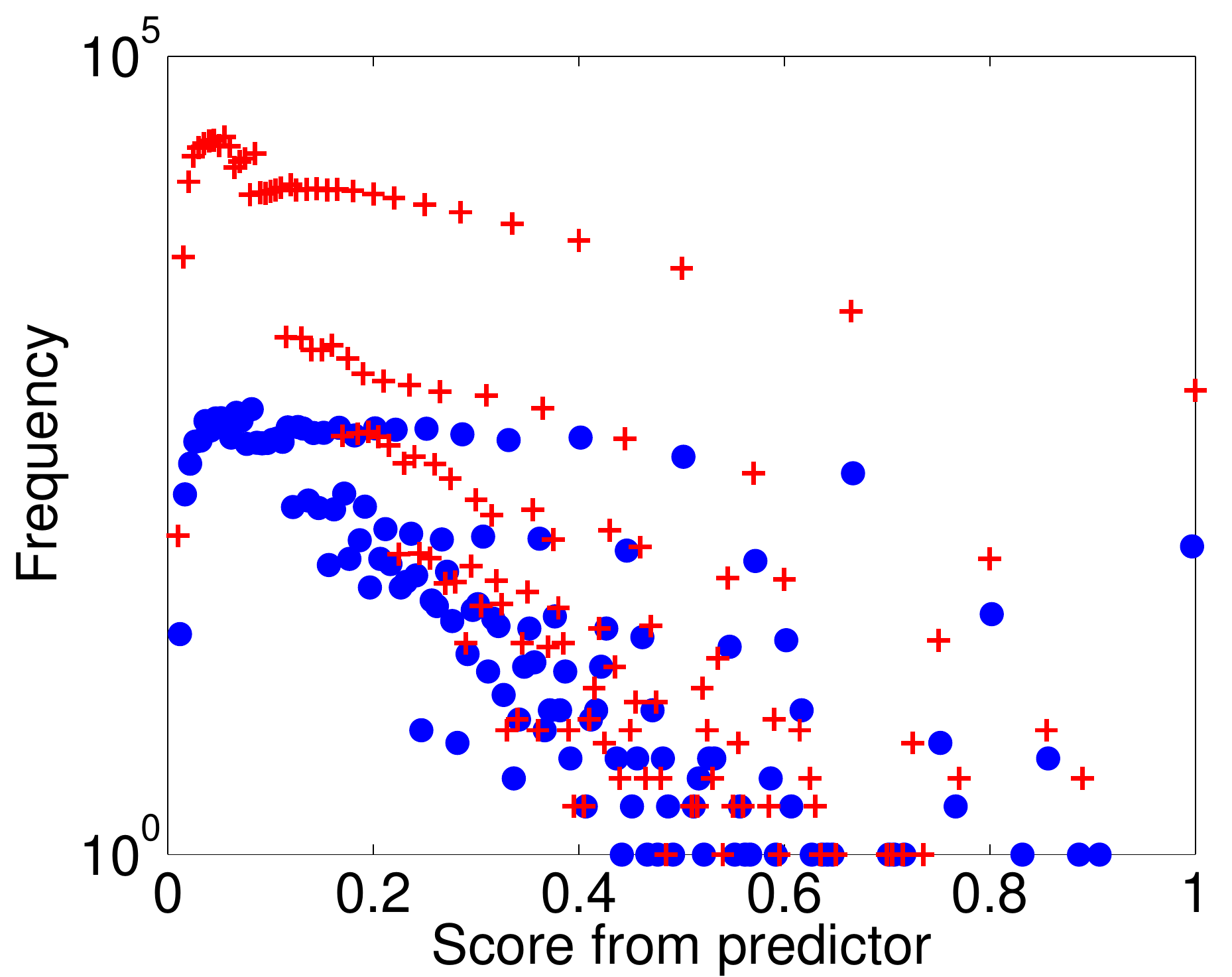}}\\
\subfigure[Twitter Id similarity]{\includegraphics[width=.24\linewidth]{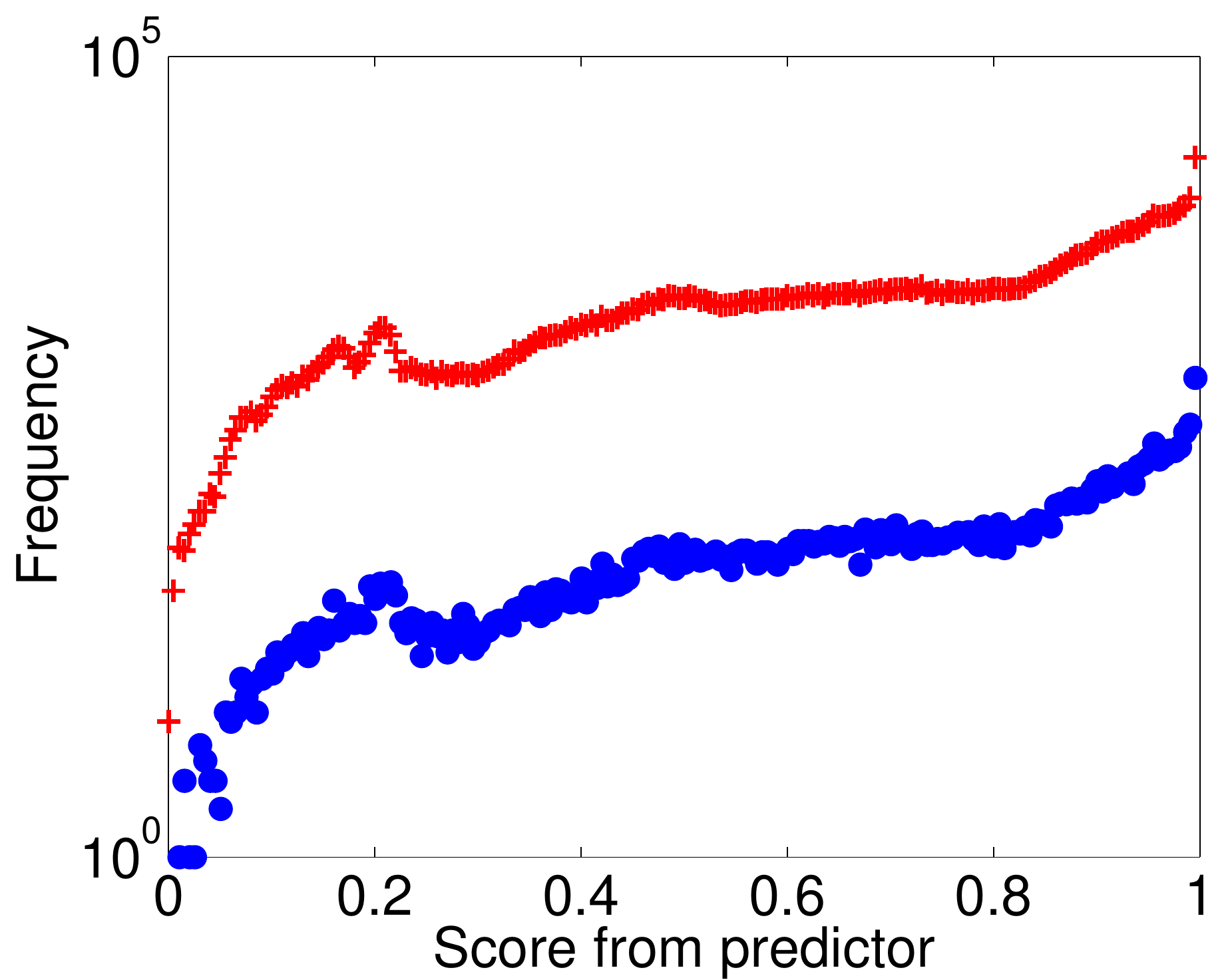}}
\subfigure[Tweet count similarity]{\includegraphics[width=.24\linewidth]{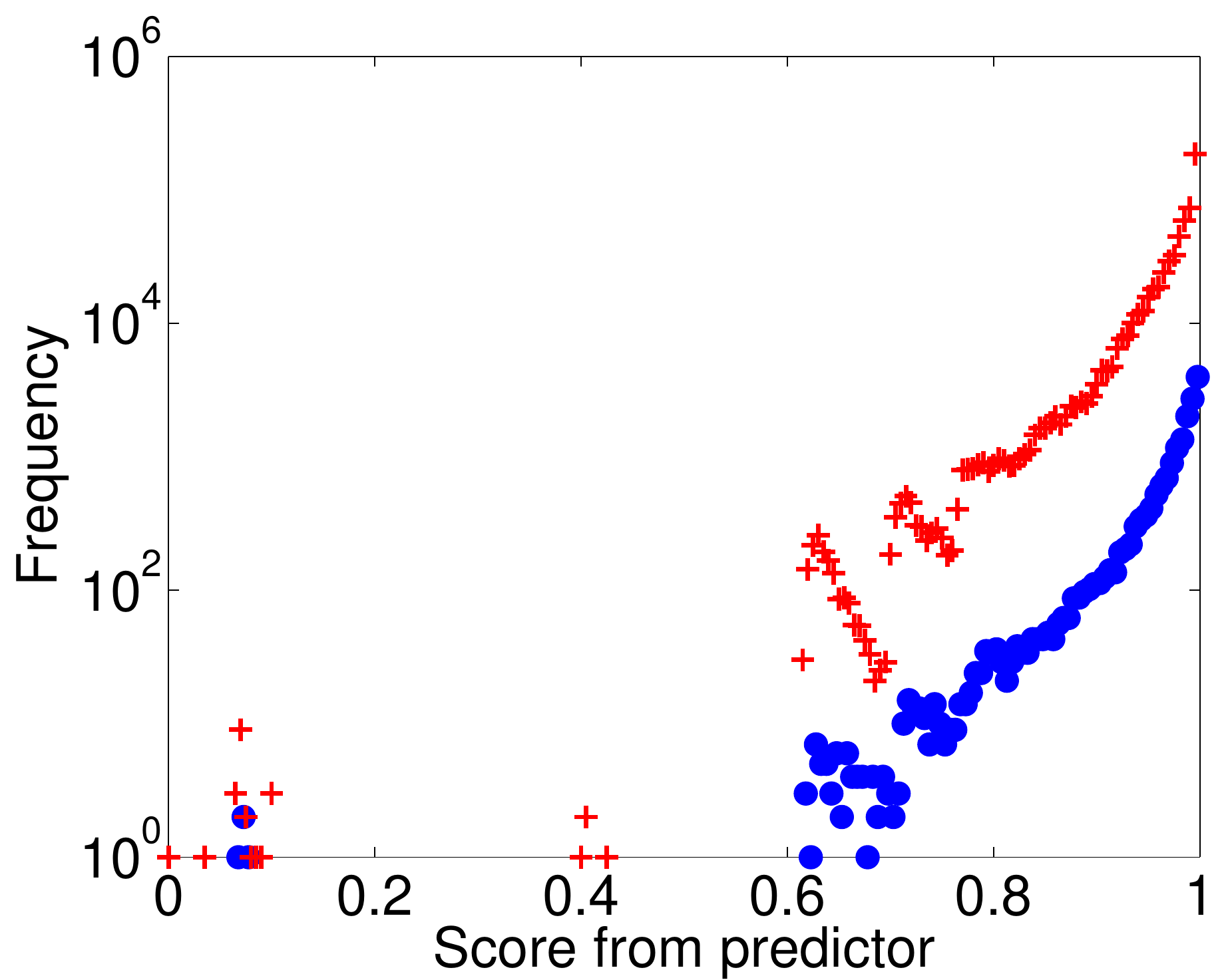}}
\subfigure[Happiness similarity]{\includegraphics[width=.24\linewidth]{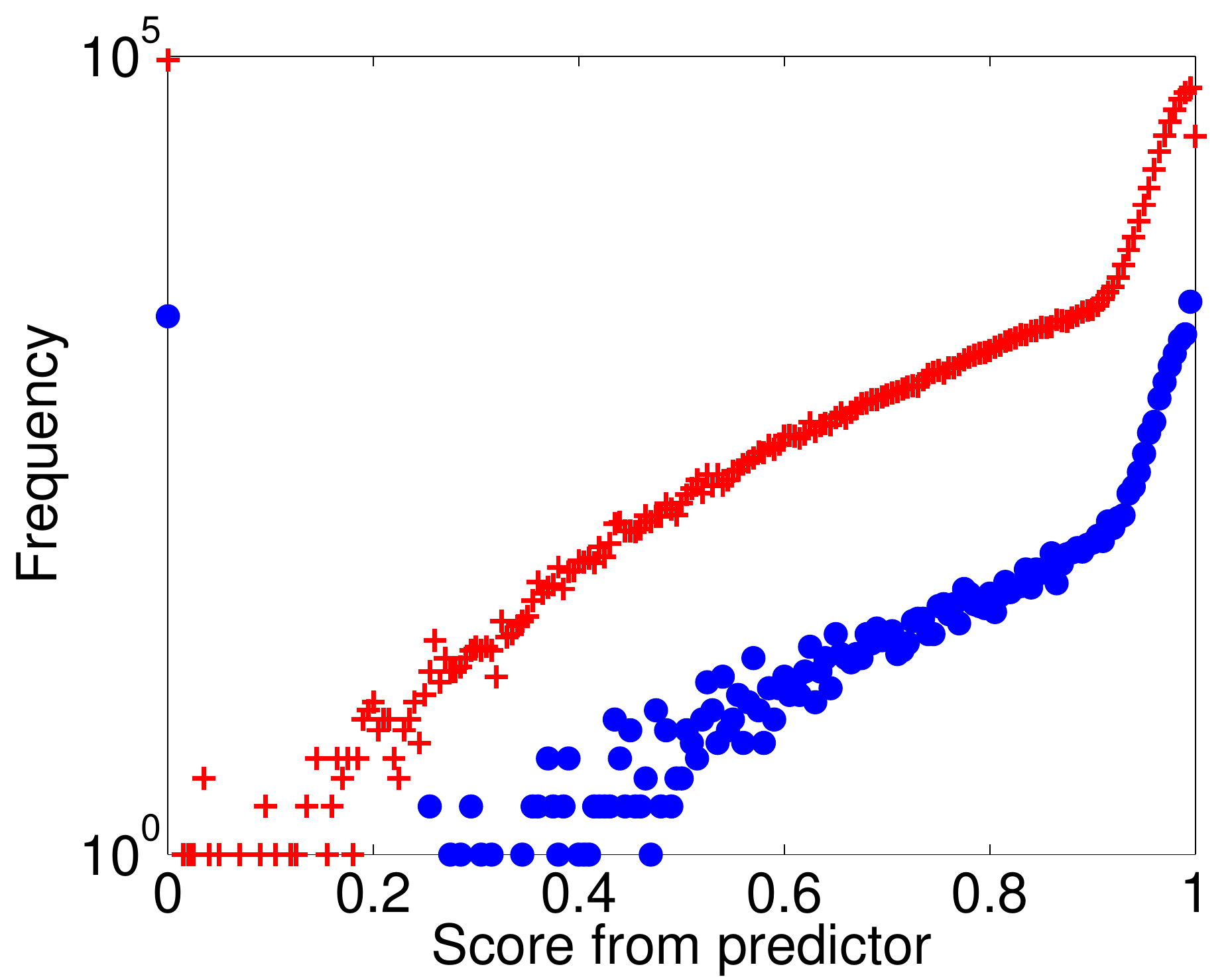}}
\subfigure[Word similarity]{\includegraphics[width=.24\linewidth]{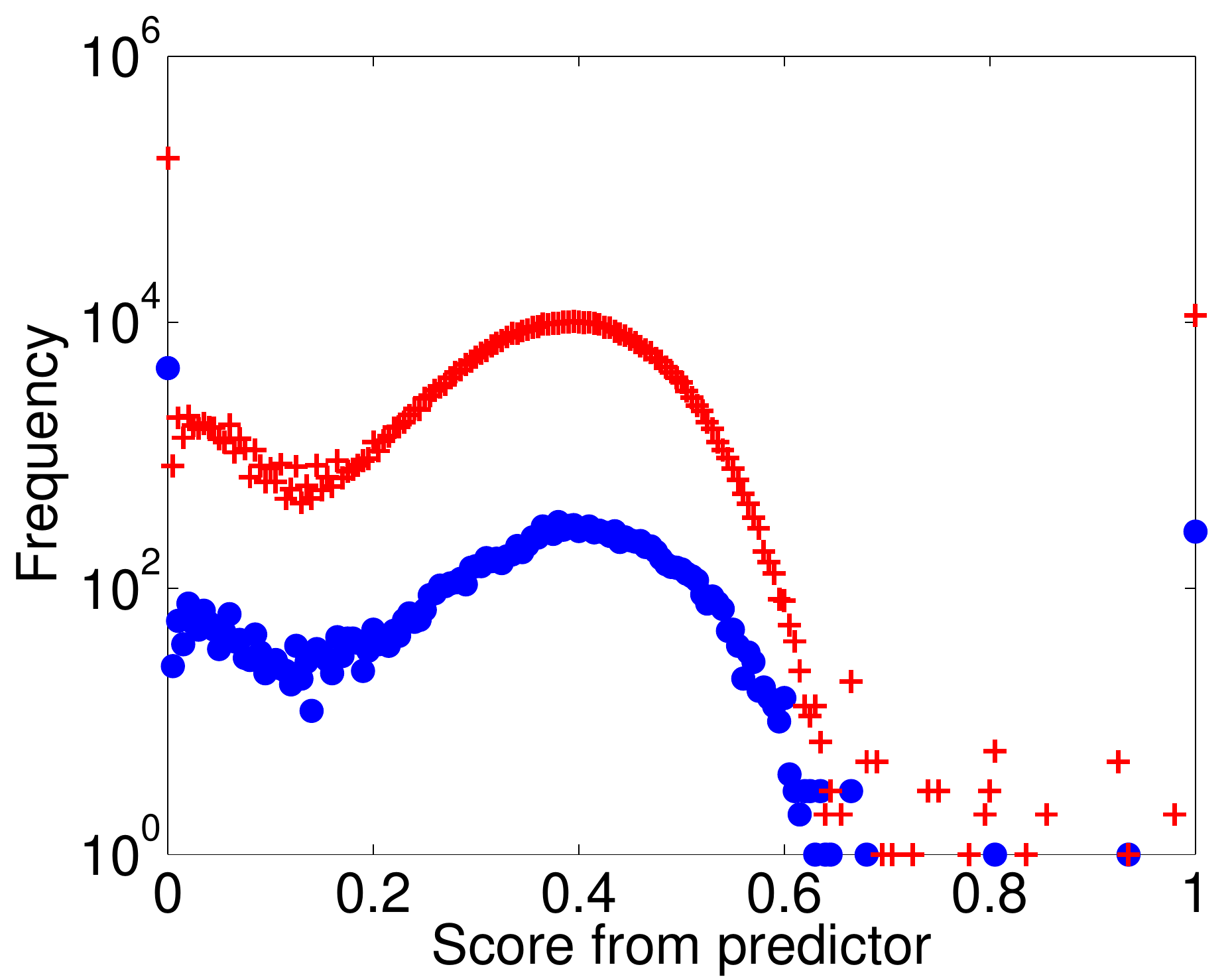}}
\caption{Scores for user-user pairs with path length two in Week 7, which exhibit a link (blue) and which did not (red) in Week 8. A higher score means that the user-user pair is more similar. For many indices, there are more ``duds'' than ``links'' for a given score. Indices for which there are ``links'' scoring higher than ``duds'' tend to exhibit a large, positive evolved coefficient (e.g., Adamic-Adar). }
\label{ref:signal_raw}
\end{figure*}
\begin{figure*}
\centering
\includegraphics[width=.7\textwidth]{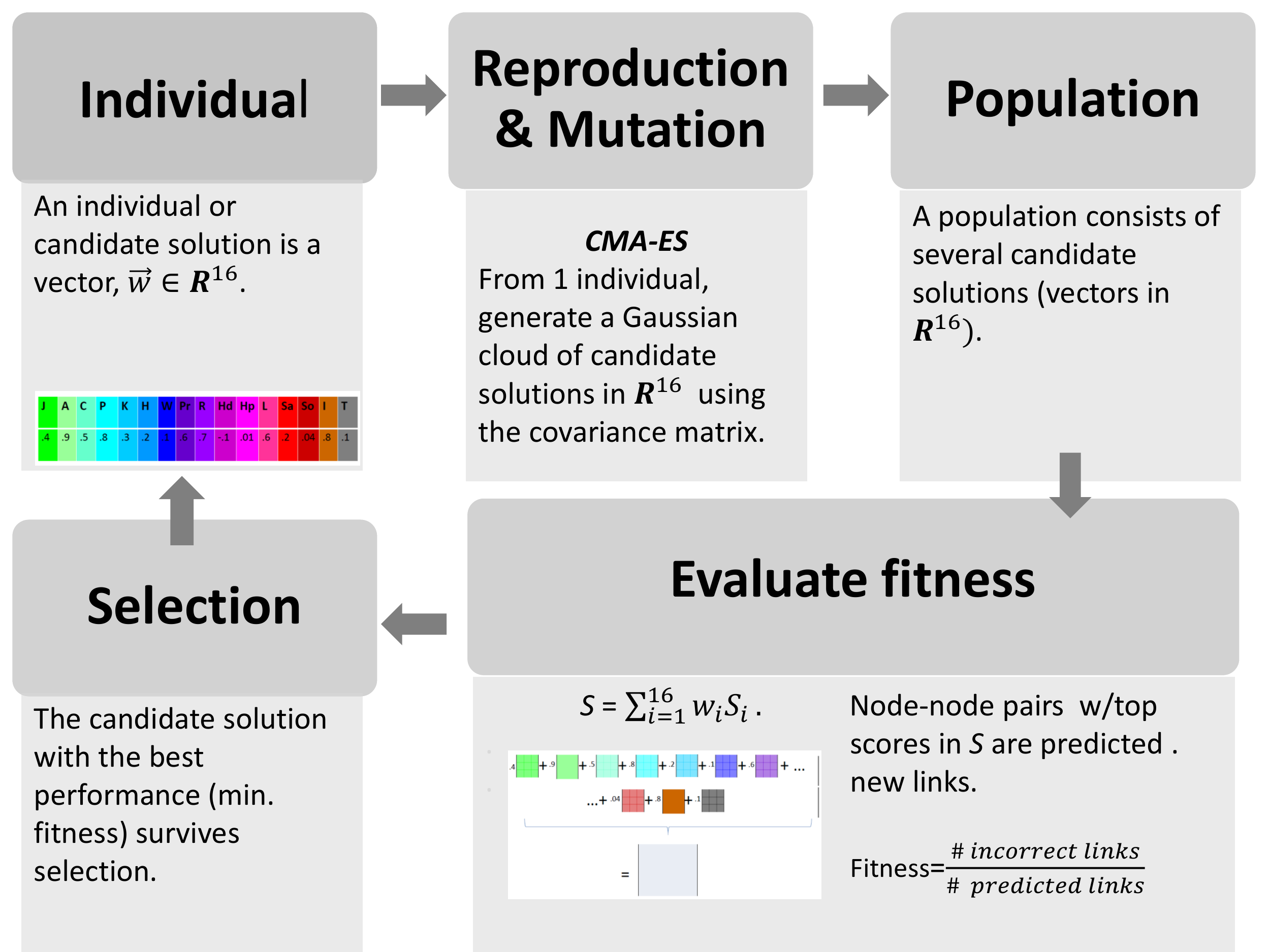}\\
\caption{Link prediction with CMA-ES. An individual (or candidate solution) is a vector, $\vec{w} \in \mathbb{R}^{n}$, where $n$ represents the number of indices used to constructor the predictor. We chose 16 such similarity indices. The initial individual is $\vec{w}_0$ where each entry is initialized between 0 and 1. From one individual, a Gaussian cloud of points in $\mathbb{R}^{16}$ is generated from the covariance matrix. This step mimics reproduction and mutation and creates a population of candidate solutions. Fitness is calculated for each candidate as the proportion of links incorrectly predicted, where a new link $e_{ij}$ is predicted if $s_{ij}$ is one of the top entries in matrix $S$. Selection occurs by taking the best candidate solution, $\vec{w} \in \mathbb{R}^{16}$. This one individual survives the generation and the cycle is repeated.}
\label{fig:linear_combo_visual}
\end{figure*}
Topological similarity indices may be characterized by local, quasi-local, or global measures. Since global similarity measures (i.e., Katz, SimRank, and Matrix Forest Index) are computationally laborious for large networks~\cite{lu2011link}, we forgo these measures in lieu of local topological indices. For node similarity we calculate four indices: Twitter Id similarity, tweet count similarity, word similarity and happiness similarity. All of these indices are described in Table~\ref{tab:indices}. We then rescale the computed scores to range from 0 to 1, inclusive, and store as $N \times N$ sparse matrices, hereafter referred to as $S_i$, for $i=1,2,\hdots,16$. 

We depict frequency plots for the computed similarity indices in Figure~\ref{ref:signal_raw}. These plots demonstrate that none of the similarity indices separate the newly formed ``links'' (user-user pairs who are separated by a minimal path of length 2 at $t$ and a path of length 1 at $t+1$) and ``duds'' (user-user pairs who are separated by a minimal path of length 2 at $t$ and a path of length $\delta \neq 1$ at $t+1$). This lack of separation is one indication that a predictor which combines information from several indices may improve link prediction efforts. Figure~\ref{ref:signal_raw} also reveals that the manner in which the predictors should be combined is not as straightforward as one might envision. For example, some similarity indices, such as Adamic-Adar (Fig.~\ref{ref:signal_raw}b) and Resource Allocation (Fig.~\ref{ref:signal_raw}i) show potential for differentiating links and duds. Other indices, such as Twitter Id similarity (Fig.~\ref{ref:signal_raw}o) maintain a greater number of duds than links, across all scores. This is a result of the large class imbalance between the number of potential user-user pairs for new links and the actual numbers of new links formed, a common occurrence in large, sparse networks. 

\subsection{Evolutionary algorithm}
Evolutionary algorithms take inspiration from biological systems whereby individuals representing candidate solutions evolve over generational time via selection, reproduction, mutation, and recombination (Fig.~\ref{fig:linear_combo_visual}). In our task, we construct a linear combination of similarity indices, $S_i$, and use an evolutionary strategy to evolve the coefficients, $w_i$, used in computing a score matrix, $S$,  
\begin{equation}
S=\sum^{16}_{i=1} w_i S_i,
\label{eq:linear_comb}
\end{equation}
for which the minimum error in link prediction is desired. 

Our task is essentially an optimization problem. Our choice for CMA-ES stems from its efficiency in finding real valued solutions in noisy landscapes~\cite{suttorp2009efficient}. In contrast to gradient descent approaches for finding optimal solutions, CMA-ES is not reliant on assumptions of differentiability nor continuity of the fitness landscape. Our method requires no heuristics, which is an advantage over many existing supervised learning methods (e.g., SVM) that require extensive parameter tuning and kernel selection~\cite{hansen2001completely}. Additionally, our method is flexible and allows for any similarity index to be substituted into or added to the evolutionary algorithm. Ideally, the transparency of the evolved ``best'' predictors will help illustrate possible driving mechanisms behind the network's evolution. This method is also one of the best evolutionary algorithms for finding optima of real valued solutions due to its fast convergence.\footnote{Here, we refer to fast convergence in generational time. The CPU time for one generation of our CMA-ES implementation for link prediction was 13 seconds.} We refer the interested reader to~\cite{hansen2005cma} for more detail regarding the CMA-ES algorithm.

Figure~\ref{fig:linear_combo_visual} outlines our implementation of CMA-ES for link prediction. Before employing the evolutionary algorithm, all similarity indices are computed and stored as $N \times N$ sparse matrices, $S_i$ for $i=1,2,\hdots, 16$. The evolutionary algorithm begins with a candidate solution termed an ``individual'' in the language of evolutionary computation. Entries of $\vec{w}$ are initially set to real values between 0 and 1 chosen from a uniform random distribution. These values are not constrained during evolution.  Using CMA-ES with both rank-1 and rank-$\mu$ updates\footnote{Briefly, rank-1 updates utilize information about correlations between generations, which is helpful for evolution with small populations of candidate solutions. Rank-$\mu$ updates utilize information from the current generation, which helps speed up the algorithm for large populations.} we evolve $\vec{w}=\langle w_1,w_2,\hdots, w_{16}\rangle \in \mathbb{R}^{16}$ over 250 generations~\cite{hansen2001completely}.  At each generation, a population of candidate solutions is selected from a multivariate Gaussian cloud\footnote{We use the default population size of $4+\lfloor3\log(m)\rfloor$, for solutions in $\mathbb{R}^m$, from Hansen's source code available at \url{https://www.lri.fr/~hansen/cmaes_inmatlab.html} (last accessed on October 1, 2012). Increasing the population size did not improve our results.} surrounding the ``individual'' surviving the previous generation. 

Each candidate solution in the ``population'' is assessed for fitness and the individual with the best fitness survives the generation. The standard implementation of CMA-ES selects the ``best solution'' as that which minimizes fitness. As such, our fitness function\footnote{for each of four fitness functions $\text{fitness}_{20}$, $\text{fitness}_{200}$, $\text{fitness}_{2000}$, $\text{fitness}_{20000}$ where the subscript denotes the top $N$ scoring user-user pairs (e.g., predicted links). By incorporating fitness functions which operate at different scales, we investigate the sensitivity of the top $N$ on the link predictor's performance in validation.} computes the link prediction error for each $\vec{w} \in \mathbb{R}^{16}$. One of the difficulties with CMA-ES is the potential to be trapped in local optima. To avoid this, we perform 100 restarts, a technique suggested by Auger and Hansen~\cite{auger2005restart}. 

\subsection{Cross referencing links}
From the 100 best solutions evolved via CMA-ES for each of the four fitness functions (e.g., where the top 20, 200, 2000 or 20000 scores are used to predict future links) we cross-reference the top $N$ scoring user-user pairs. The user-user pairs which are most heavily cross-referenced (i.e., links which most models agree upon) are those for which we predict a link. In addition to the 400 best evolved predictors, we also feed in information from the Resource Allocation similarity index when prediction top $N<$10 because of the high performance of this index for predicting the top $10$ or fewer links on training sets.  

\section{Results}
\begin{figure}[ht!]
\includegraphics[width=\linewidth]{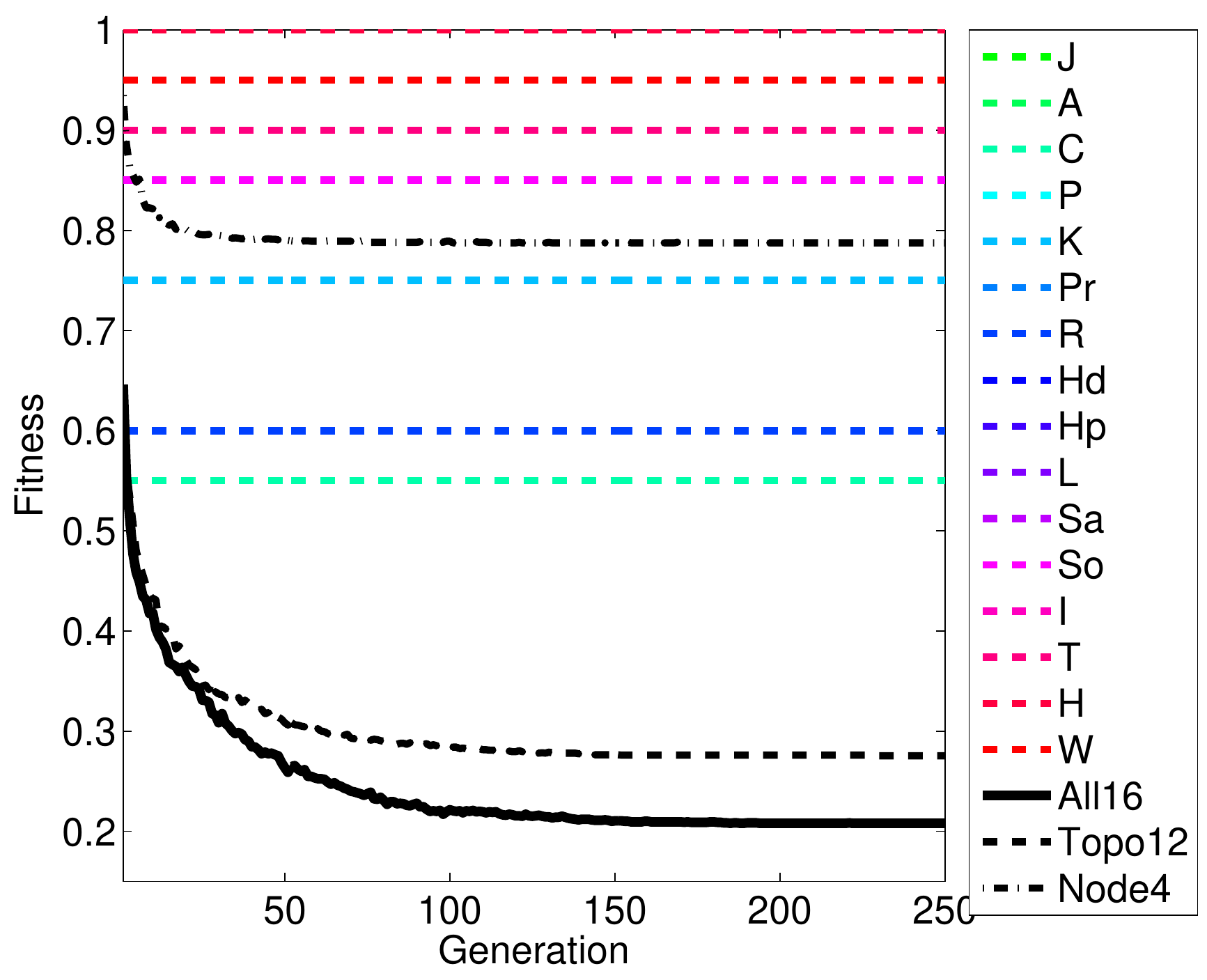}
\caption{Mean best fitness computed from 100 simulations of CMA-ES for training on the new links that occur in Week 8 (i.e., links present in Week 8 that were not present in Week 7) using $\text{fitness}_{20}$. The evolutionary algorithm seeks to minimize fitness (i.e., minimize the proportion of falsely predicted links).  We compare each individual index (shown in color), along with the three evolved predictors (shown in black): ``all16'' (all 16 indices), ``topo12'' (12 topological indices), and ``node4'' (4 individual similarity indices). The ``all16'' predictor performs the best, followed by the ``topo12'' predictor.}
\label{fig:example_fitness}
\end{figure}

Our overall finding is that the evolved predictor consisting of all sixteen similarity indices outperformed all other combined and individual indices on the training data when training occurred on a given week's RRN. In Figure~\ref{fig:example_fitness}, we present the results for $\text{fitness}_{20}$ during training on new links formed from Week 7 to Week 8. 
The solid black curve depicting the ``all16'' predictor shows that while the average fitness at generation 1 for the 100 candidates was far worse ($\approx 0.65$) than several similarity indices such as Adamic-Adar ($\approx 0.55$), Common neighbors ($\approx 0.55$) and Resource Allocation $\approx 0.60$), convergence to a far better set of solutions occurred within 100 generations ($\approx .22$). The combination of the twelve topological indices outperformed all individual indices, but was outperformed by the all16 predictor. This difference is most pronounced for the top $N$=20 cases, however this trend holds true for the other fitness functions (Appendix, Fig.~\ref{fig:fitness_week}). 
\begin{figure}[ht!]
\centering
\subfigure[100 evolved best ``inidividuals'' from CMA-ES]{\includegraphics[width=\linewidth]{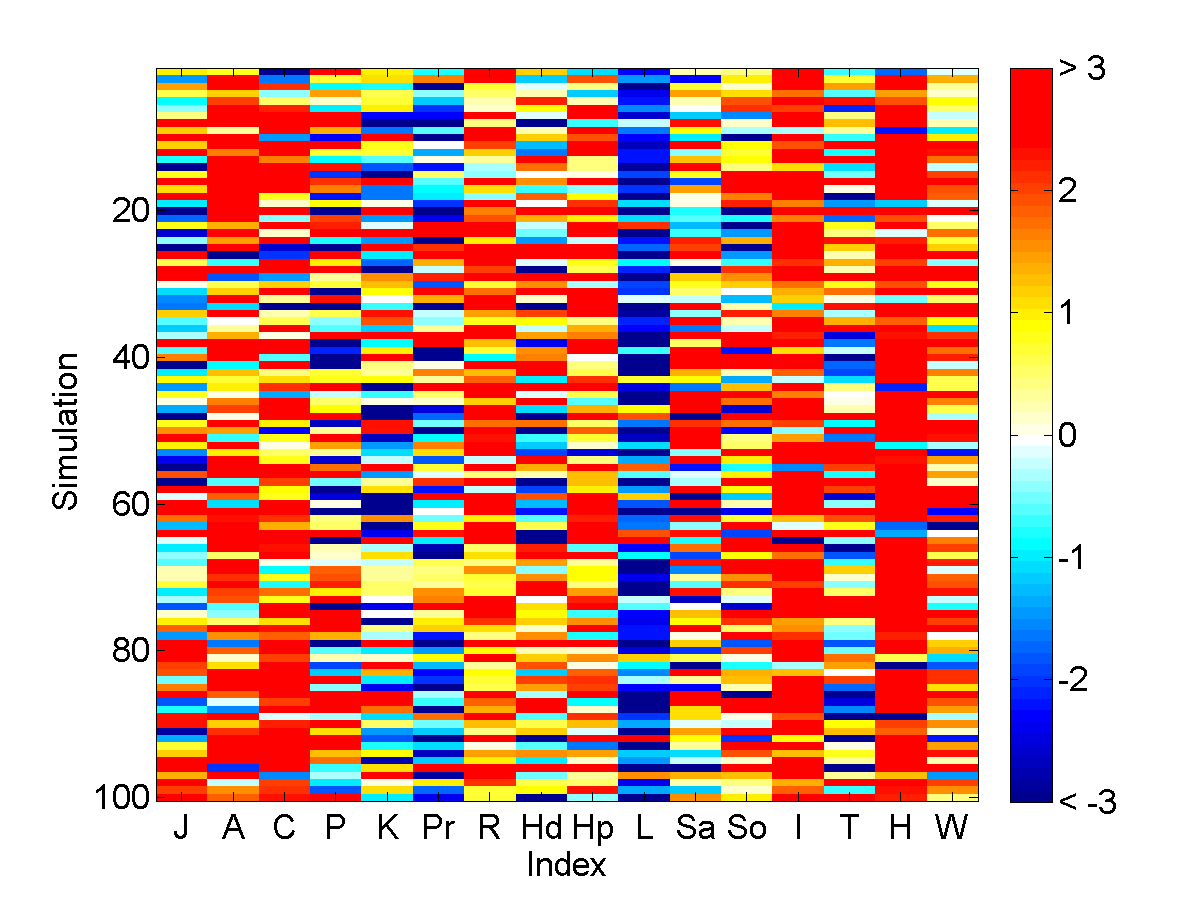}}\\
\subfigure[Frequency plot for ranked coefficients, $w_i$ corresponding to similarity indices]{\includegraphics[width=\linewidth]{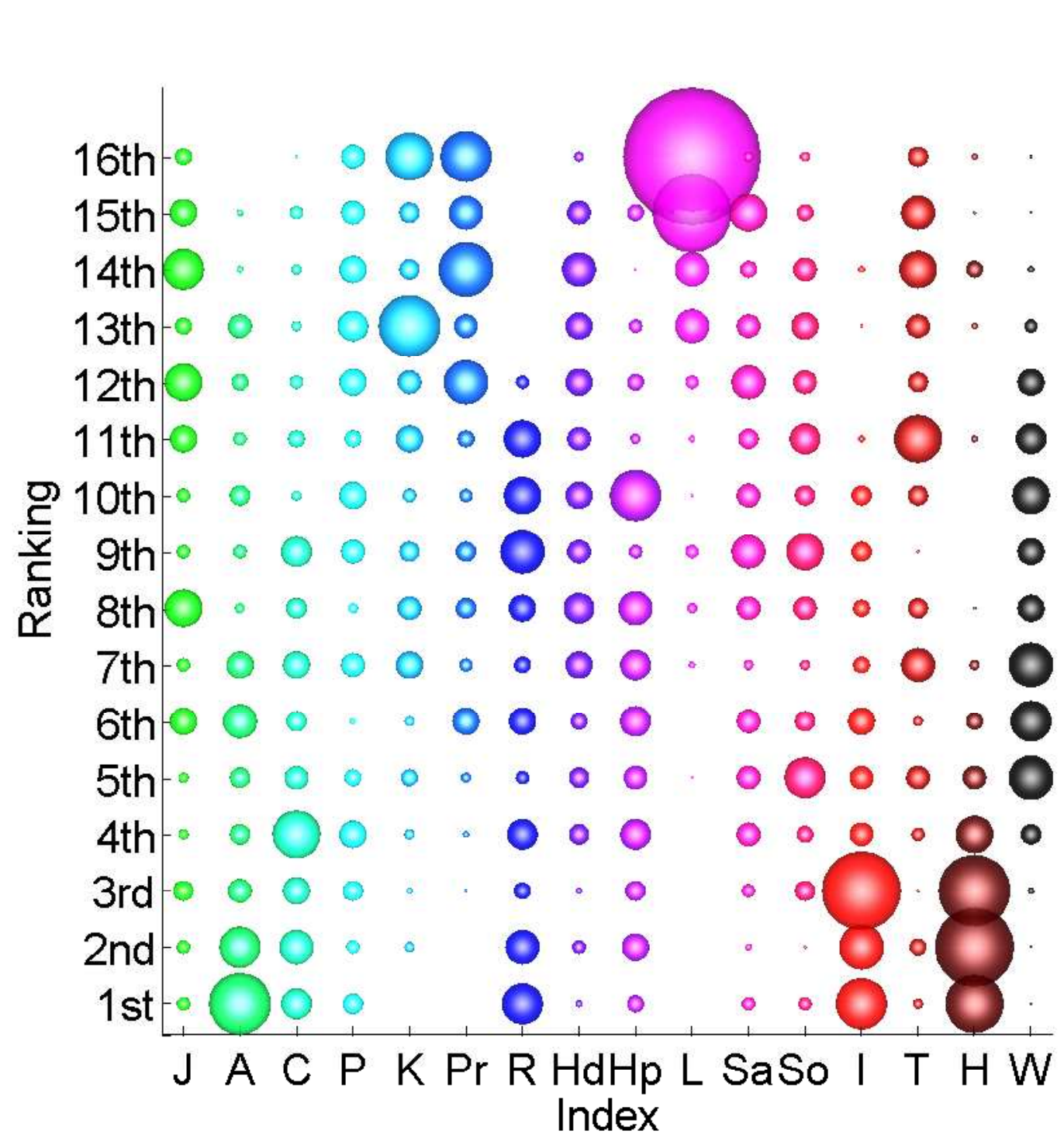}}
\caption{(a.) Presentation of the best solutions evolved from each of 100 simulations using $\text{fitness}_{20}$ and the ``all16'' predictors to predict new links that occurred from Week 7 to 8.  (b.) Frequency plot of ranked coefficients from (a.), where 1st place represents large, positive coefficients and 16th place represents large, negative coefficients. Disk size indicates the fraction of times an index received a given ranking. Adamic-Adar, Happiness similarity, Resource Allocation and Twitter Id similarity were the most commonly occurring indices ranked 1st (largest, positive) coefficient, and LHN often evolved to the largest, negative coefficient. This suggests possible mechanisms which may have been driving the evolution of the network during this time period. J=Jaccard, A=Adamic-Adar, C=Common neighbors, P=Paths, K=Katz, Pr=Preferential attachment, R=Resource allocation, Hd=Hub depressed, Hp=Hub promoted, L=Leicht-Holme-Newman, Sa=Salton, So=Sorenson, I=Twitter id similarity, T=Tweet count similarity, H=Happiness similarity, W=Word similarity.}
\label{fig:all_solns_week_78_top20}
\end{figure}

Our interest extends beyond an analysis of the proportion of links correctly predicted. We reveal the constituents of our link predictor ($\vec{w} \in \mathbb{R}^{16}$) as a means to gain an (initial) understanding of the mechanisms which may be driving the evolution of Twitter RRNs. In this spirit, we present two visualizations which capture this information. For illustration purposes, we highlight the results from Week 8, using a fitness function which selects the top 20 scores as new links, in Figure~\ref{fig:all_solns_week_78_top20}. 

Figure~\ref{fig:all_solns_week_78_top20}a shows all 100 solutions which evolved after 250 generations of CMA-ES, $\vec{w}$, as horizontal rows. The $i$th column signifies the $w_i$ coefficient used in the linear combination of the weights. The color axis reveals the value of $i$th coefficient. Several trends are worth noting here. First, there is considerable variability between the 100 evolved best candidates. Second, despite this variability, Adamic-Adar, Common neighbors, Resource Allocation, Happiness, and Twitter Id similarity columns have many more positive values than negative. On the other hand, the coefficient for the Leicht-Holme-Newman index often evolved to a large negative weight. This signifies that user-user pairs which had high scores for the indices which evolved large, positive weights (e.g., Adamic-Adar, Common neighbors, Resource Allocation, Happiness, and Id similarity) and low scores for the indices which evolve large, negative weights (e.g., Leicht-Holme-Newman) were more likely to exhibit a future link.

We also visualize the relative ranking of the indices by their coefficients the Fig.~\ref{fig:all_solns_week_78_top20}b (and corresponding plots in the Appendices~\ref{ref:all_solns_weeks13}--\ref{ref:all_solns_weeks79}). Ordering the coefficients from greatest (most positive in 1st place) to least (most negative in 16th place) reveals that Adamic-Adar, Common neighbors, Resource Allocation, Happiness, and Twitter Id similarity often occupied the 1st-4th rankings (i.e., indices with the largest positive contribution, whereas LHN was often in 16th place (the largest negative weight). Other indices showed considerable variability in their ranking. We explore the implications of these findings in our discussion.
\begin{figure}[ht!]
\centering
\includegraphics[width=\linewidth]{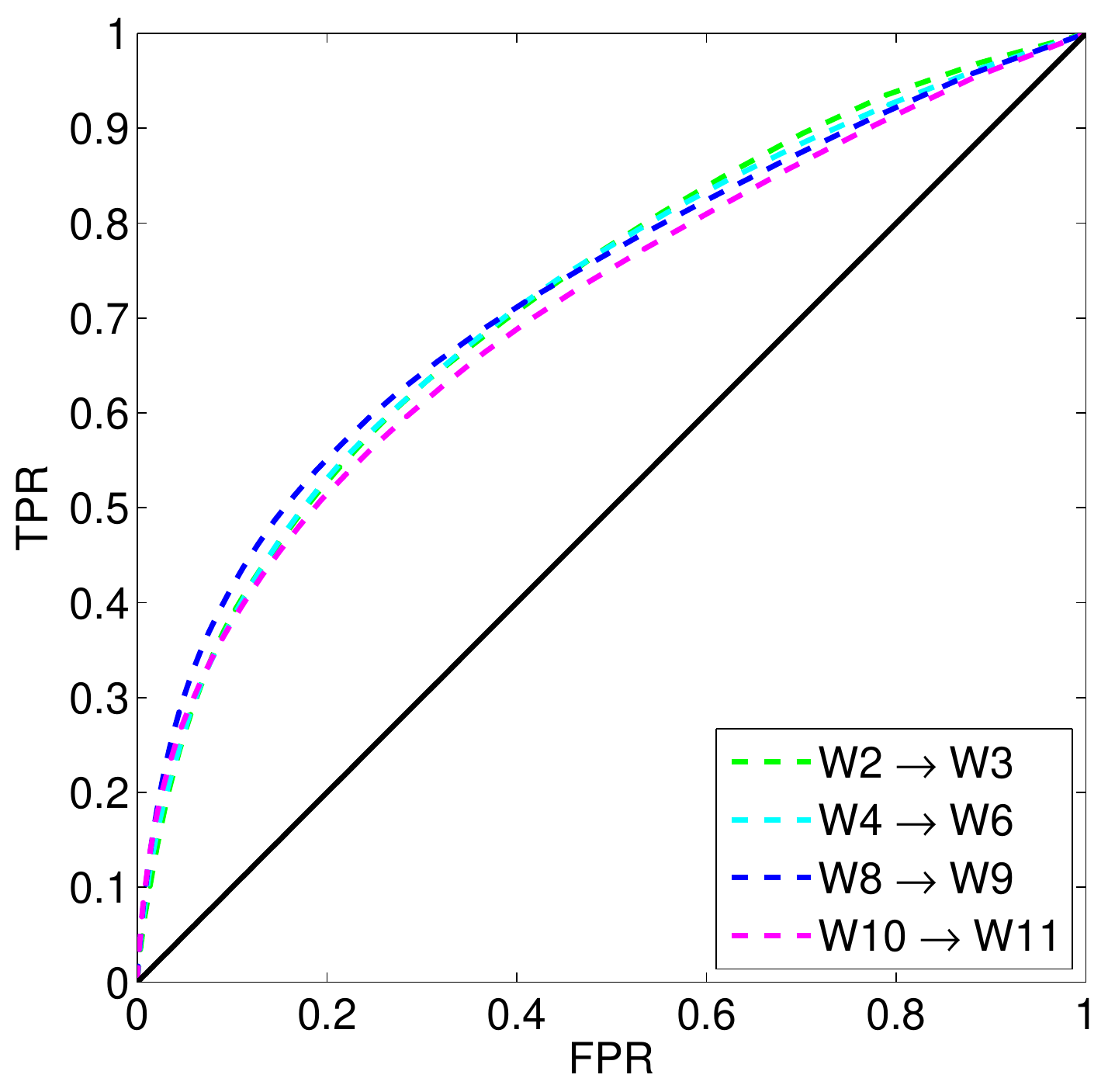} 
\caption{Receiver Operating Curve (ROC) for the ``all16'' predictors evolved from CMA-ES with $\text{fitness}_{20000}$. $AUC_{\text{Week 2} \longmapsto 3}=.723, AUC_{\text{Week 4} \longmapsto 5}=.721, AUC_{\text{Week 8} \longmapsto 9}=.726,$, and  $AUC_{\text{Week 10} \longmapsto 11}=.707$.}
\label{fig:ROC_weekly_20000}
\end{figure}

The ROC curve demonstrates that the true positive rate is considerably larger than the false positive rate ($TPR>FPR$) (Fig.~\ref{fig:ROC_weekly_20000}). We find AUC scores greater than 0.7 for all weeks in the validation set, suggesting that our predictor performs quite well, especially compared to other work with Twitter follower networks which did not suffer from missing data issues~\cite{Rowe2012}. We discuss these implications further in Section 4.

For large, sparse networks, the negative class is often much larger than the positive class. In our case, the number of new links (positive class) is on the order of $10^4$, whereas the number of potential links which do not exhibit future links (negative class) is on the order of $10^8$. Given this imbalance, measures such as accuracy, negative predictive value, and specificity will be very close to 1, even for random link predictors.  As suggested by Wang et al.~\cite{Wang2011}, more emphasis should be placed on recall and precision due to the large class imbalance between positives and negatives. The tunable parameter $\beta$ allows for unequal weighting on recall vs. precision:
\begin{equation}
F_{\beta}=(1+\beta^2) \cdot \frac{\text{precision} \cdot \text{recall}}{(\beta^2 \cdot \text{precision}) + \text{recall}}.
\label{eq:F_beta}
\end{equation}
In some applications, false positives (``false alarms'') may be relatively costless, whereas false negatives (``misses'') may pose an imminent threat. In these cases, recall is much more important than precision and setting $\beta >1 $ will weight recall more heavily in the $F_{\beta}$ score. In contrast, other applications may involve scenarios where false positives are costly to explore and a small number of links, for which we are fairly certainly about, is highly prized. In these cases, one can set $\beta < 1$ to place more importance on precision. 
\begin{figure}[!ht]
\centering
\includegraphics[width=.9\linewidth]{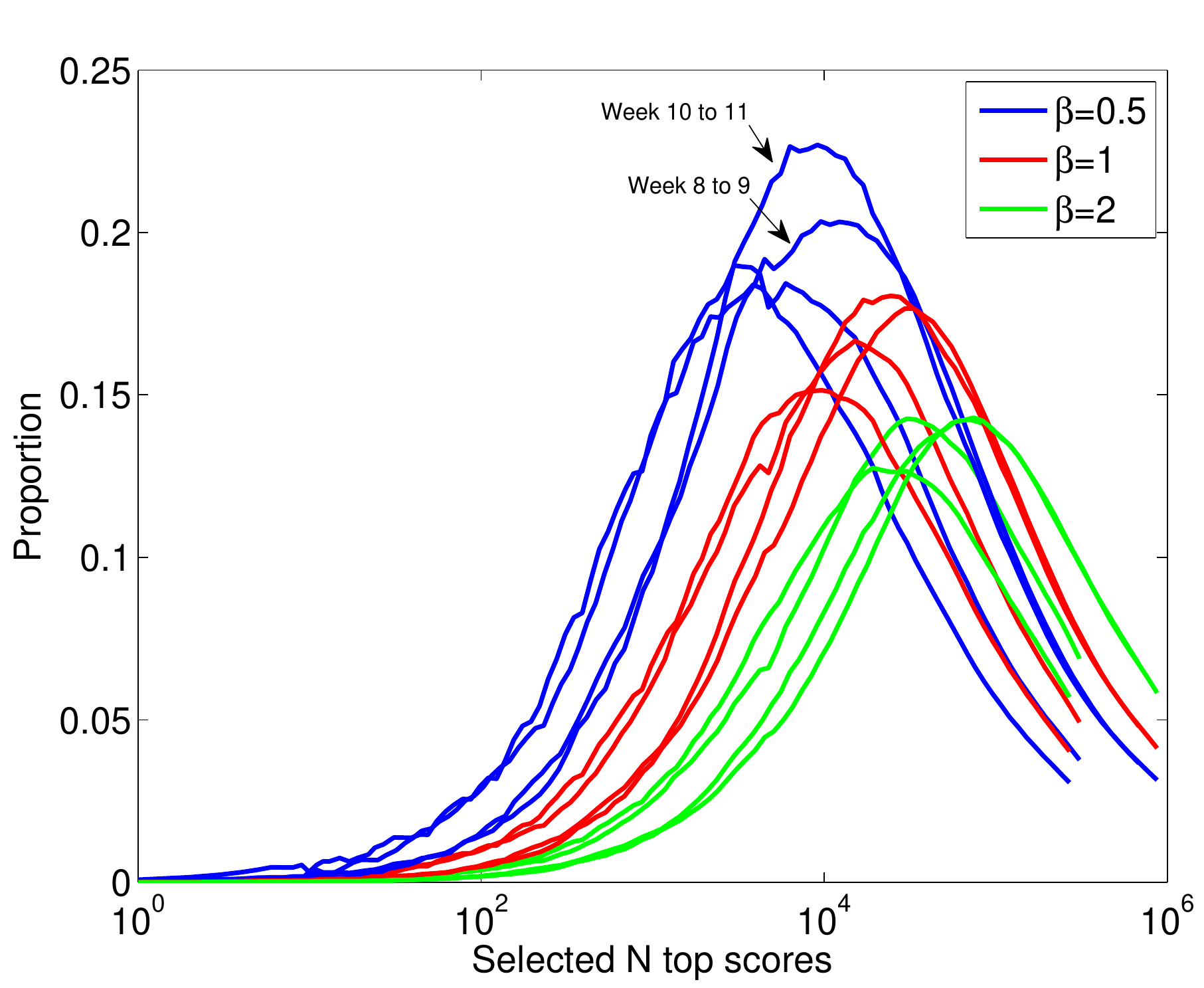}
\caption{$F_{\beta}$ scores for each of the validation sets ($W2 \rightarrow 3, W4 \rightarrow W5, W8 \rightarrow 9, W10 \rightarrow 11$) encode information about the performance of the link predictor with respect to precision and recall. When $\beta=1$, precision and recall are weighted equally. $\beta>1$ weights recall ($TPR=\frac{TP}{TP+TP+FN})$, whereas $\beta<1$ places more importance on precision ($PPV=\frac{TP}{TP+FP}$). Our predictor performs better with respect to precision and peaks for values on the order of $10^3$. The standard $F_1$ score peaks around $10^4$ and compares favorably with the work of~\cite{Rowe2012}. The highest $F_{\beta}$ scores are found for $W10 \rightarrow 11$. }
\label{fig:FB_scores_weekly}
\end{figure}

Tuning $\beta$ to one of 0.5, 1 or 2, we find that the $F_1$ peaks around top $N \approx 10^4$ (Fig.~\ref{fig:FB_scores_weekly}). $F$-scores are higher for weeks during which we received a higher percentage of tweets from the Twitter API service. For example, $F_{0.5}=0.203, F_{1}=.177, F_{2}=.142$, and $F_{0.5}=0.226, F_{1}=.181, F_{2}=.143$ for links which occurred from Weeks 8 to 9 and Weeks 10 to 11, respectively. In Week 5, we received a far smaller percentage of tweets. $F$-scores for new links occurring from Weeks 4 to 5 are $F_{0.5}=0.184, F_{1}=.152, F_{2}=.128$.
\begin{figure}[!ht]
\centering
\includegraphics[width=.95\linewidth]{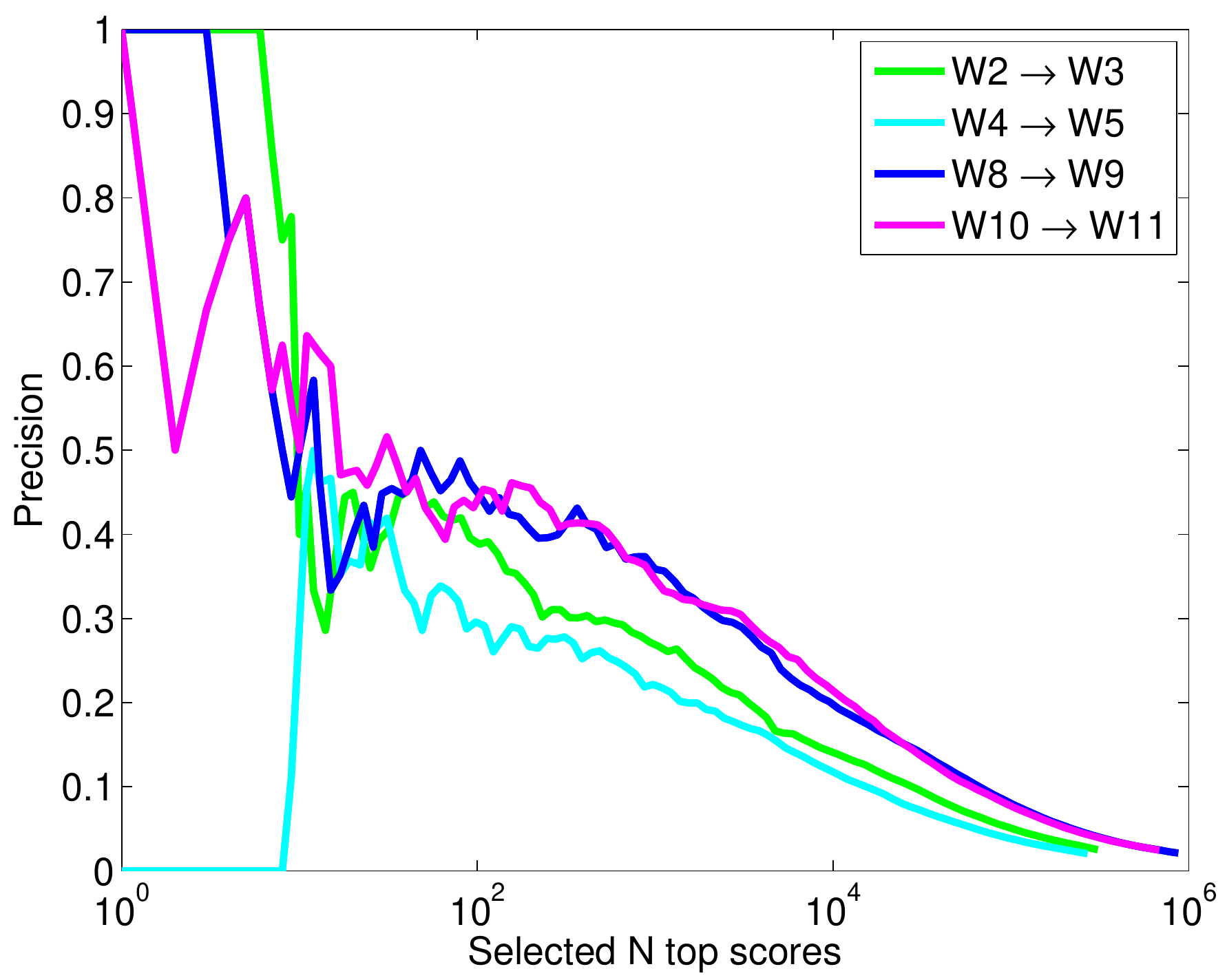}
\caption{Precision ($\frac{TP}{TP+FP}$) for the predicted links in the validation sets ($W2 \rightarrow 3, W4 \rightarrow W5, W8 \rightarrow 9, W10 \rightarrow 11$). High precision is achieved for top$N<20$, which is often the region of interest. The precision for predicted links in $W4 \rightarrow W5'$ is lower than the other weeks and this may be due to missing data for those weeks (see Table A2). }
\label{fig:precision}
\end{figure}
Figure~\ref{fig:precision} depicts the precision of the predicted links as a function of the top $N$ scoring user-user pairs. High precision is achieved for the fitness function which operates by selecting the top 20 scoring user-user pairs, which is often the region of interest. Precision is lower for predicted links from Week 4 to 5, a week in which we received a very low percentage of tweets from the Twitter API service, and higher for predicted links from Week 8 to 9 and Week 10 to 11, weeks for which we received a higher percentage of tweets from the Twitter API service (see Table A1). We also compute negative predictive value, and find this is consistently close to 1 due to the large true negative class. Specificity and accuracy are close to 1 for nearly all values of top $N$ links predicted, except for particularly large $N$ ($>10^4$). This is due to the large class imbalance of true negatives $(TN$), which dominate the numerator and denominator of these calculations.

\subsection{Exploring the impact of missing data}
During the twelve week period from September 9, 2008 - Dec 1, 2008 we received approximately 40\% of all tweets from Twitter's API service (Table A1). There are therefore both individuals and interactions that are unaccounted for in our training and validation period. Consequently, there are individuals who are connected by a path of length two in the true network, but which appear to be connected by a longer path because we have not captured interactions for intermediaries.

We explore the potential impact of missing tweets on our predictor by randomly selecting 50\% of our observed tweets and constructing the reciprocal reply subnetworks for Weeks 1 through 12. The evolutionary algorithm trains and validates on these subnetworks. For clarity, we denote $G$ for our observed networks and $G^s$, for our subnetworks. We identify the percent of links which are labeled as false positives in $G^s$ and true positive in $G$. This occurs precisely because our link predictor suggested a link which was actually correct, but for which an incomplete data set caused the link to be classified as a false positive. As such, we are underestimating the success of our link prediction method. Given a more complete data set, our results would most likely be better than we report here. 
\begin{figure}[!ht]
\centering
\includegraphics[width=\linewidth]{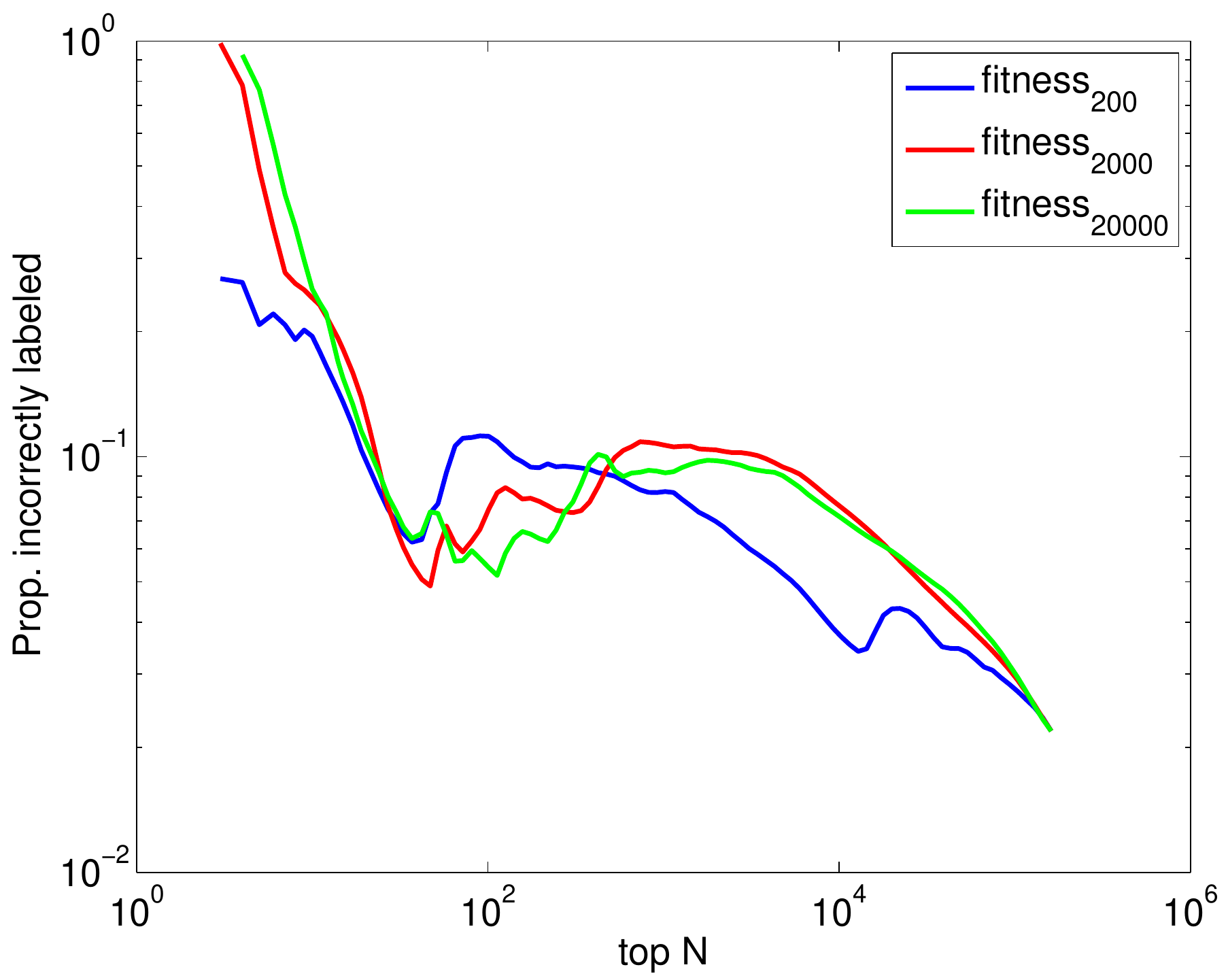} 
\caption{The proportion of incorrectly labeled false positives due to missing data when 50\% of our observed tweets were hidden from view and networks were recreated using this subsample of the data for Week 7 to 8. }
\label{fig:falsies}
\end{figure}

We next investigate the effects of missing data on our predictor, under the condition that 50\% of the Tweets have been removed. We observe that the number of correctly predicted links is hindered by the missing data, and the proportion of links which are incorrectly termed ``false-positive" because they are actually links in the weekly network containing a more complete data set is roughly 10\% (Fig.~\ref{fig:falsies}). This result from bootstrapping suggests that the performance of our predictors is a lower bound on performance, i.e., true precision and recall are most likely better than we report.

\subsection{Comparison to other methods}
Other studies in the area of link prediction have reported the factor improvement over random link prediction~\cite{Liben-Nowell2007, Wang2011}. We follow suit and compute the factor improvement of our predictor over a randomly chosen pair of users. The probability that a randomly chosen pair of individuals who are not connected in week $i$ become connected in week $i+1$ is  $\frac{\left|\text{Edges}_{\text{new}}\right|}{\binom{|V(G)|}{2}- \left| \text{Edges}_{\text{old}} \right| }$. There are 44,439 nodes in the validation set and, as a sample calculation, 71,927 edges in week 7. There are 53,722 new links that occur from Week 7 to 8. Thus, the probability of a randomly chosen pair of nodes from Week 7 exhibiting a link in Week 8 is approximately $\frac{53,722}{\binom{44,439}{2}-71,927} \approx .0054 \% $. 
\begin{figure*}[!ht]
\centering
\subfigure[$W2 \rightarrow W3$]{\includegraphics[width=.48\linewidth]{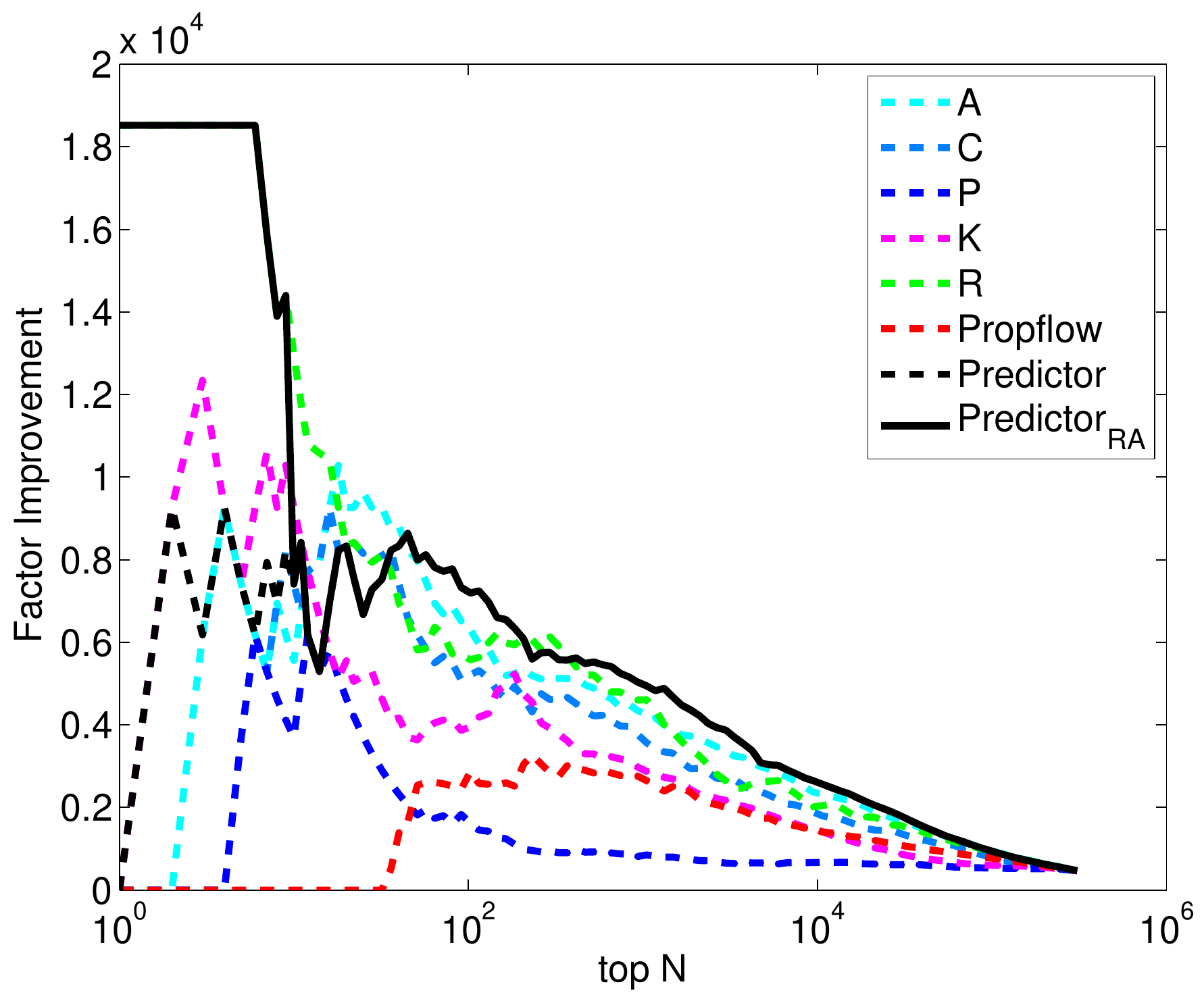}}
\subfigure[$W4 \rightarrow W5$]{\includegraphics[width=.48\linewidth]{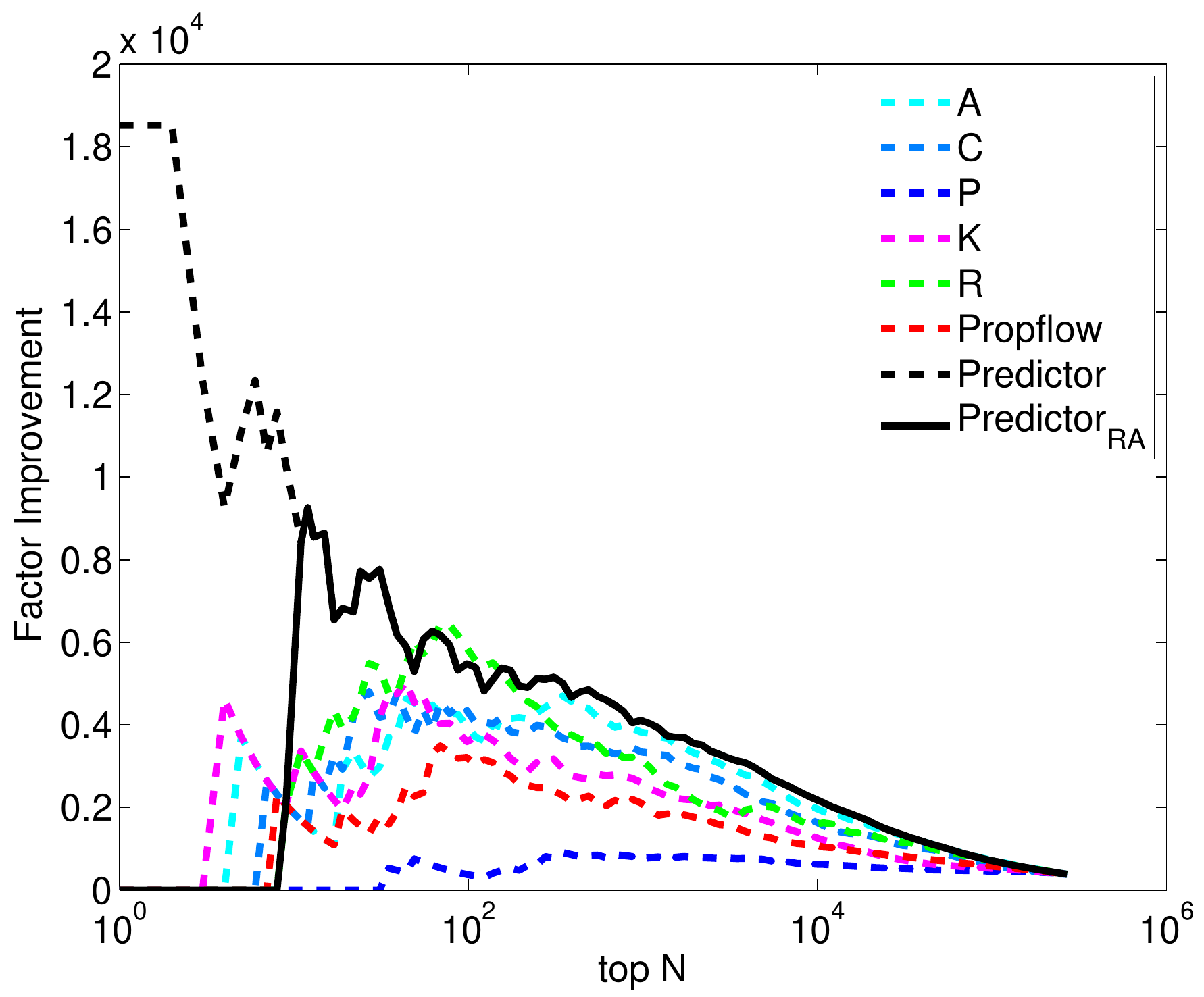}}\\
\subfigure[$W8 \rightarrow W9$]{\includegraphics[width=.48\linewidth]{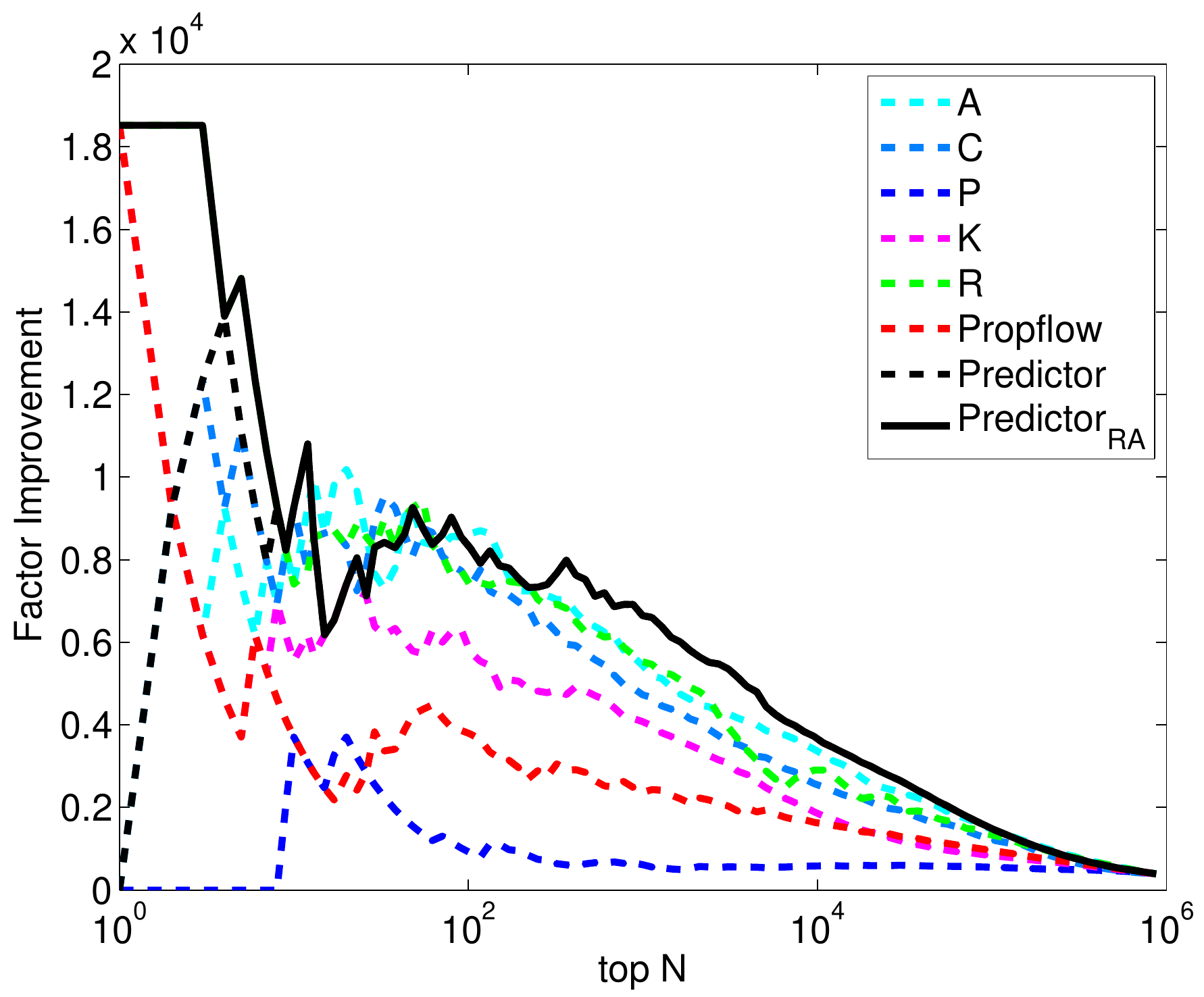}}
\subfigure[$W10 \rightarrow W11$]{\includegraphics[width=.48\linewidth]{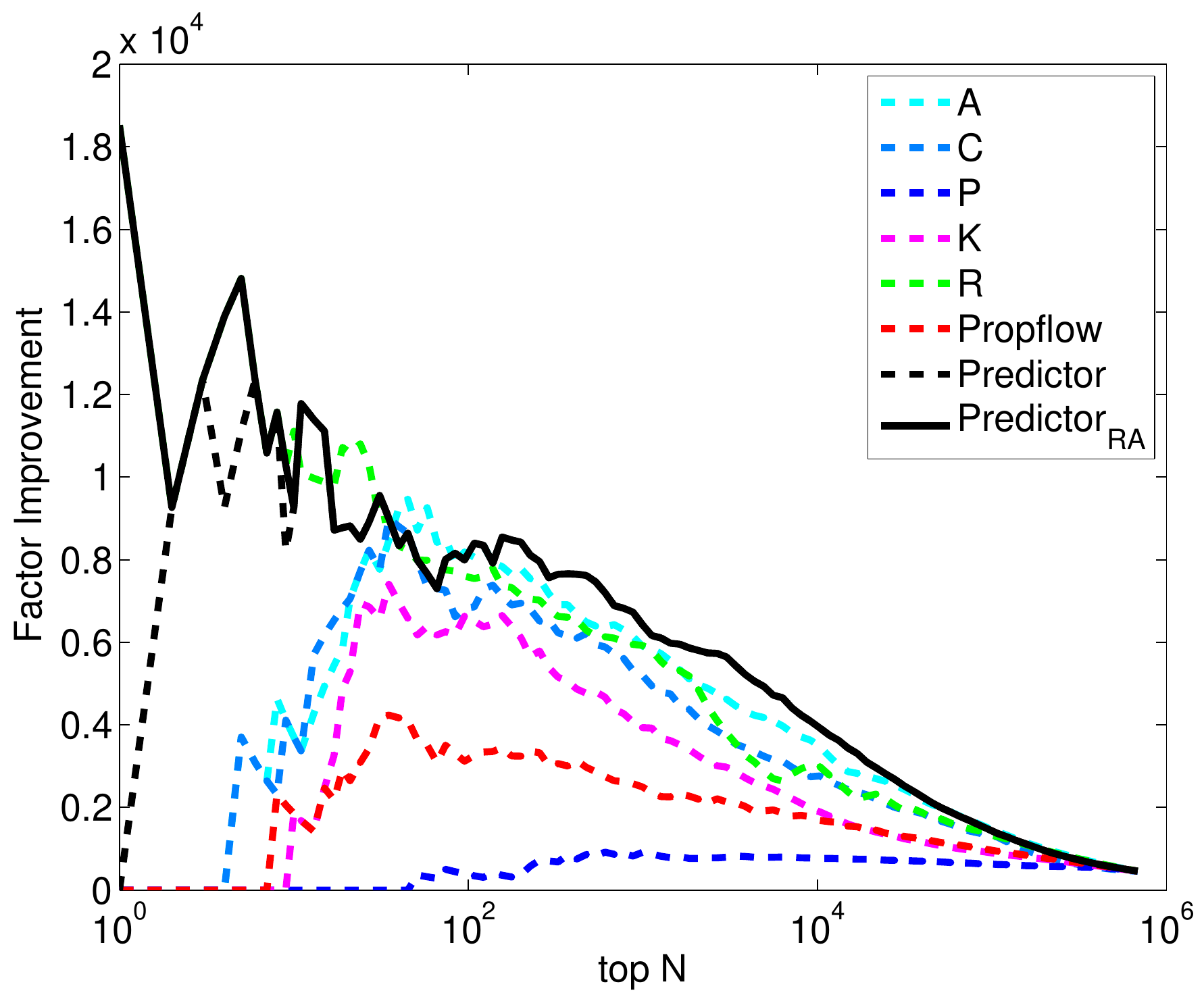}}
\caption{Factor improvement over randomly selected user-user pair is depicted. Large factor improvements are exhibited for predicting the top $N$ links, with notable peaks for $N<$100. The combined predictor outperforms the Common neighbors, Adamic-Adar, Paths, Katz, and Resource Allocation indices used in isolation over most choices for the top $N$ links predicted. }
\label{fig:factor_improve}
\end{figure*}

We observe significant factors of improvement over randomly selected new links, usually on the order of $10^{4}$ for top $N<$20 (Fig.~\ref{fig:factor_improve}). We notice that Resource Allocation outperforms other similarity indices when used in isolation to select the top 5 links during training and have included this in the cross-validation ($\text{Predictor}_{RA}$) step for selecting the top 10 (or fewer) links. We observe that the combined predictor outperforms indices used in isolation most choices of top $N$ link prediction. Due to the recent interest in using network flow measures, we also compare our predictors  
to \textit{propflow} restricted to a path of length two, a method proposed by Lichtenwalter et al.~\cite{lichtenwalter2010new}. Our method strongly outperforms this index.
\begin{table} [!ht]
\centering
\begin{tabular}{|l|c|c|}
\hline
& Binary Decision Tree & CMA-ES \\
\hline
Accuracy & 0.9555 & 0.9741 \\
Precision & 0.0894 & 0.2131 \\
Recall (true pos. rate) & 0.0694 & 0.0858\\
False positive rate & 0.0197 & 0.0068 \\
False discovery rate & 0.9106 & 0.7869 \\
\hline
\end{tabular}
\caption[Comparison of binary decision trees vs. CMA-ES for top$N$ link prediction.]{Comparison of binary decision trees vs. CMA-ES for top$N$ link prediction. CMA-ES (with top$N$=7417) slightly outperforms binary decision trees trained on new links that form from Week 7 to Week 8. We note that unbalanced classes are used in both cases.}
\label{table:binary_decision_tree}
\end{table}

Lastly, we compare our results to those obtained by training a binary decision tree classifier.\footnote{We use Matlab's implementation of binary classification trees to train on new links that form from Week 7 to Week 8.} Typically, balanced classes are used in training binary decision trees in order to overcome problems associated with unbalanced classes~\cite{lichtenwalter2010new, chawla2004editorial, cieslak2008learning}. We note that since our method for link prediction operates on all node-node pairs separated by path length two (e.g., highly unbalanced classes), we train our binary decision tree on unbalanced classes to avoid confounding our comparison with issues related to balanced and unbalanced classes. Furthermore, we set our method to select the top$N$=7417 links, which provides for roughly the same number of true positives as identified by the binary decision tree classifier.  Table~\ref{table:binary_decision_tree} reveals the results of this comparison. With this choice of top$N$, our approach performs slightly better across several indicators, such as accuracy and recall. Most notably, our precision is nearly three times as great as that obtained from our binary decision tree. Our false discovery rate is lower than that obtained for binary decision trees and this may be simply due to our taking a top$N$ approach to link prediction, which inherently limits the number of false positives by tuning the top$N$ links to predict. We discuss these results in more detail in the next section.

\section{Discussion}
Several studies have suggested that the combination of topological similarity indices and node-specific similarity indices may greatly enhance link prediction efforts~\cite{Aiello2012, Lu2010, Rowe2012, hutto2013longitudinal, Wang2011, romero2013, hasan2006}. We find support for this claim in our work with Twitter reciprocal reply networks. For experiments in which training occurred on a given week, we find that the combined ``all16'' predictor outperforms the topological only predictor ``topo12'' and find that this difference is most pronounced for top $N<20$. 

Our measures perform quite well in comparison to other researchers working in the area of link prediction for Twitter. Rowe, Stankovic, and Alani~\cite{Rowe2012} explore topological and individual specific similarity indices (words and topic similarity) in an effort to predict following behavior. They find an $AUC<0.6$ whereas we find $AUC>0.7$ for all experiments. Yin, Hong, and Davison~\cite{Yin2011} develop a structure based link prediction model and report $F$-scores on the order of $F=.190$ for Twitter follower networks. These networks do not suffer from incomplete data in the same way that Twitter reciprocal reply networks do. Our predictor performs comparatively well, with scores ranging from $F_1=$0.152 for validation on new links occurring from Week 4 to 5, a week for which we obtained approximately 24\% of all tweets, to $F_1=$0.181 for validation on new links occurring from Week 10 to 11, a week for which we obtained approximately 48\% of all tweets. 

We have developed a meaningful link predictor for Twitter reciprocal reply networks, a social subnetwork consisting of individuals who demonstrate active and ongoing engagement. We were able to achieve a factor of improvement over random link selection on the order of $10^4$ for the top 20 (or fewer) links predicted and $10^3$ over several orders of magnitude for the top $N$ links predicted. 

Wang et al.~\cite{Wang2011} examine a social network constructed from mobile phone call data and find a factor improvement of approximately $1.5 \times 10^3$. To compare our work, however, one must standardize for the number of nodes in the network.\footnote{These researchers report 579,087,610 potential new links and a factor improvement of 1500. Rescaling the factor improvement for networks of the same size amounts to computing the probability of a randomly predicted link being correct.} Upon doing so, we find our factor improvement is an order of magnitude higher. 

We compare our results to other approaches, such as \textit{propflow} and binary decision trees. As suggested by others and observed here, link prediction in large, sparse networks suffers from problems related to unbalanced classes. As such, we caution the interpretation of our results in comparision to industry standards, such as binary decision trees. Future work may improve upon our methods by using balanced classes in the evolution of coefficients over generational time in CMA-ES. Incorporating these strategies and others may allow for more insightful comparisons between our methods and other supervised learning approaches. 

One of the most intriguing aspects of this work is the detection of similarity indices which evolve to have large, positive weights in our link predictors. Perhaps the most notable similarity index for which this is the case is the Resource Allocation index. Resource allocation considers the amount of resource one node has and assumes that each node will distribute its resource equally among all neighbors~\cite{zhou2009predicting}. Considering the limits to time and attention an individual has, this may be suggestive of a mechanism by which users limit their interaction, a result suggested by Gon{\c{c}}alves et al.~\cite{gonccalves2011modeling} and also noted by~\cite{Bliss2012} in Twitter RRNs.

In addition to suggesting that our work is comparable to or an improvement upon other work which combines measures via supervised learning, we present a method which is transparent and transferable. Future work may involve the inclusion of geospatial data~\cite{frank2013} or community structure to predict links. Efforts to consider the persistence or decay of links over time, or inconsistencies in flow rates~\cite{bagrow2013shadow} could also prove fruitful.

\section{Acknowledgments}
The authors acknowledge the Vermont Advanced Computing
Core which is supported by NASA (NNX-08AO96G) at the University
of Vermont for providing High Performance Computing resources
that have contributed to the research results reported within this
paper. CAB and PSD were funded by an NSF CAREER Award to PSD (\# 0846668). CMD, PSD, and MRF were funded by a grant from the MITRE Corporation. The authors thank Brian Tivnan and Maggie J. Eppstein for their helpful suggestions.

\bibliographystyle{model1-num-names}
\bibliography{sigproc}{}  
\FloatBarrier
\section{Appendix}

  \renewcommand{\thefigure}{A\arabic{equation}}    
  \renewcommand{\thetable}{A\arabic{equation}}    

\setcounter{equation}{1}
\begin{table*}[ht!]
\label{table:datawehave}
\begin{center}
\begin{tabular}{lllllll}
\hline
\hline                                                  
Week & Start date	& \# Obsvd.	Msgs. & \# Total Msgs.	& \% Obsvd. & \# Replies	& \%  Replies  \\
& & $\times 10^6$ & $\times 10^6$ &  $\left( \frac{\# Obsvd.}{\# Total} \times 100 \right)$  & $\times 10^6$ &  $\left( \frac{\# Replies}{\# Obsvd.}  \times 100\right)$  \\[1ex]
\hline\hline                            
 1 & 09.09.08 & 3.14 & 7.26 & 43.2 &  0.88 & 28.1 \\ 
 2 & 09.16.08 & 3.36 & 8.31 & 40.4 &  0.90 & 26.9 \\ 
 3 & 09.23.08 & 3.43 & 8.89 & 38.6 &  0.90 & 26.2 \\ 
 4 & 09.30.08 & 3.33 & 9.06 & 36.8 &  0.89 & 26.6 \\ 
 5 & 10.07.08 & 2.33 & 9.38 & 24.8 &  0.64 & 27.5 \\ 
 6 & 10.14.08 & 4.39 & 9.87 & 44.4 &  1.24 & 28.3 \\ 
 7 & 10.21.08 & 4.70 & 10.01 & 47.0 &  1.35 & 28.8 \\
 8 & 10.28.08 & 5.74 & 10.34 & 55.5 &  1.64 & 28.5 \\ 
 9 & 11.04.08 & 5.58 & 11.14 & 50.1 &  1.63 & 29.3 \\ 
10 & 11.11.08 & 4.70 & 9.88 & 47.6 &  1.42 & 30.2 \\ 
11 & 11.18.08 & 5.48 & 11.34 & 48.3 &  1.67 & 30.5 \\ 
12 & 11.25.08 & 5.71 & 11.47 & 49.8 &  1.73 & 30.2 \\ 
\hline
\end{tabular}

\end{center}
\caption{The number of ``observed'' messages in our database comprise a fraction of the total number of Twitter message made during period of this study (September 2008 through November 2009). While our feed from the Twitter API remains fairly constant, the total \# of tweets grows, thus reducing the \% of all tweets observed in our database. We calculate the total $\#$ of messages as the difference between the last message id and the first message id that we observe for a given month. This provides a reasonable estimation of the number of tweets made per month as message ids were assigned (by Twitter) sequentially during the time period of this study. We also report the number observed messages that are replies to specific messages and the percentage of our observed messages which constitute replies. }
\end{table*}

\setcounter{equation}{2}
\begin{table*}[ht!]
\label{table:netstats}
\begin{center}
\begin{tabular}{rrrrrrrrr}
\hline 
Week & Start date &$N$  & $<k>$ & $k_{\max}$ & $C_G$ & Assort & \# Comp. & $S$  \\[1ex]
\hline\hline                            
 1 & 09.09.08 & 95647 & 2.99 & 261 & 0.10 & 0.24 & 10364 & 0.71   \\ 
 2 & 09.16.08 & 99236 & 2.95 & 313 & 0.10 & 0.24 & 11062 & 0.71   \\ 
 3 & 09.23.08 & 99694 & 2.90 & 369 & 0.09 & 0.13 & 11457 & 0.70   \\ 
 4 & 09.30.08 & 100228 & 2.87 & 338 & 0.09 & 0.13 & 11752 & 0.69   \\ 
 5 & 10.07.08 & 78296 & 2.60 & 241 & 0.09 & 0.21 & 11140 & 0.63   \\ 
 6 & 10.14.08 & 122644 & 3.20 & 394 & 0.09 & 0.14 & 12221 & 0.74   \\ 
 7 & 10.21.08 & 130027 & 3.30 & 559 & 0.08 & 0.09 & 12420 & 0.75   \\ 
 8 & 10.28.08 & 144036 & 3.56 & 492 & 0.08 & 0.14 & 12319 & 0.78   \\ 
 9 & 11.04.08 & 145346 & 3.54 & 330 & 0.08 & 0.19 & 12597 & 0.78   \\ 
10 & 11.11.08 & 136534 & 3.35 & 441 & 0.08 & 0.12 & 12972 & 0.76   \\ 
11 & 11.18.08 & 153486 & 3.46 & 444 & 0.08 & 0.13 & 13594 & 0.77   \\ 
12 & 11.25.08 & 155753 & 3.46 & 1244 & 0.06 & 0.00 & 14122 & 0.77   \\
\hline
\end{tabular}
\end{center}
\caption{Network statistics for reciprocal-reply networks by week. As Twitter popularity grows, so does the number of users ($N$) in the observed reciprocal-reply network. The average degree ($<k>$), degree assortativity, the number of nodes in the giant component (\# Comp.), and the proportion of nodes in the giant component ($S$) remain fairly constant, whereas the maximum degree ($k_{\max}$) shows a great deal of variability from month to month. Clustering ($C_G$) shows a slight decrease over the course of this period.}
\end{table*}
\setcounter{equation}{1}
\begin{figure*}[ht!]

\includegraphics[width=\linewidth]{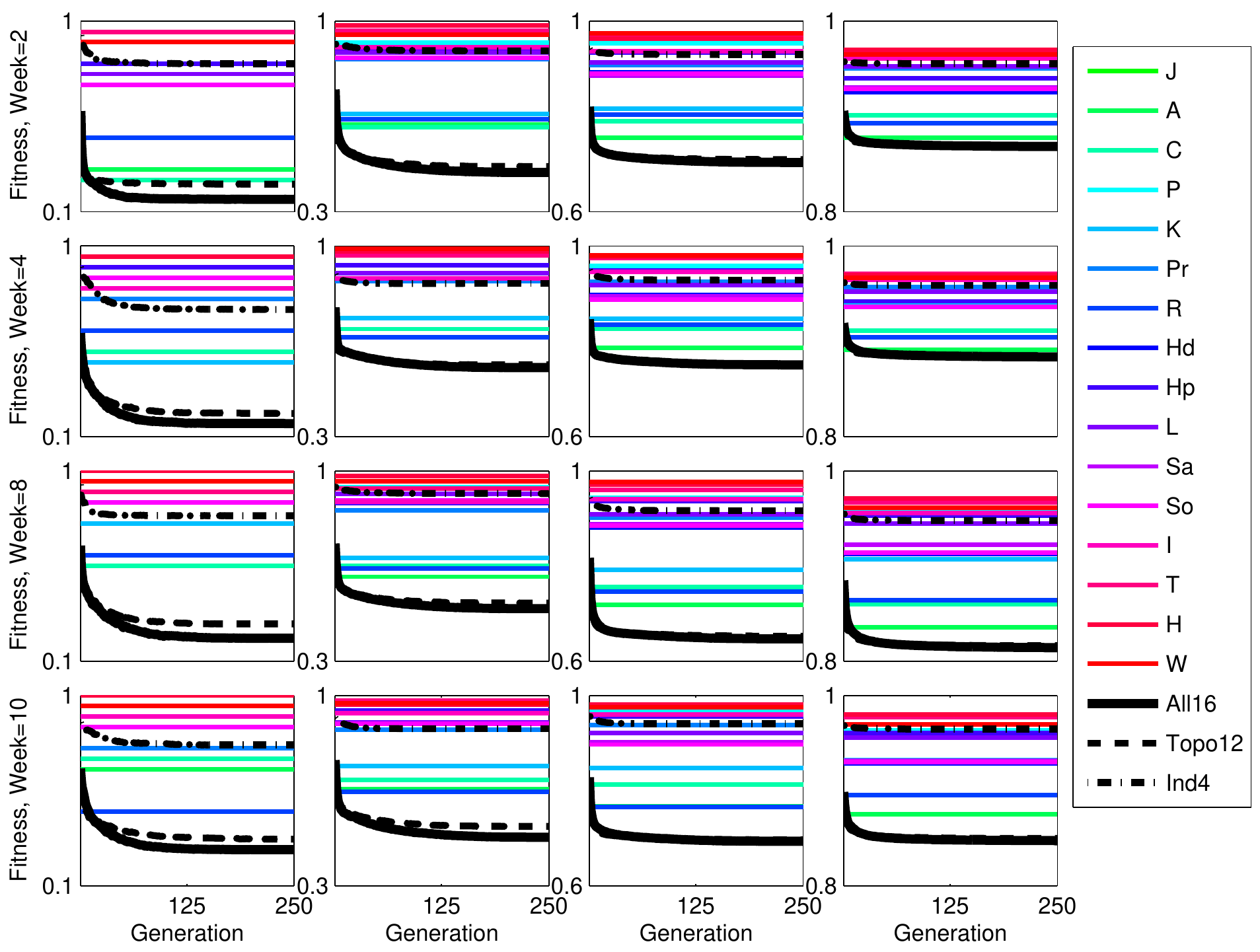}\\
\caption{Mean fitness computed from 100 simulations of CMA-ES for training on the new links that occur in a given week for each of (columns left to right) top $N$=20, top $N$=200, top $N$=2000 and top $N$=20,000. We compare each individual index, along with ``all16'' (evolved predictor consisting of all 16 indices), ``topo12'' (evolved predictor consisting of only the 12 topological indices), and ``node4'' (evolved predictor consisting of only the 4 node similarity indices). To show detail, the axes are not uniformly scaled between each column.}
\label{fig:fitness_week}
\end{figure*}

\setcounter{equation}{2}
\begin{figure*}[ht!]
\centering
\subfigure[Week 1 $\mapsto$ 2, N=20]{\includegraphics[width=.24\linewidth]{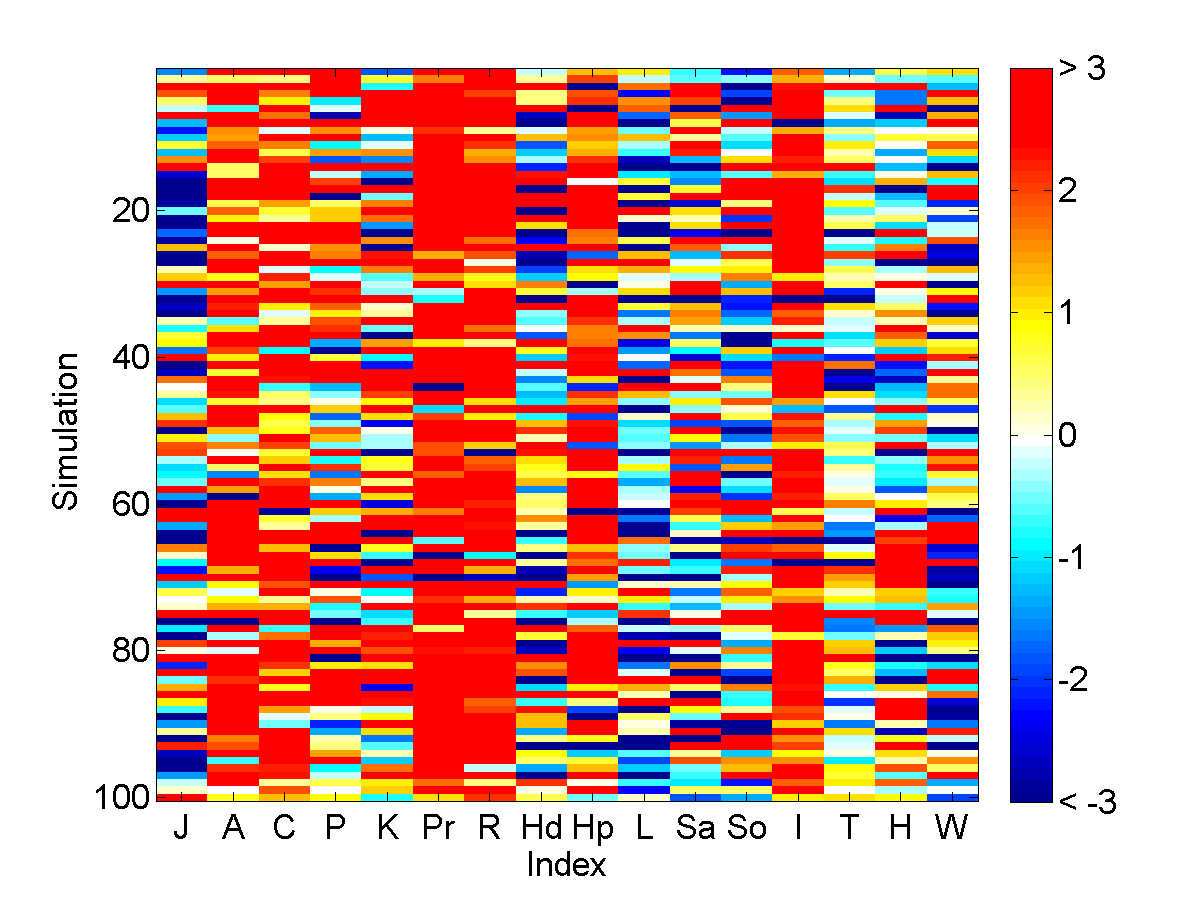}}
\subfigure[Week 1 $\mapsto$ 2, N=200]{\includegraphics[width=.24\linewidth]{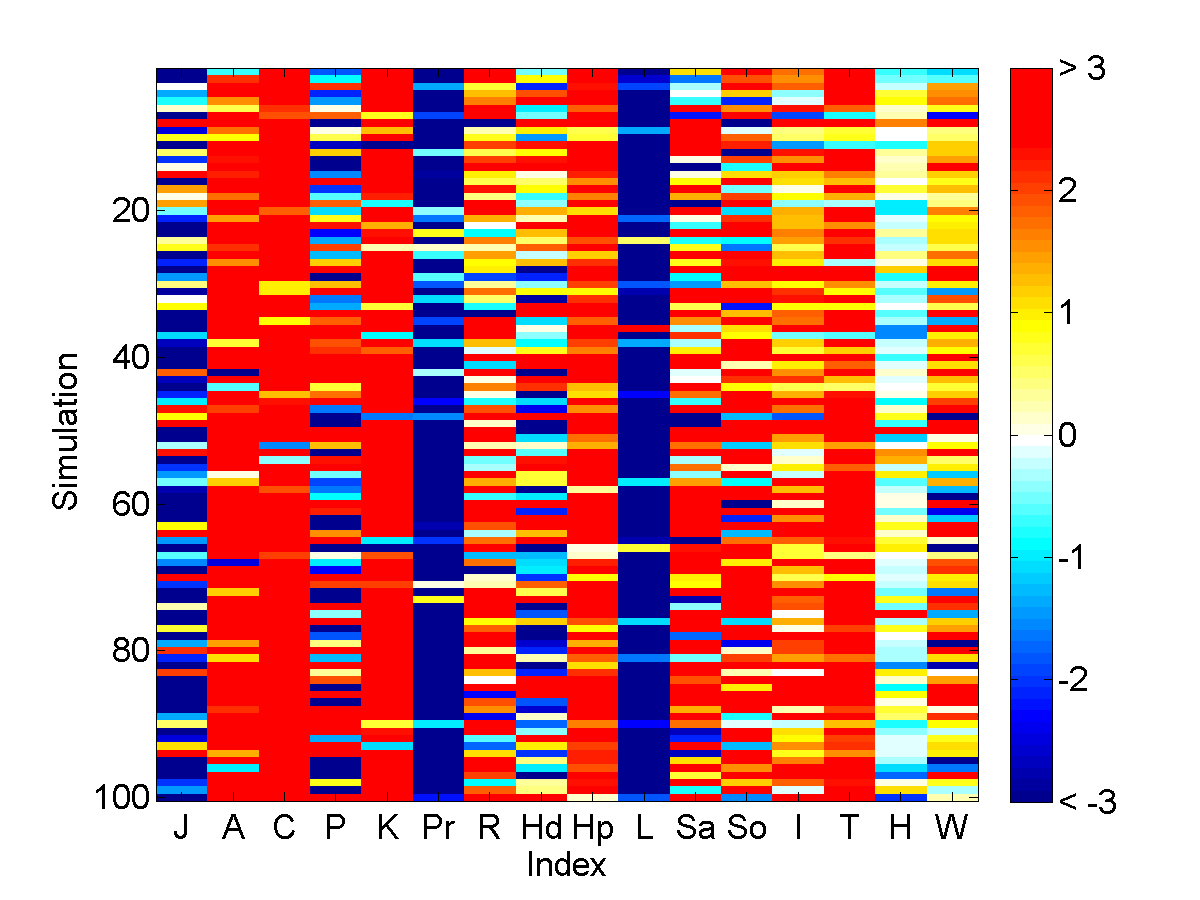}}
\subfigure[Week 1 $\mapsto$ 2, N=2000]{\includegraphics[width=.24\linewidth]{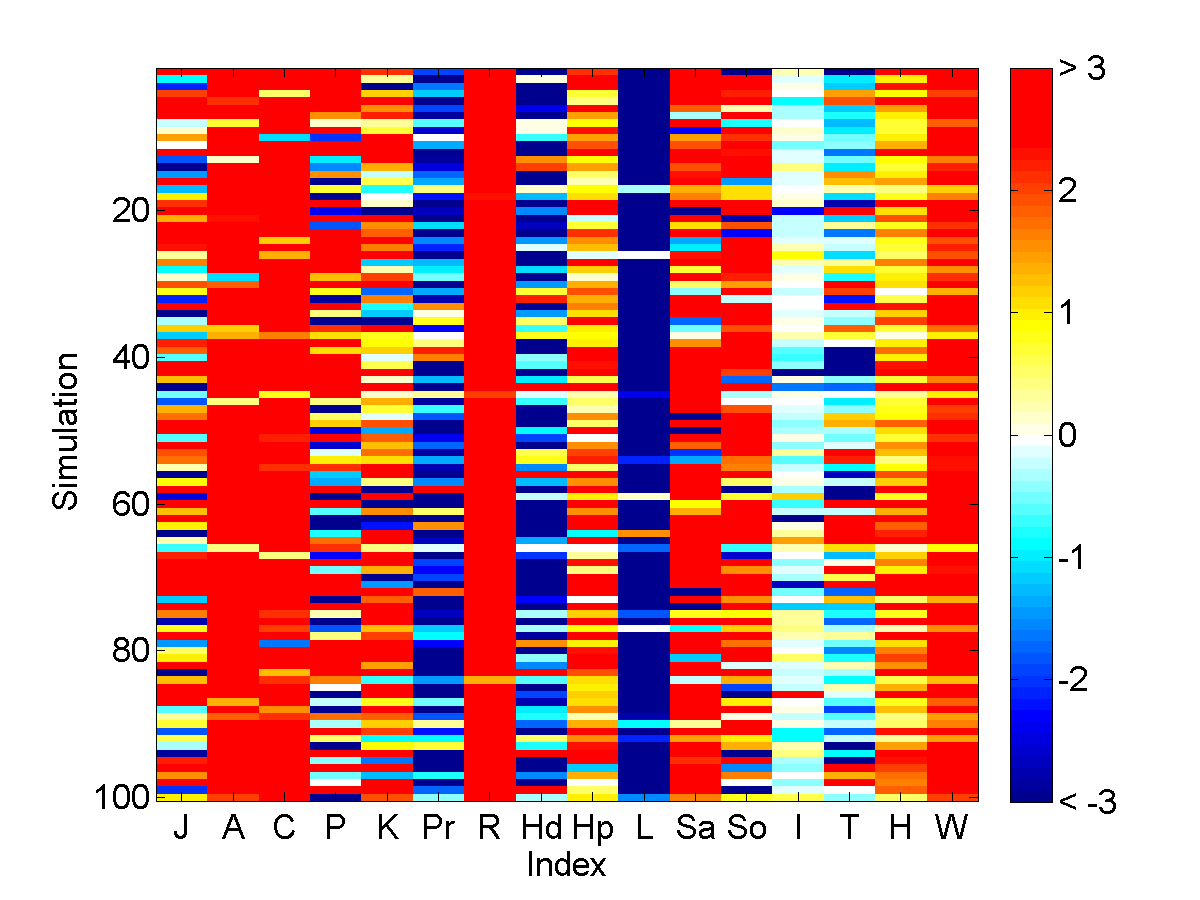}}
\subfigure[Week 1 $\mapsto$ 2, N=20000]{\includegraphics[width=.24\linewidth]{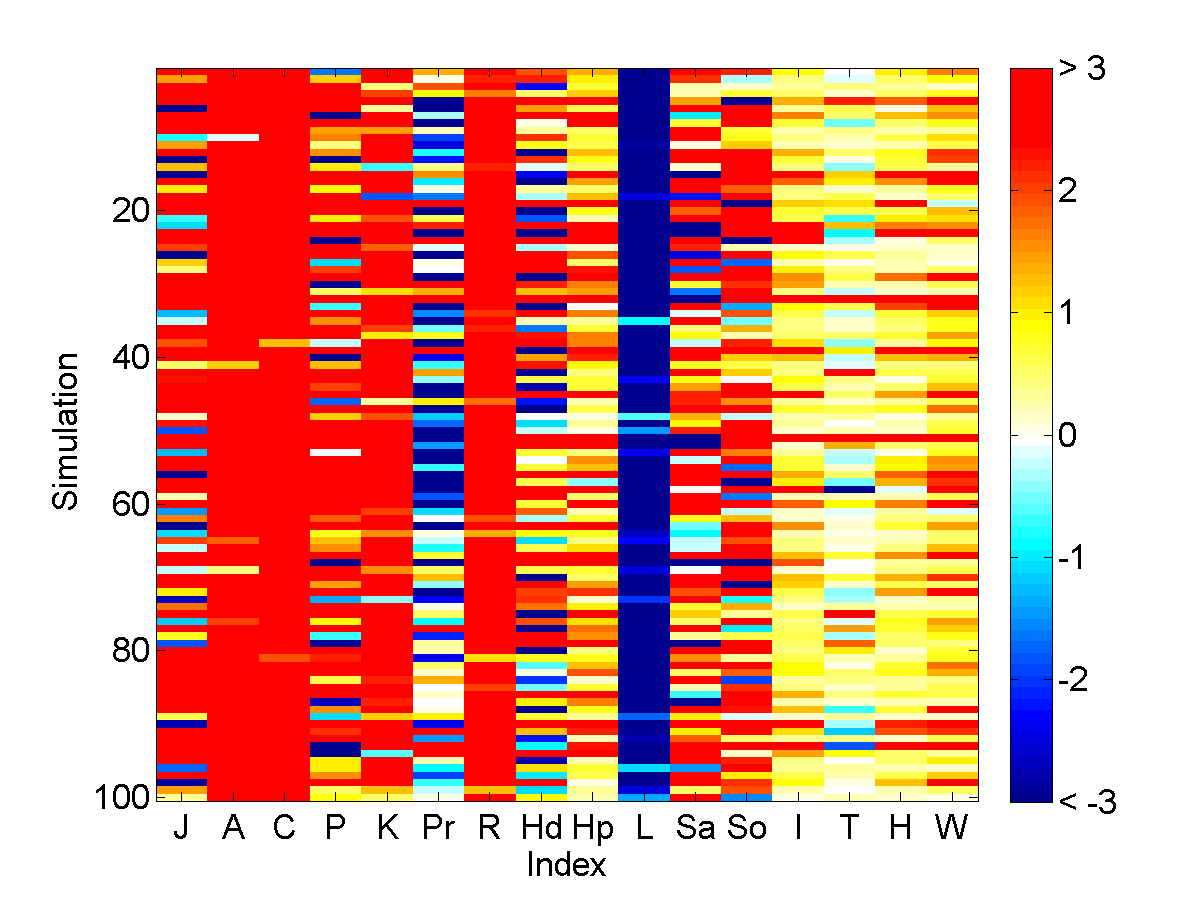}}\\
\subfigure[Week 1 $\mapsto$ 2, N=20]{\includegraphics[width=.24\linewidth]{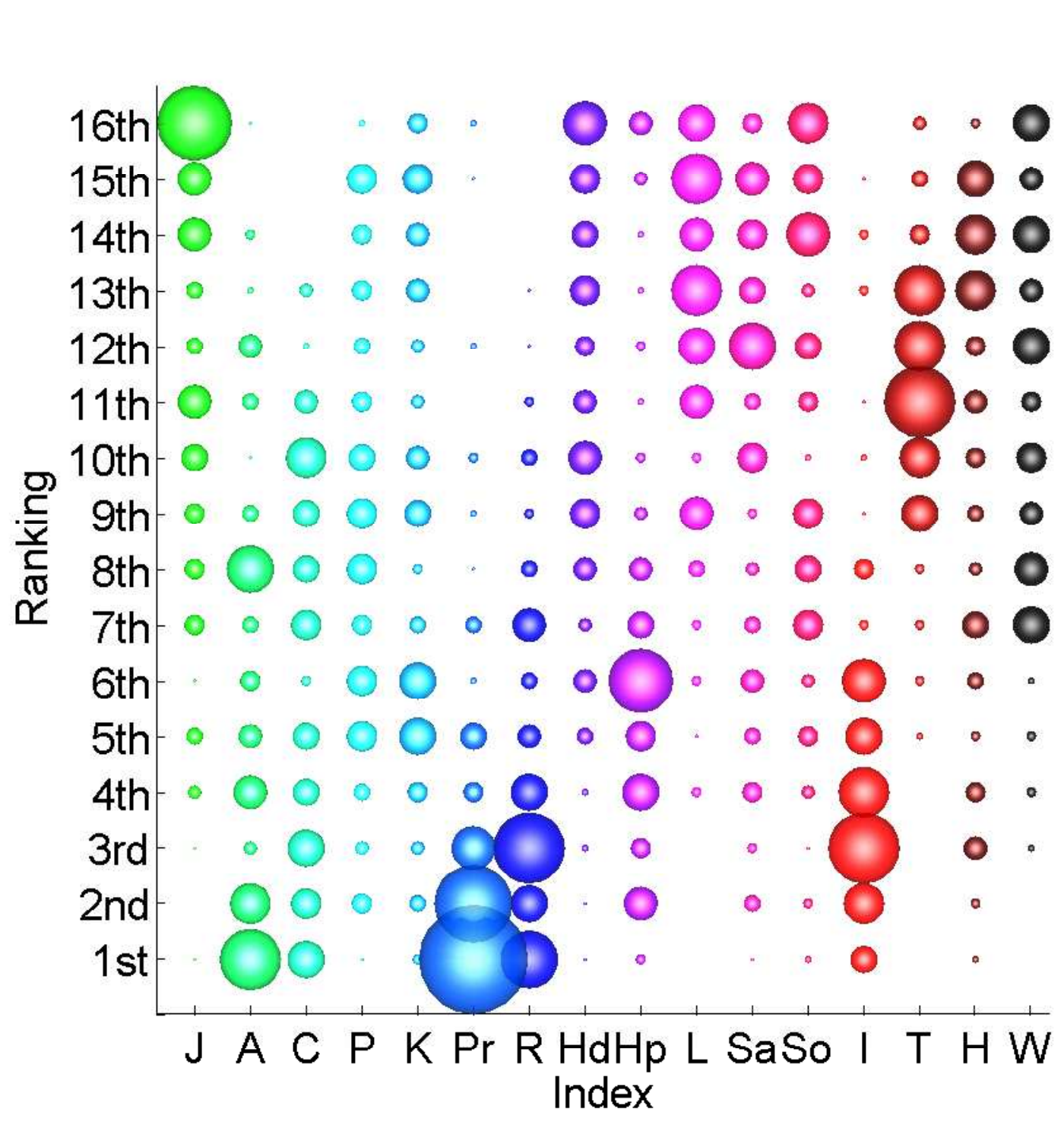}}
\subfigure[Week 1 $\mapsto$ 2, N=200]{\includegraphics[width=.24\linewidth]{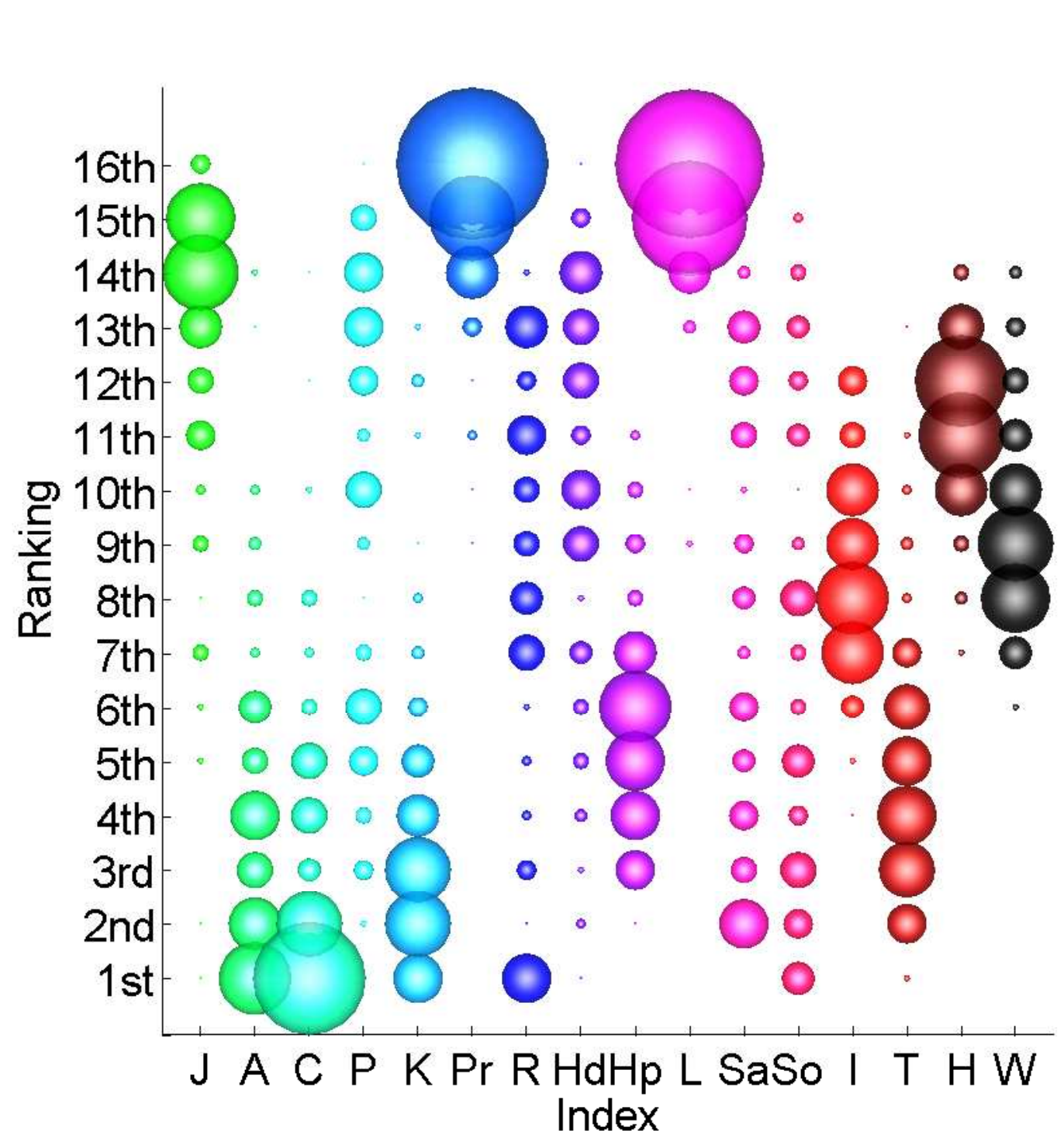}}
\subfigure[Week 1 $\mapsto$ 2, N=2000]{\includegraphics[width=.24\linewidth]{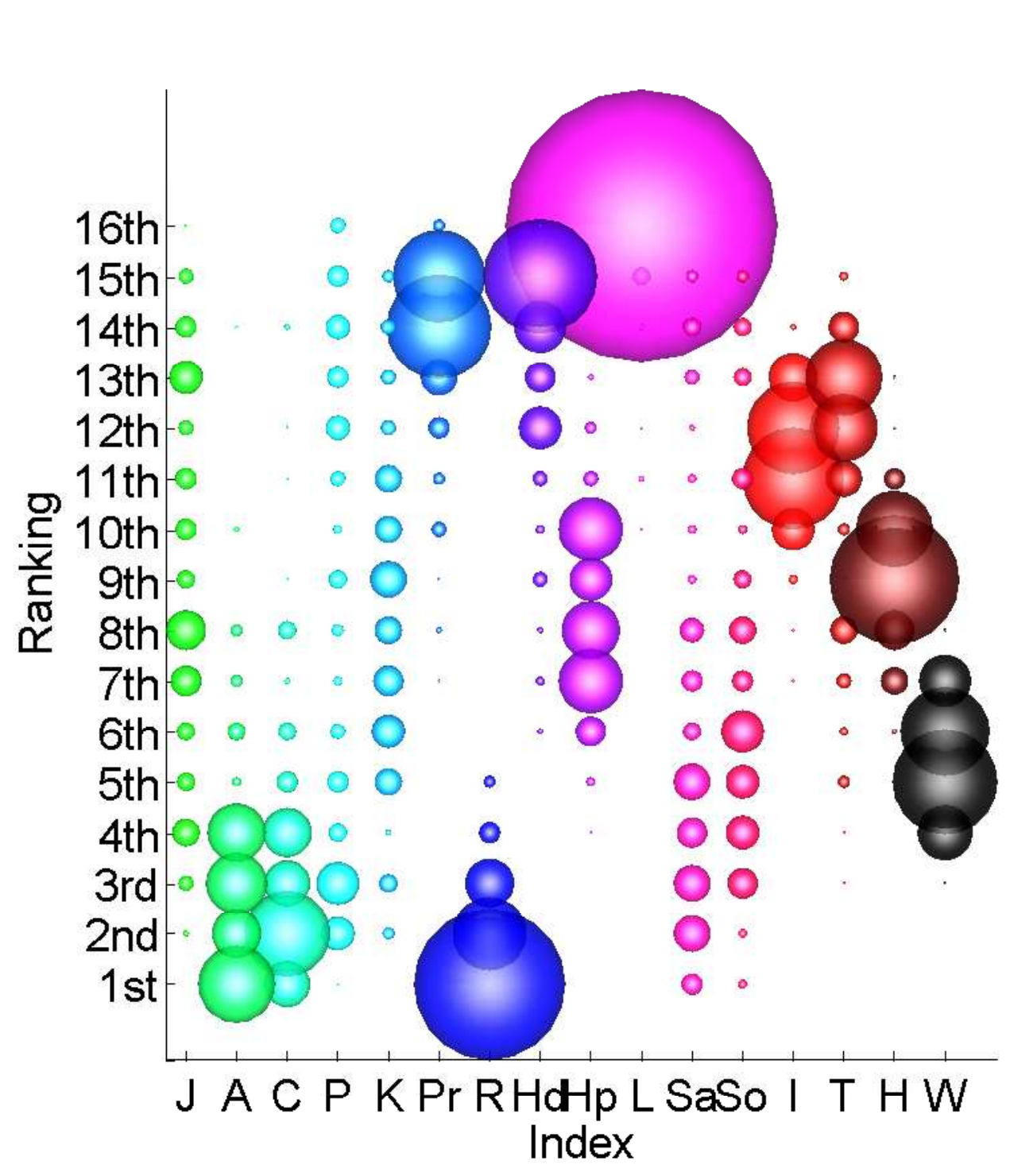}}
\subfigure[Week 1 $\mapsto$ 2, N=20000]{\includegraphics[width=.24\linewidth]{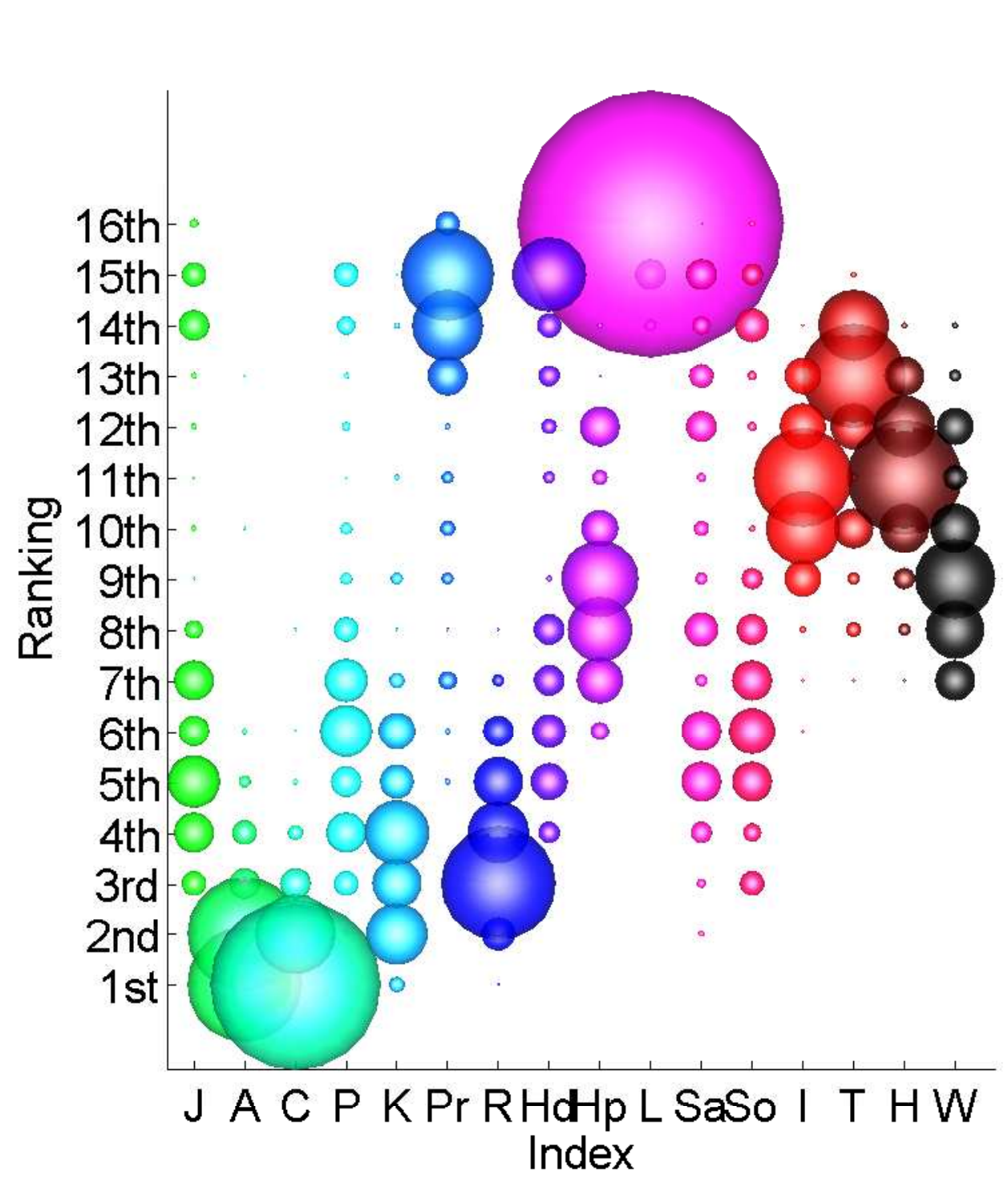}}\\
\subfigure[Week 3 $\mapsto$ 4, N=20]{\includegraphics[width=.24\linewidth]{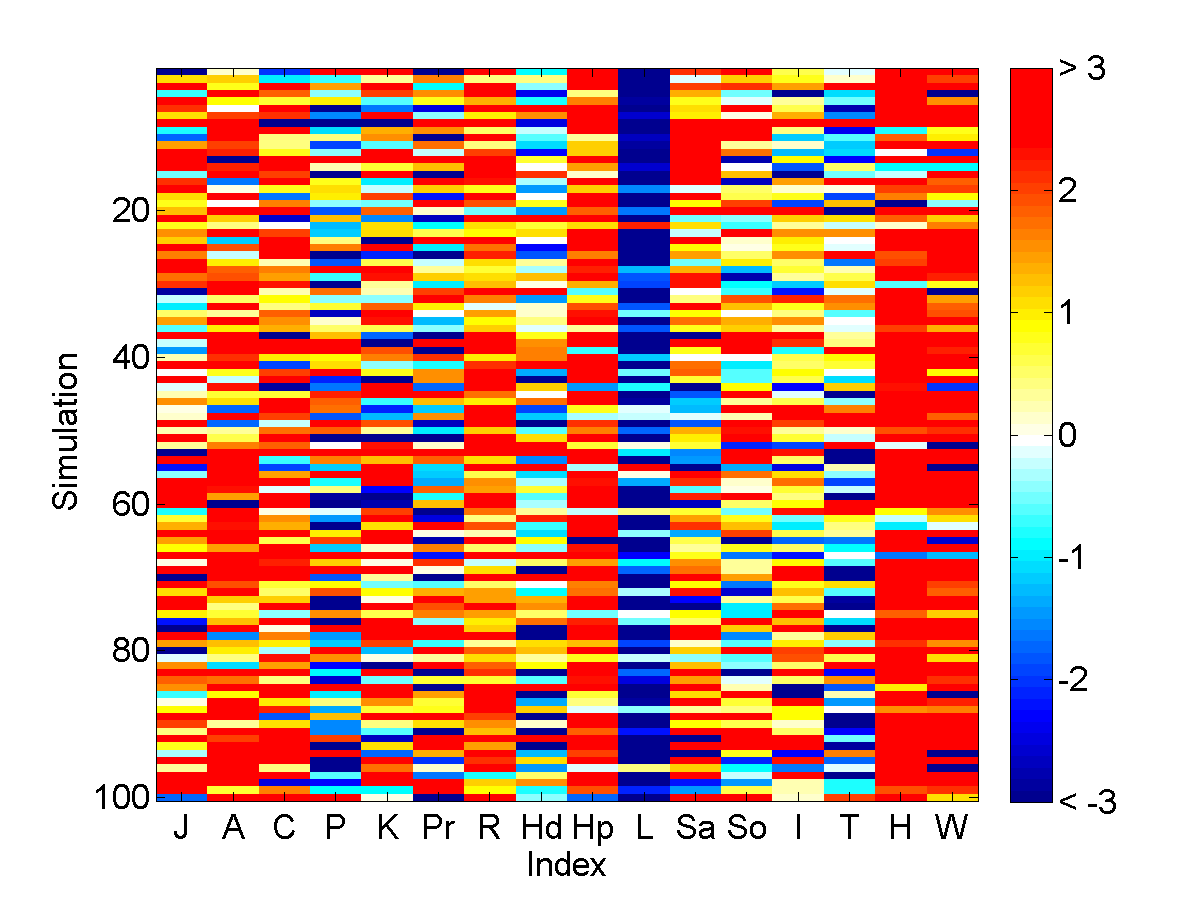}}
\subfigure[Week 3 $\mapsto$ 4, N=200]{\includegraphics[width=.24\linewidth]{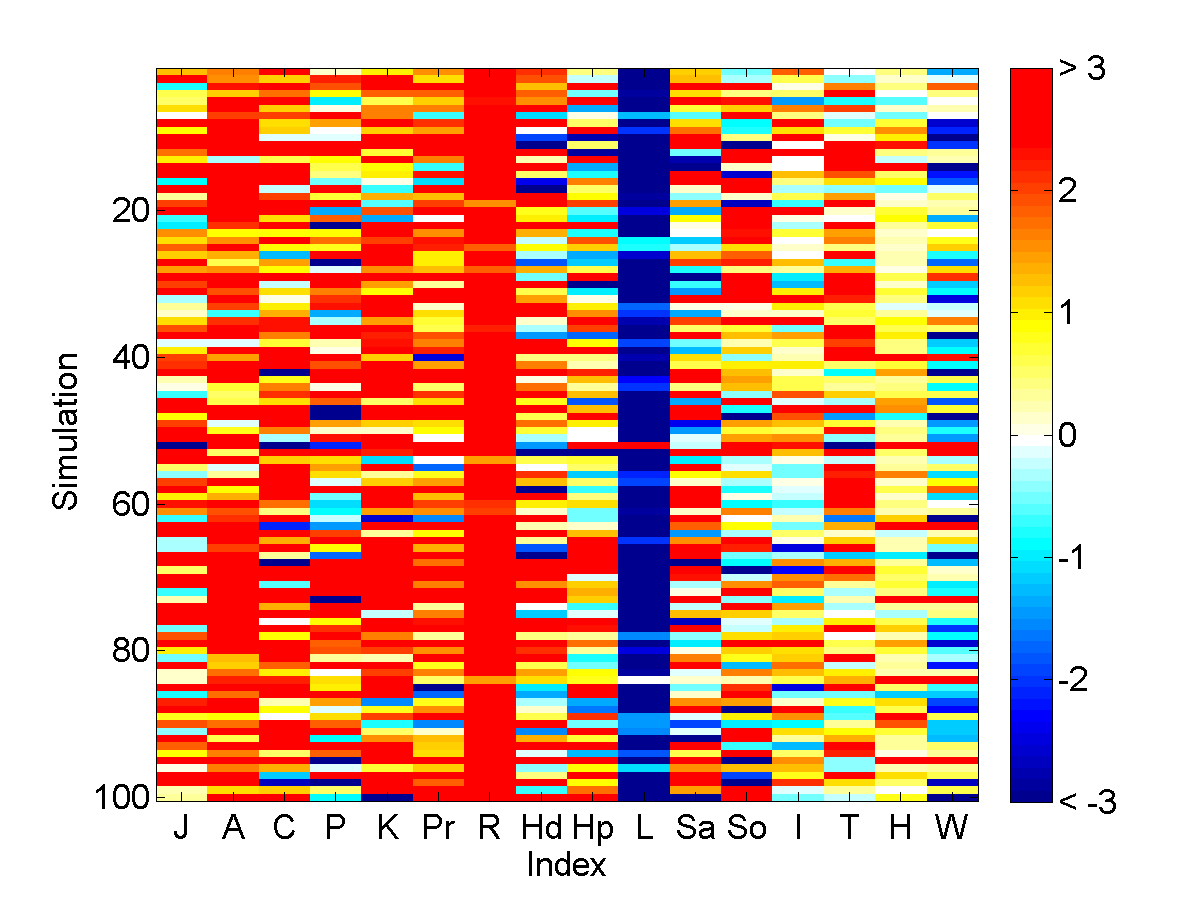}}
\subfigure[Week 3 $\mapsto$ 4, N=2000]{\includegraphics[width=.24\linewidth]{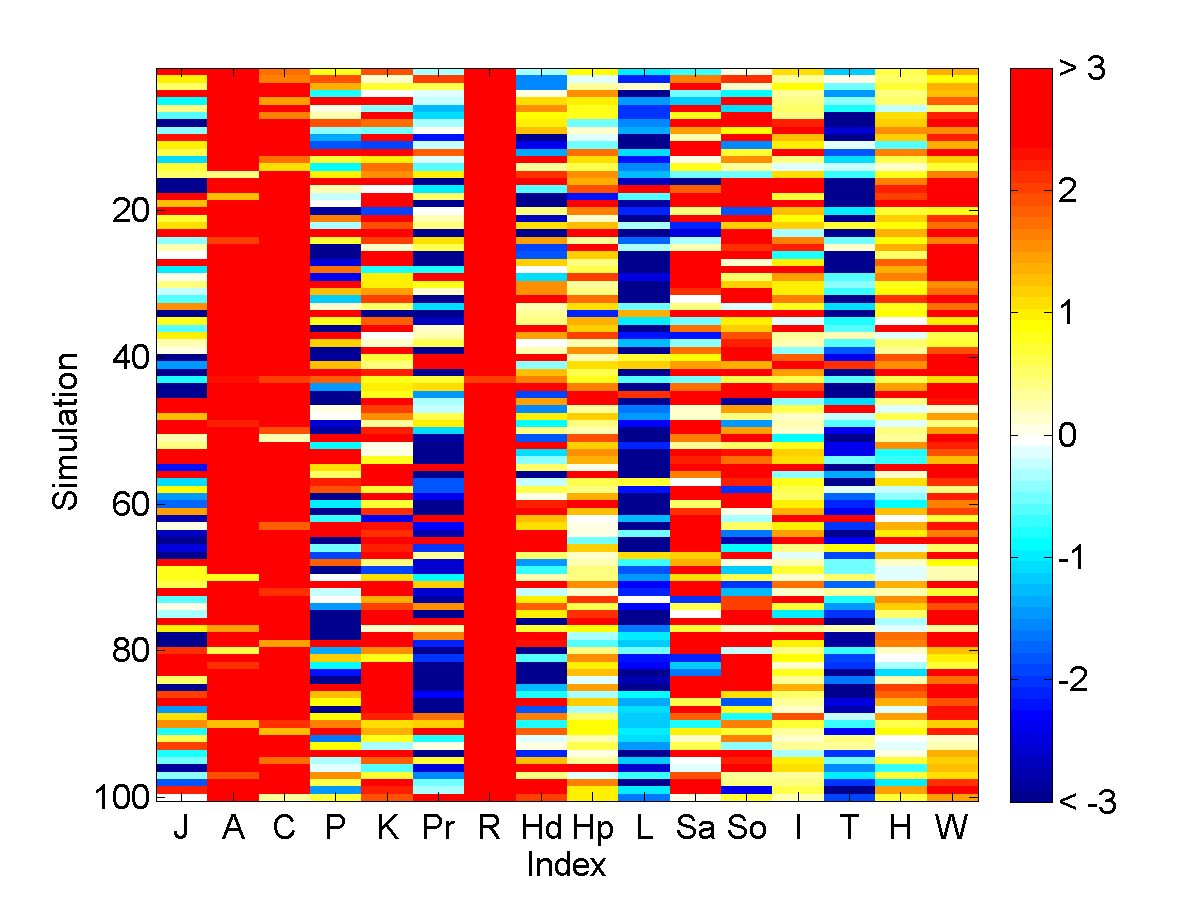}}
\subfigure[Week 3 $\mapsto$ 4, N=20000]{\includegraphics[width=.24\linewidth]{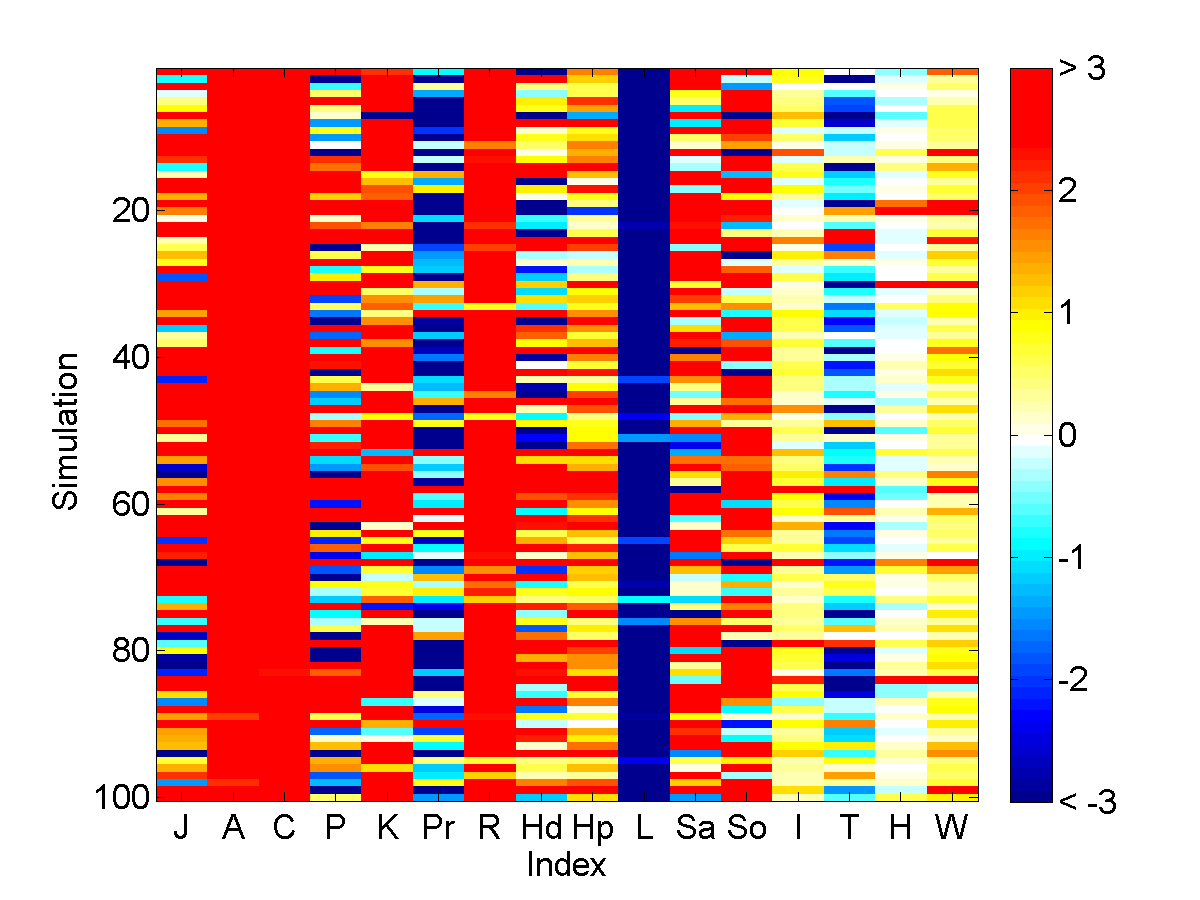}}\\
\subfigure[Week 3 $\mapsto$ 4, N=20]{\includegraphics[width=.24\linewidth]{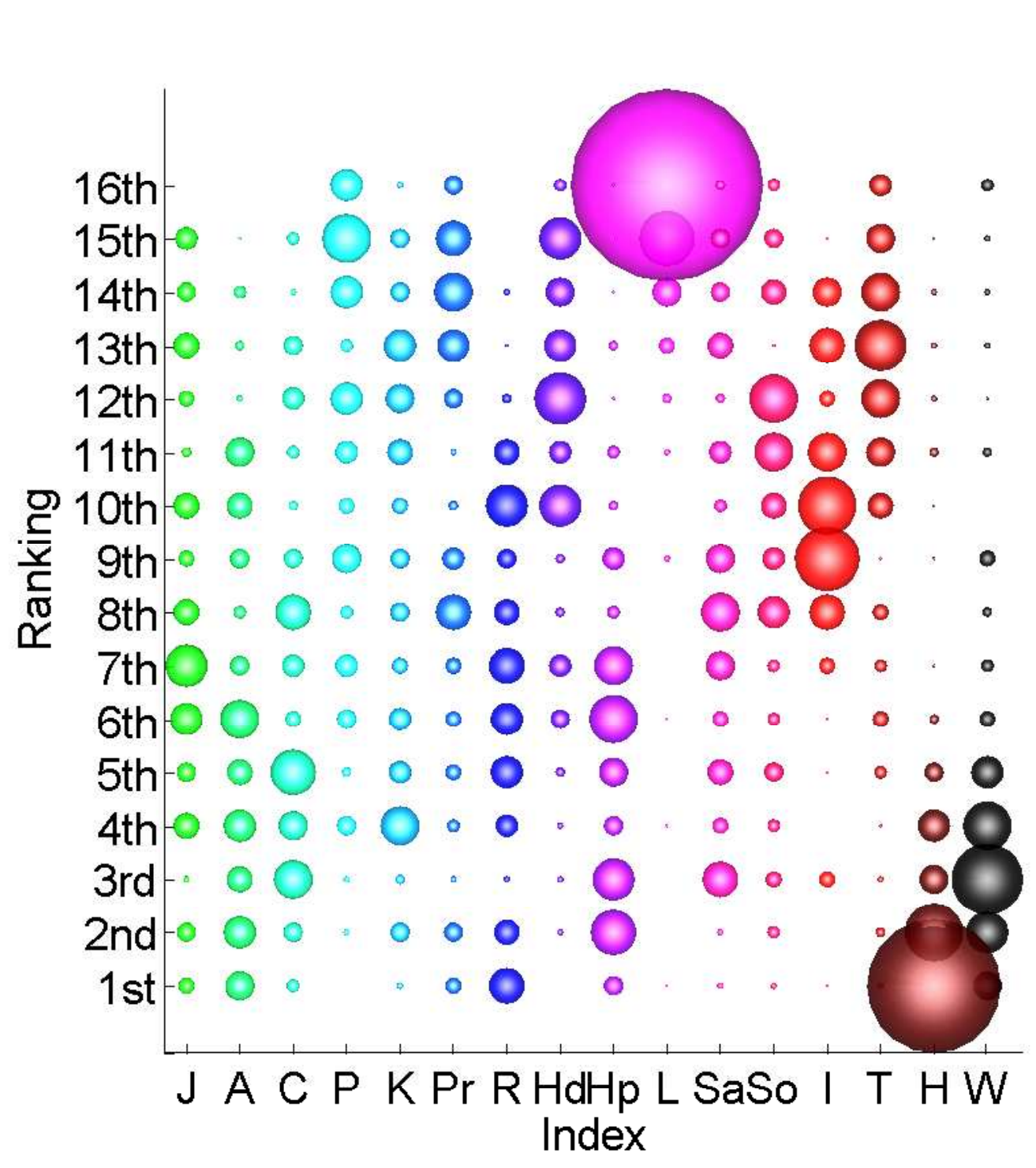}}
\subfigure[Week 3 $\mapsto$ 4, N=200]{\includegraphics[width=.24\linewidth]{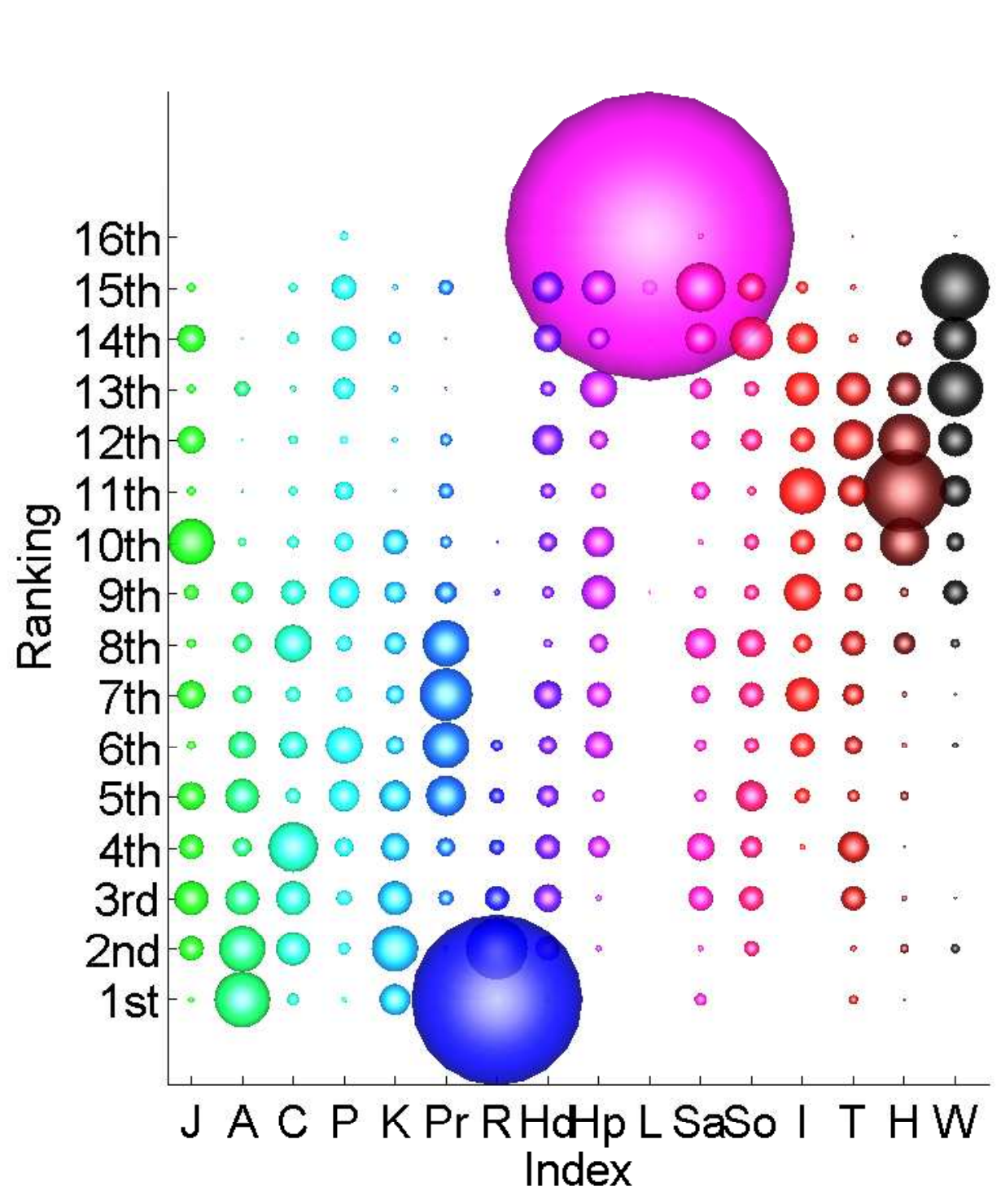}}
\subfigure[Week 3 $\mapsto$ 4, N=2000]{\includegraphics[width=.24\linewidth]{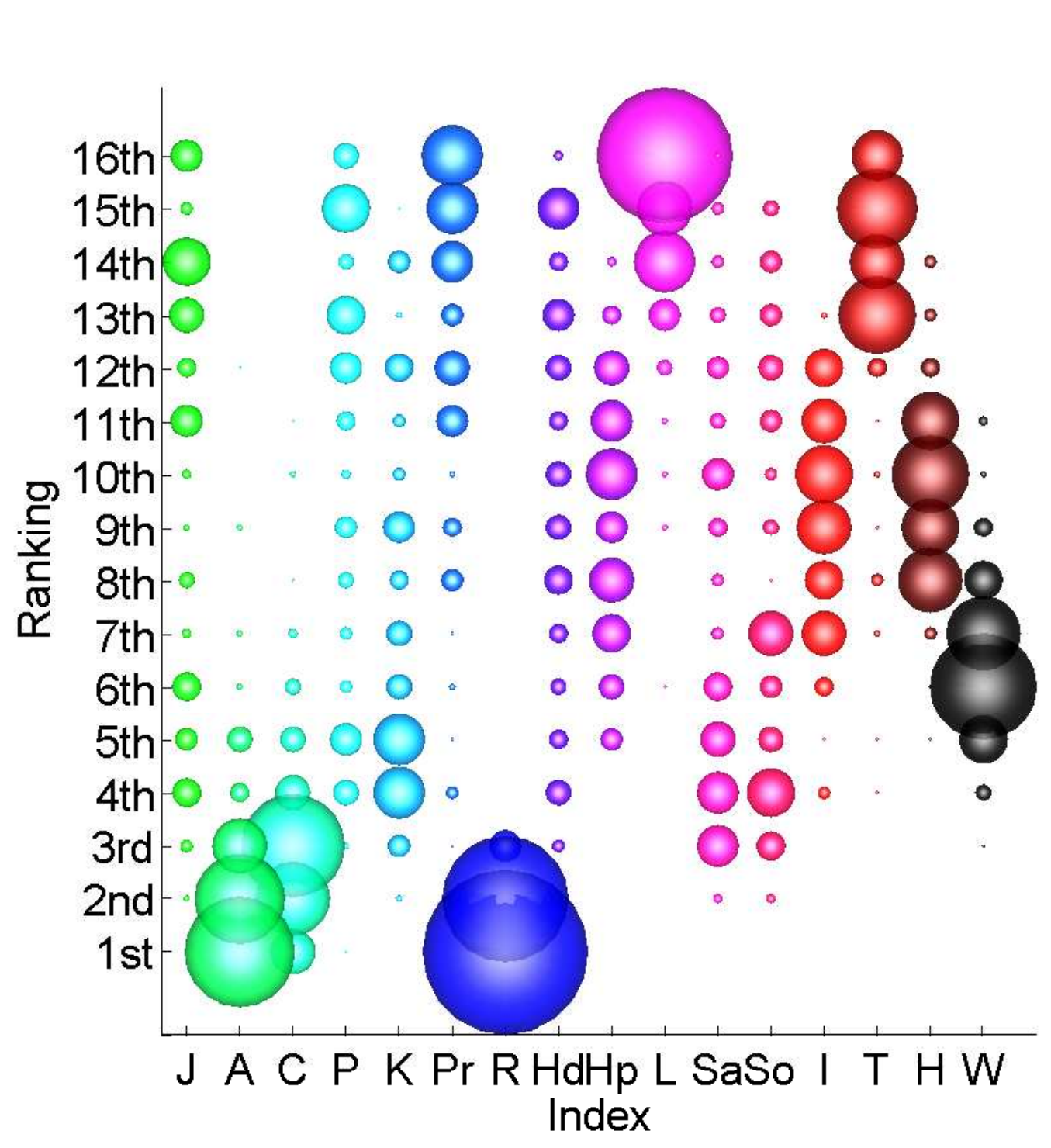}}
\subfigure[Week 3 $\mapsto$ 4, N=20000]{\includegraphics[width=.24\linewidth]{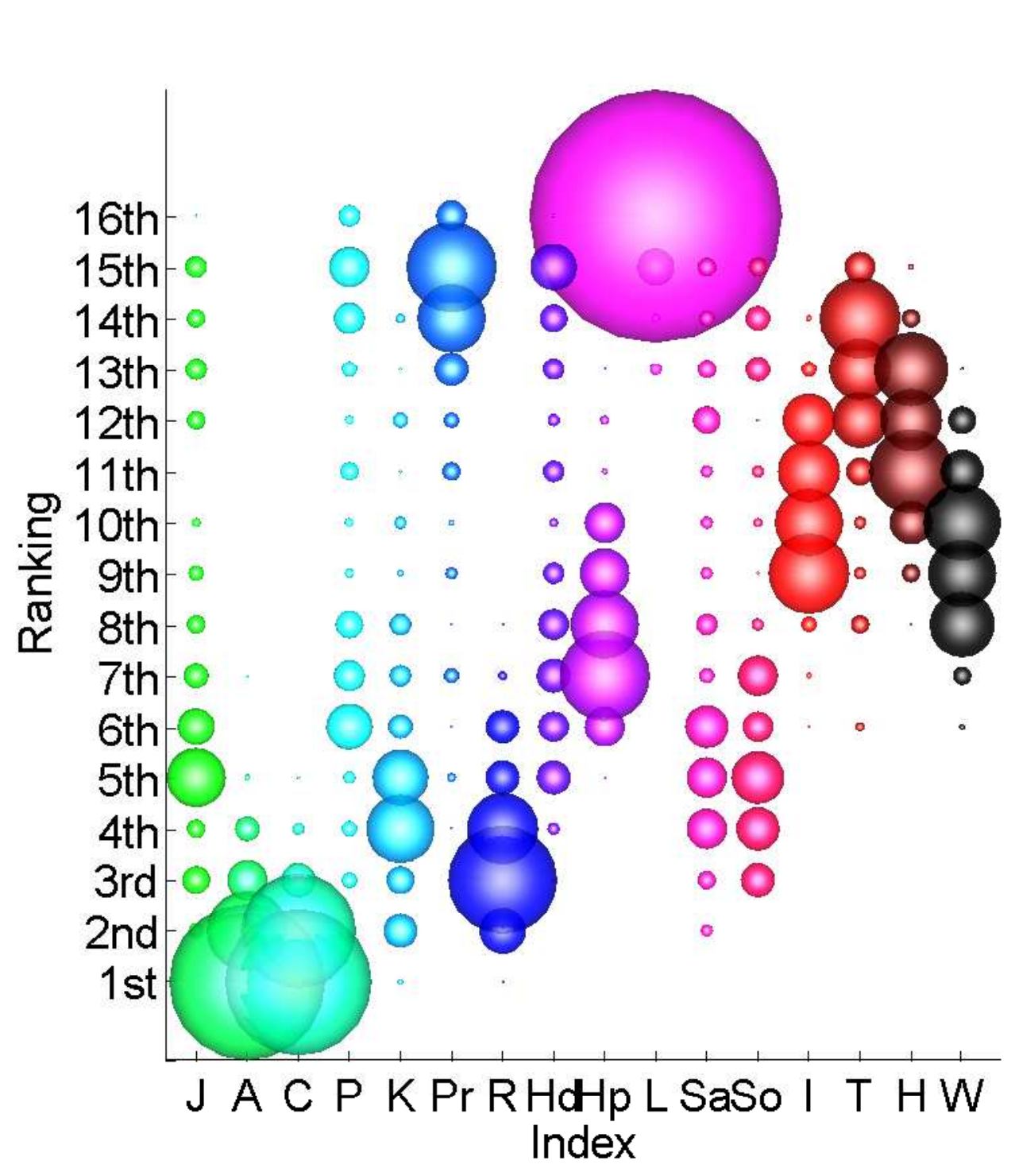}}\\
\caption{Ranking of the value of the evolved coefficients from each of 100 CMA-ES runs when fitness is based on the percent of correctly predicted links from the top $N$ scores. Adamic-Adar is the most frequently chosen top ranking (i.e.,  heavily weighted) index, followed by common neighbors and resource allocation. The lowest ranking index was LHN. Individual similarity indices, such as happiness, word similarity, Twitter user Id and Tweet count were ranked intermediate. J=Jaccard, A=Adamic-Adar, C=Common neighbors, P=Paths, K=Katz, Pr=Preferential attachment, R=Resource allocation, Hd=Hub depressed, Hp=Hub promoted, L=Leicht-Holme-Newman, Sa=Salton, So=Sorenson, I=Twitter Id similarity, T=Tweet count similarity, H=Happiness similarity, W=word similarity. }
\label{ref:all_solns_weeks13}
\end{figure*}

\setcounter{equation}{3}
\begin{figure*}[ht!]
\centering
\subfigure[Week 7 $\mapsto$ 8, N=20]{\includegraphics[width=.24\linewidth]{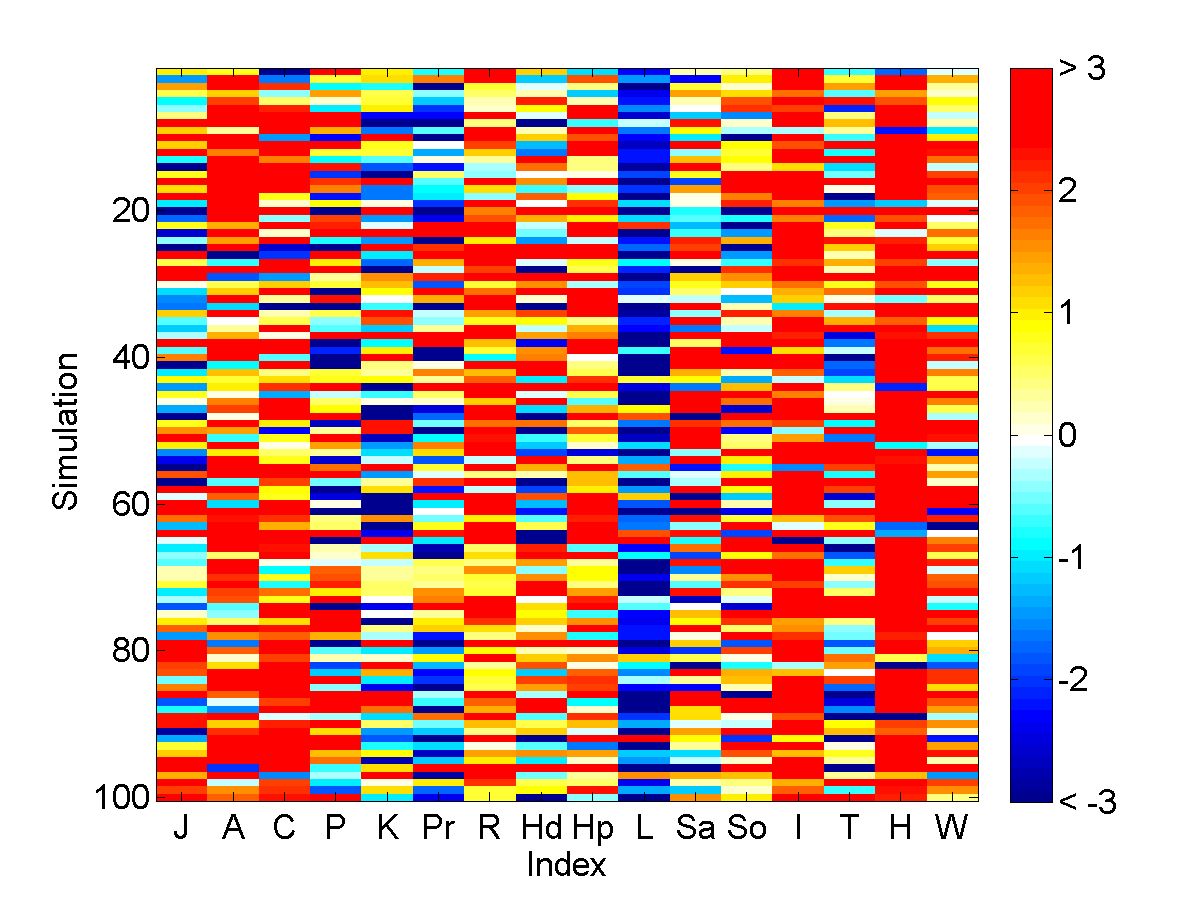}}
\subfigure[Week 7 $\mapsto$ 8, N=200]{\includegraphics[width=.24\linewidth]{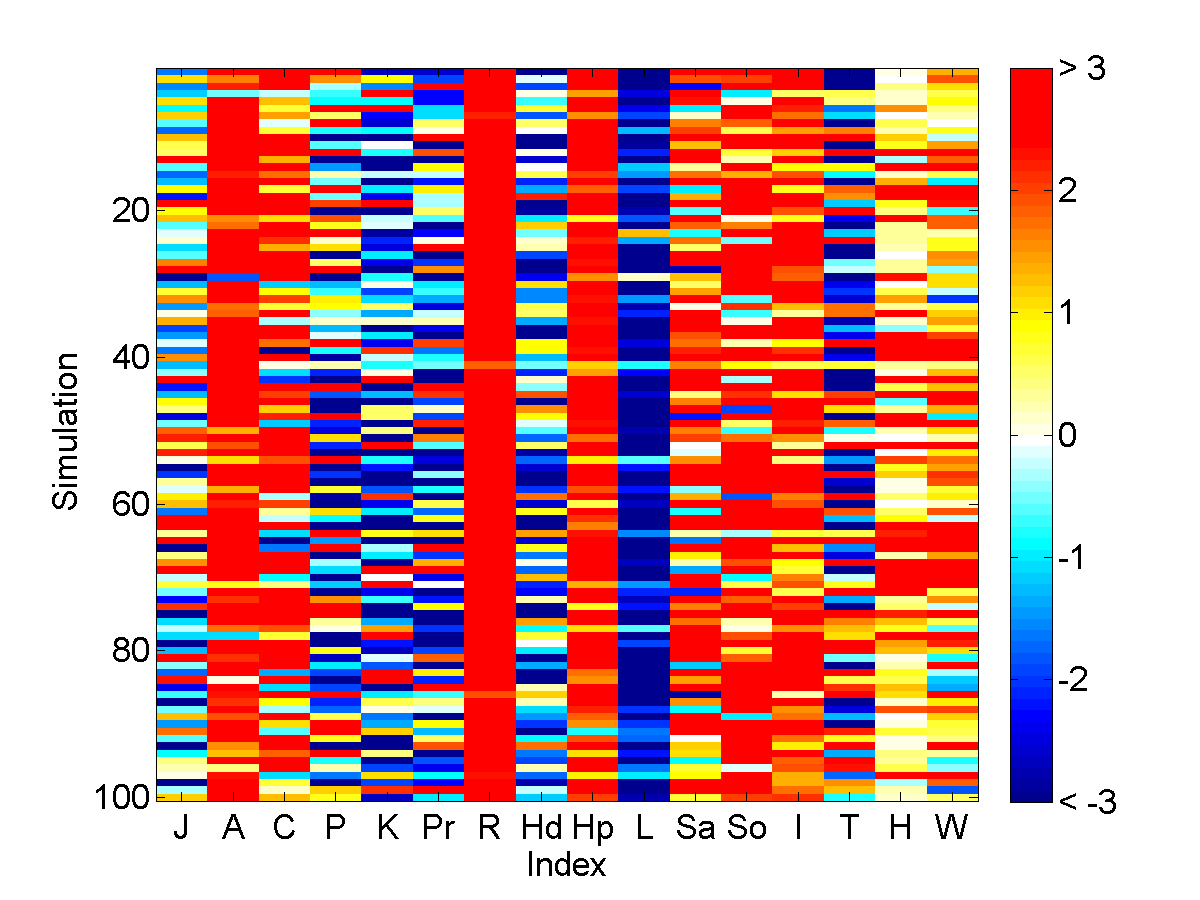}}
\subfigure[Week 7 $\mapsto$ 8, N=2000]{\includegraphics[width=.24\linewidth]{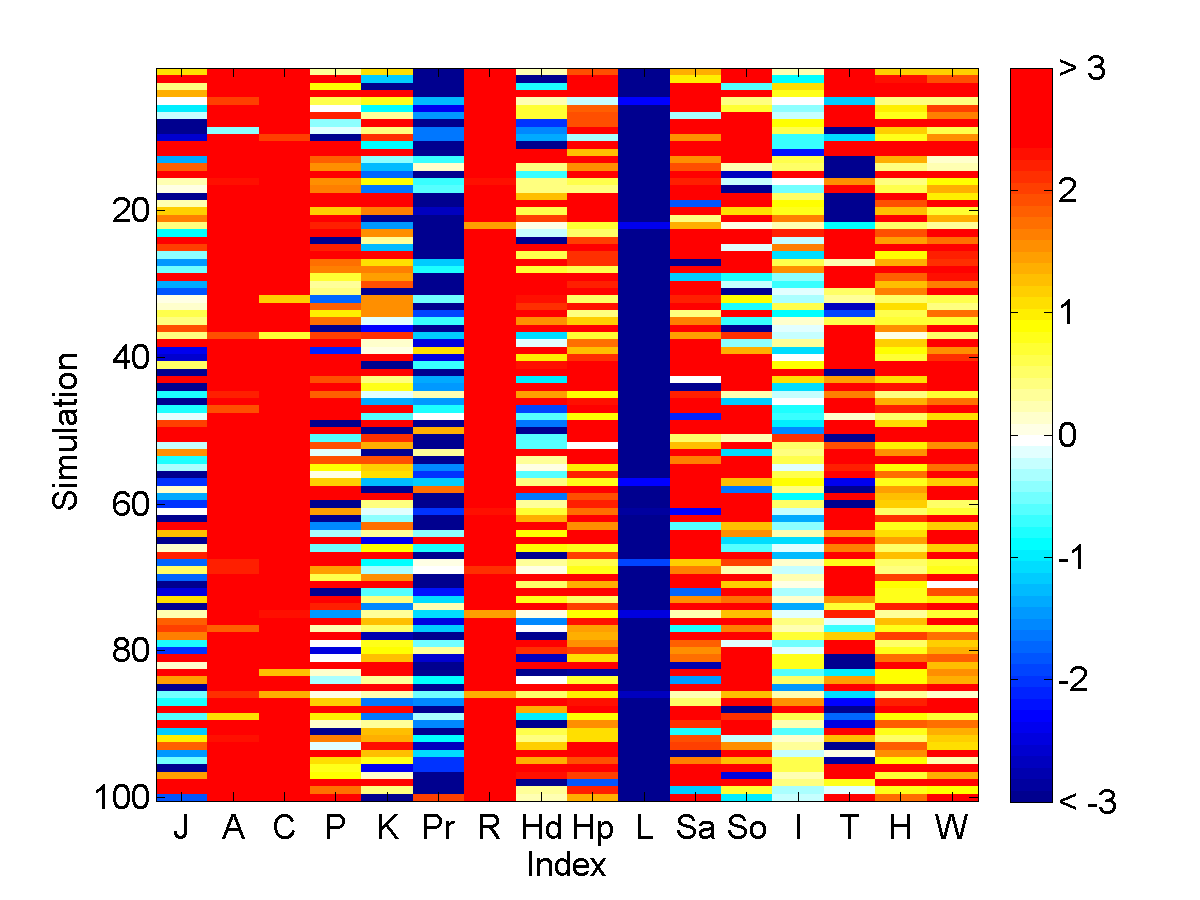}}
\subfigure[Week 7 $\mapsto$ 8, N=20000]{\includegraphics[width=.24\linewidth]{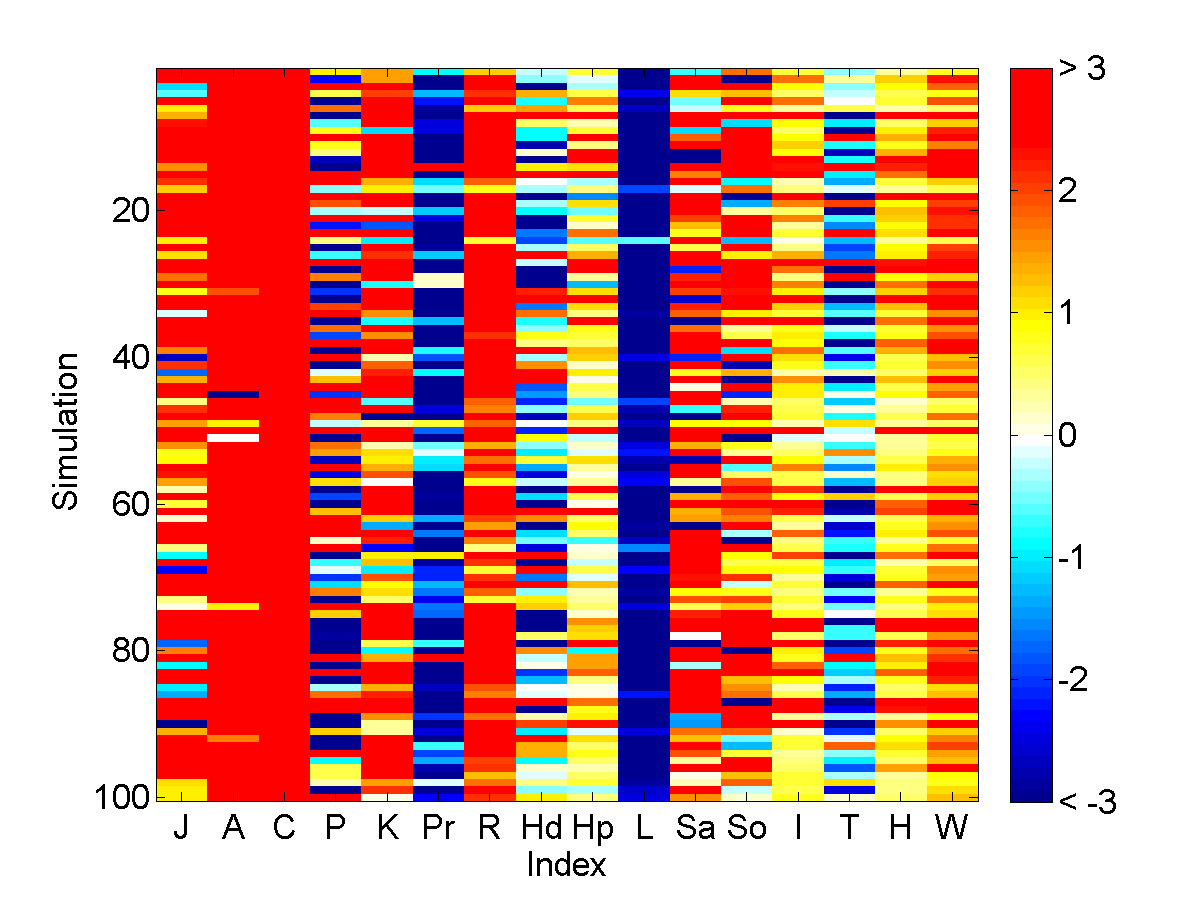}}\\
\subfigure[Week 7 $\mapsto$ 8, N=20]{\includegraphics[width=.24\linewidth]{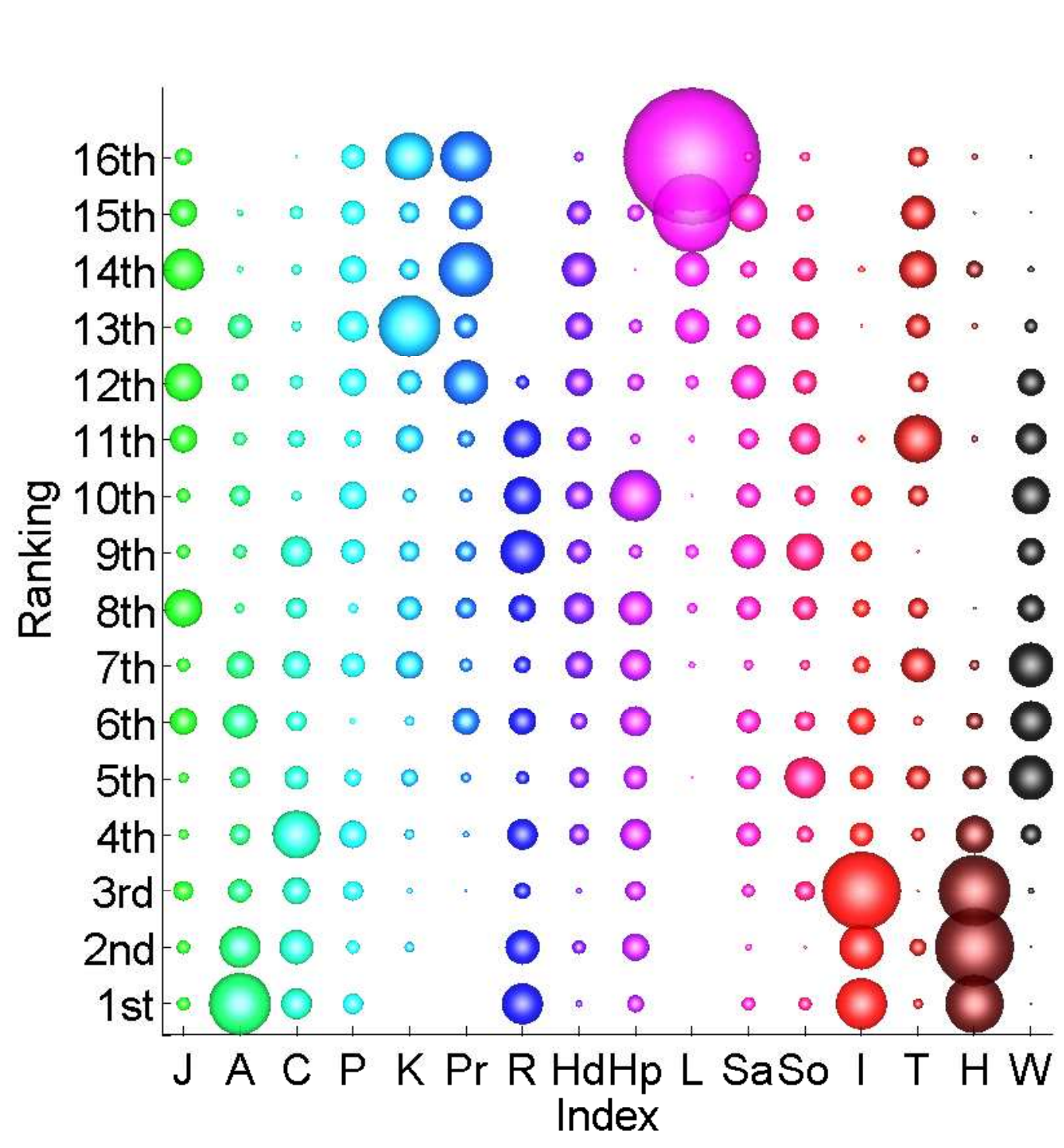}}
\subfigure[Week 7 $\mapsto$ 8, N=200]{\includegraphics[width=.24\linewidth]{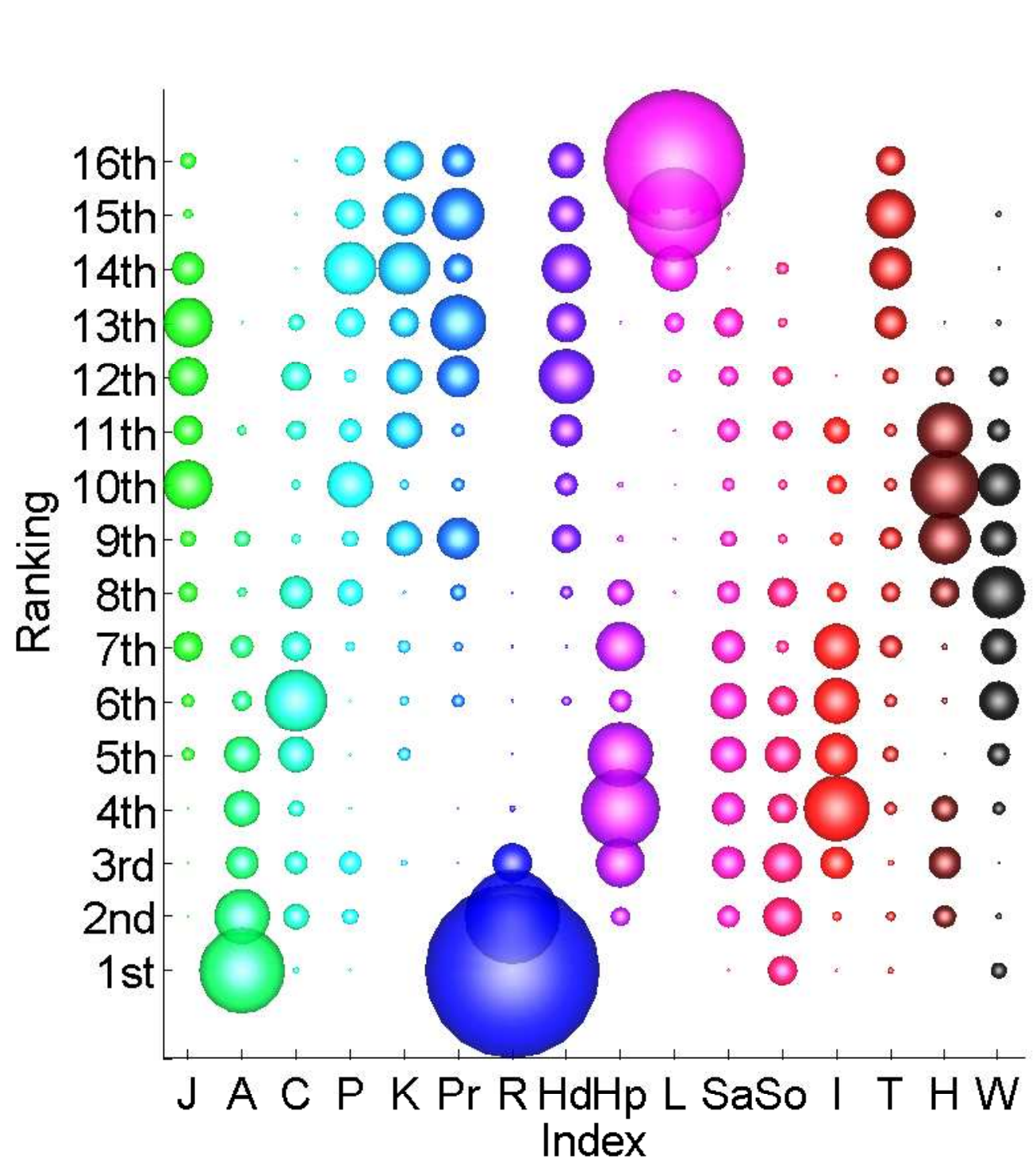}}
\subfigure[Week 7 $\mapsto$ 8, N=2000]{\includegraphics[width=.24\linewidth]{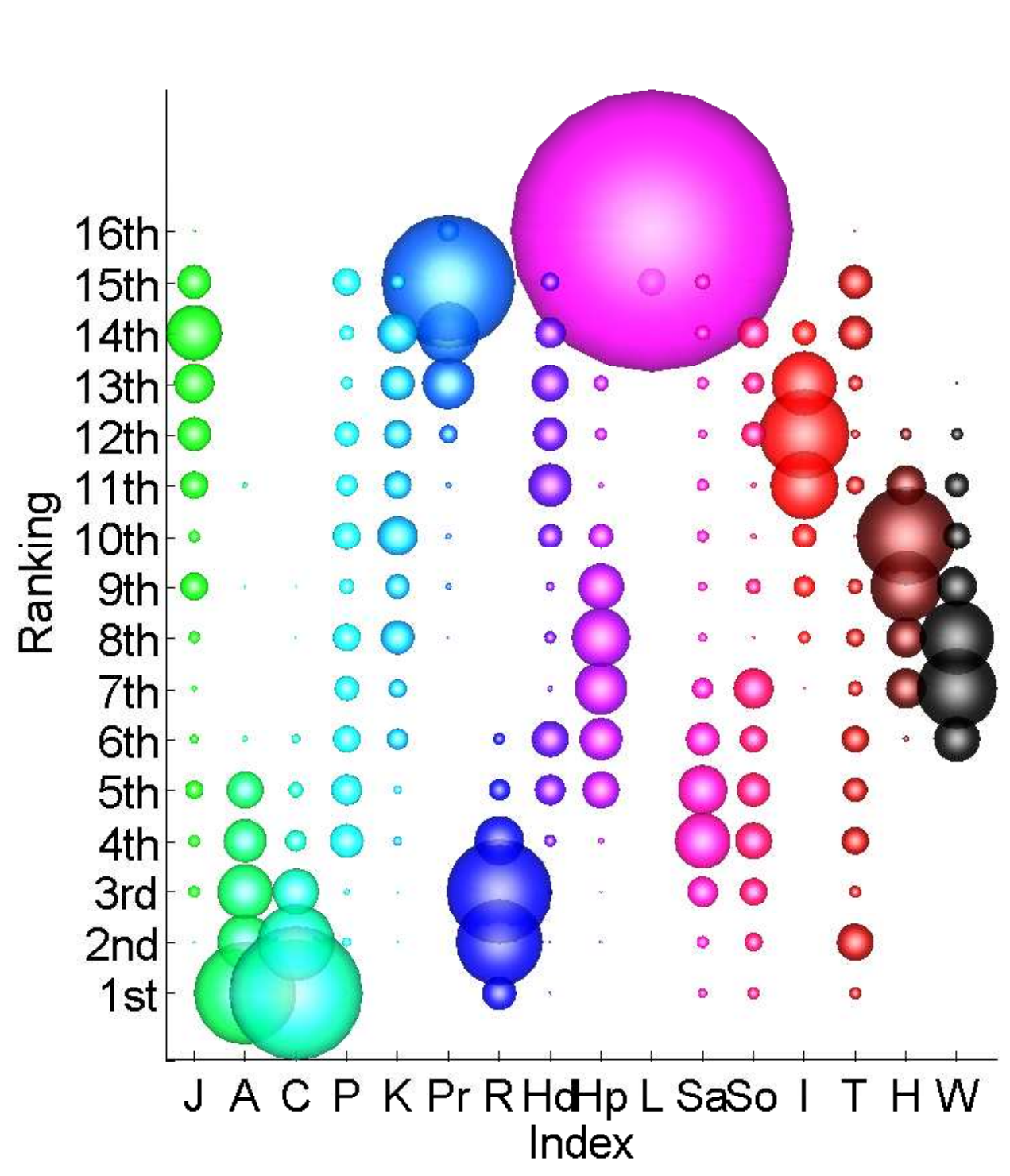}}
\subfigure[Week 7 $\mapsto$ 8, N=20000]{\includegraphics[width=.24\linewidth]{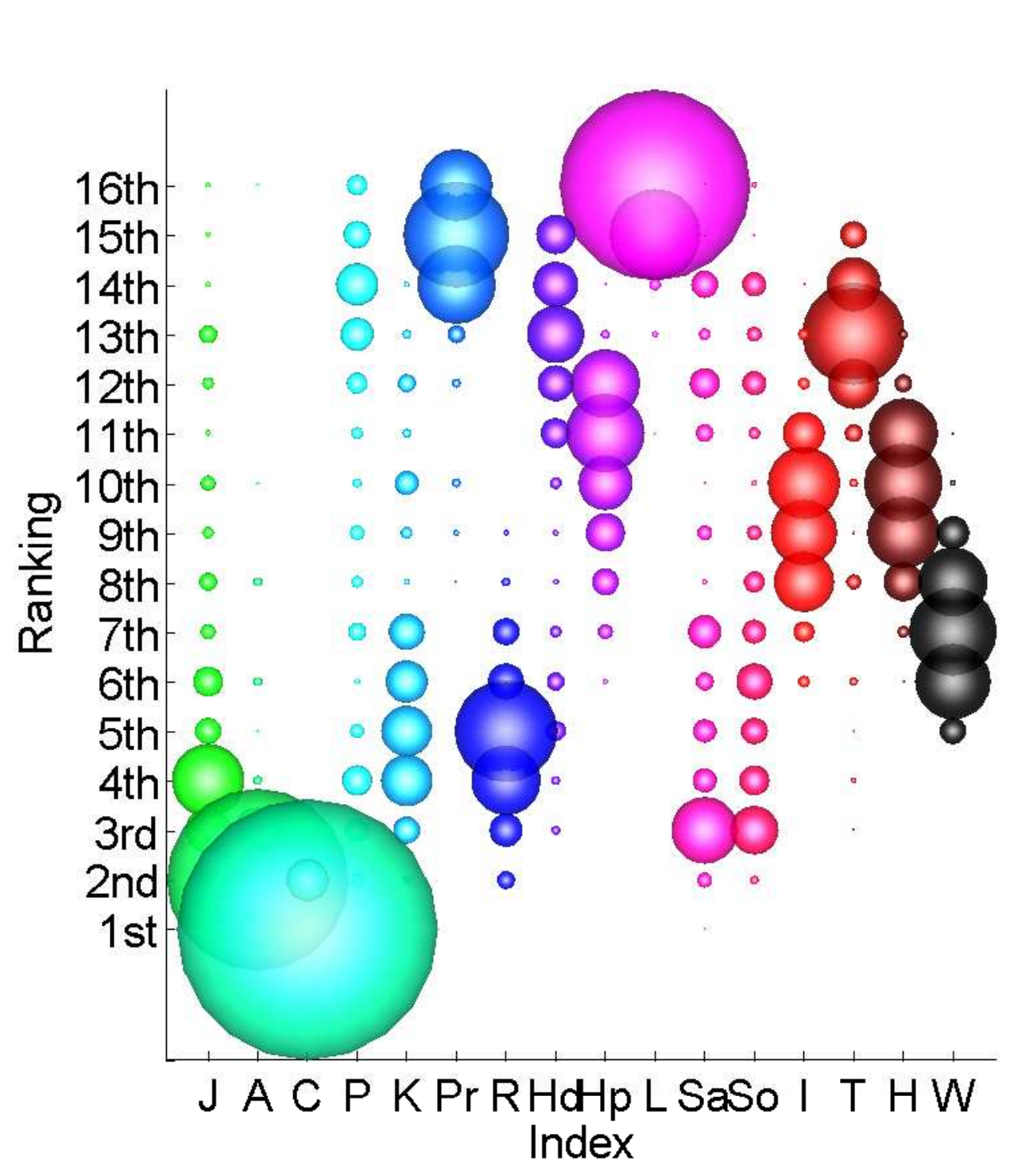}}\\
\subfigure[Week 9 $\mapsto$ 10, N=20]{\includegraphics[width=.24\linewidth]{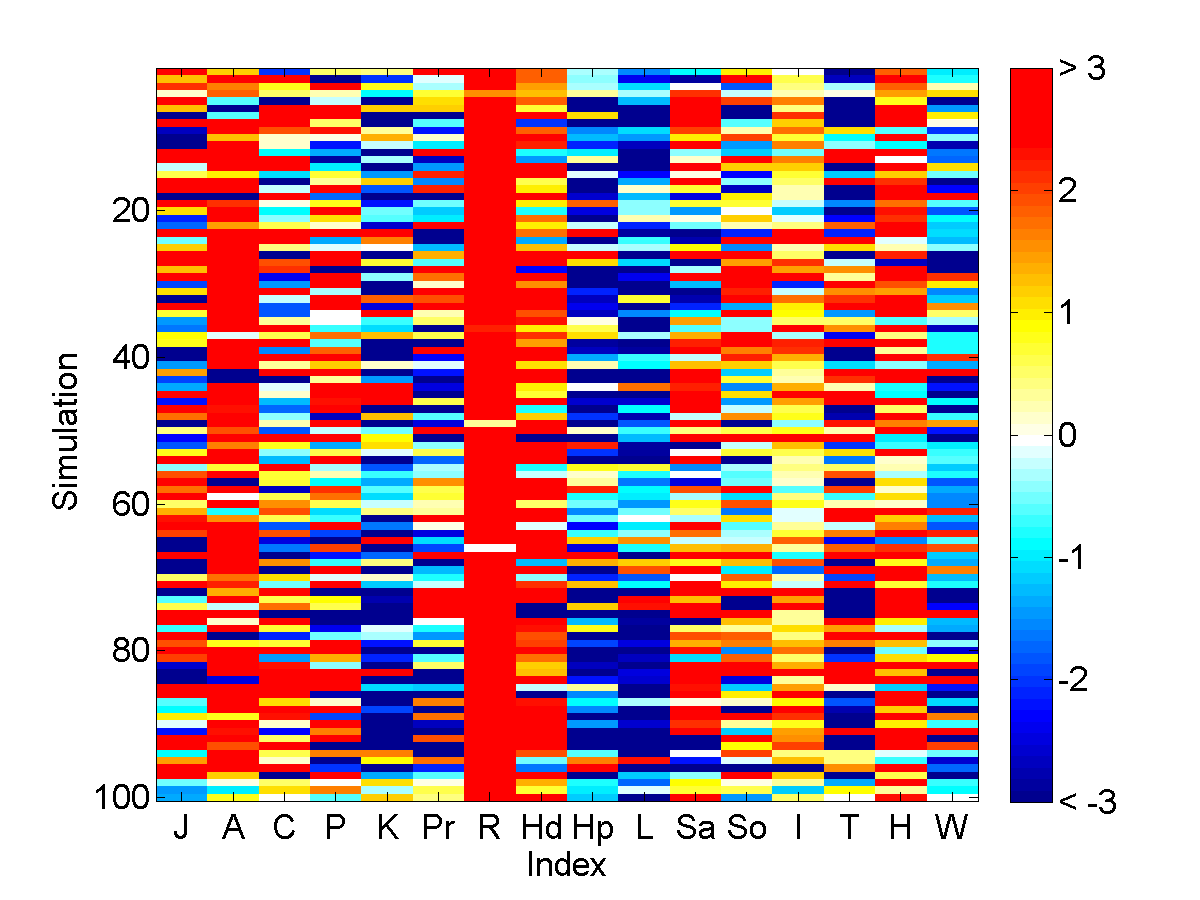}}
\subfigure[Week 9 $\mapsto$ 10, N=200]{\includegraphics[width=.24\linewidth]{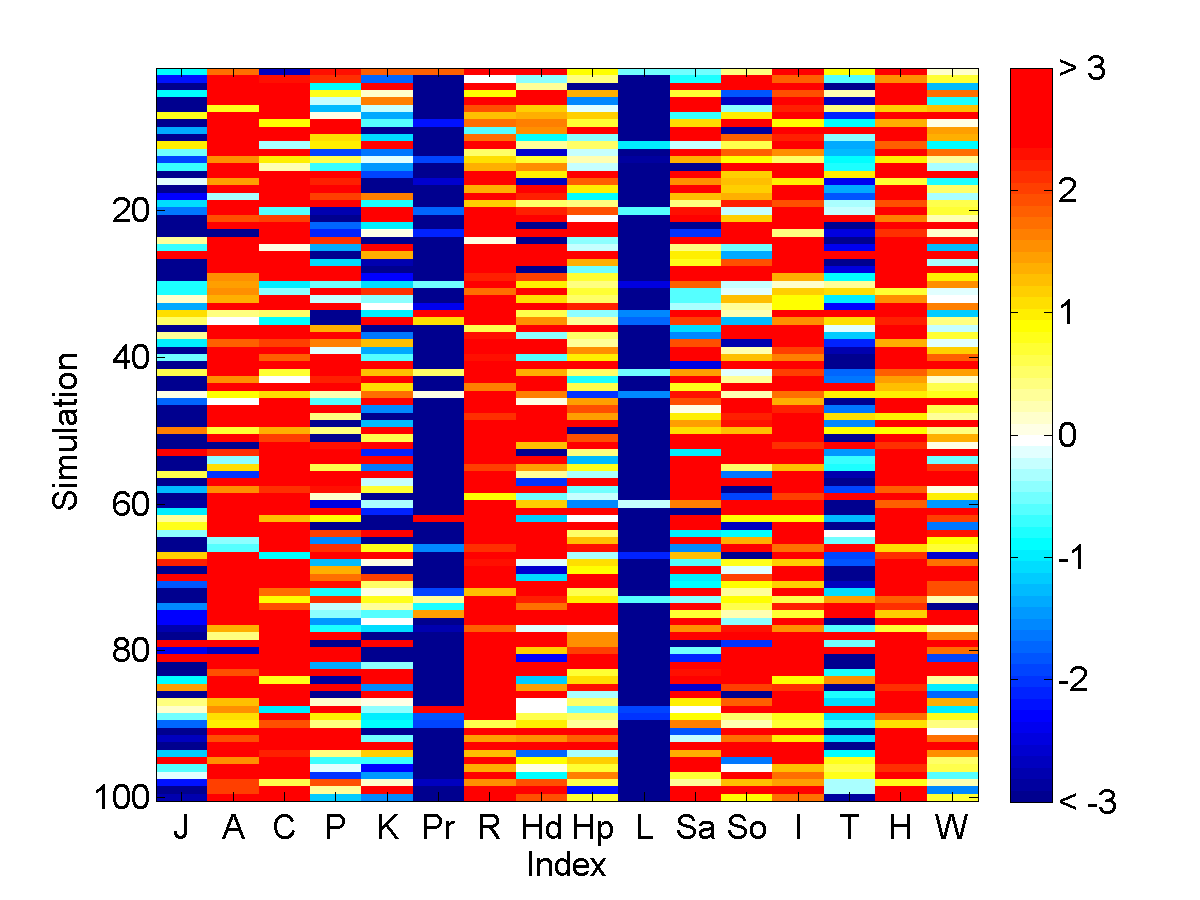}}
\subfigure[Week 9 $\mapsto$ 10, N=2000]{\includegraphics[width=.24\linewidth]{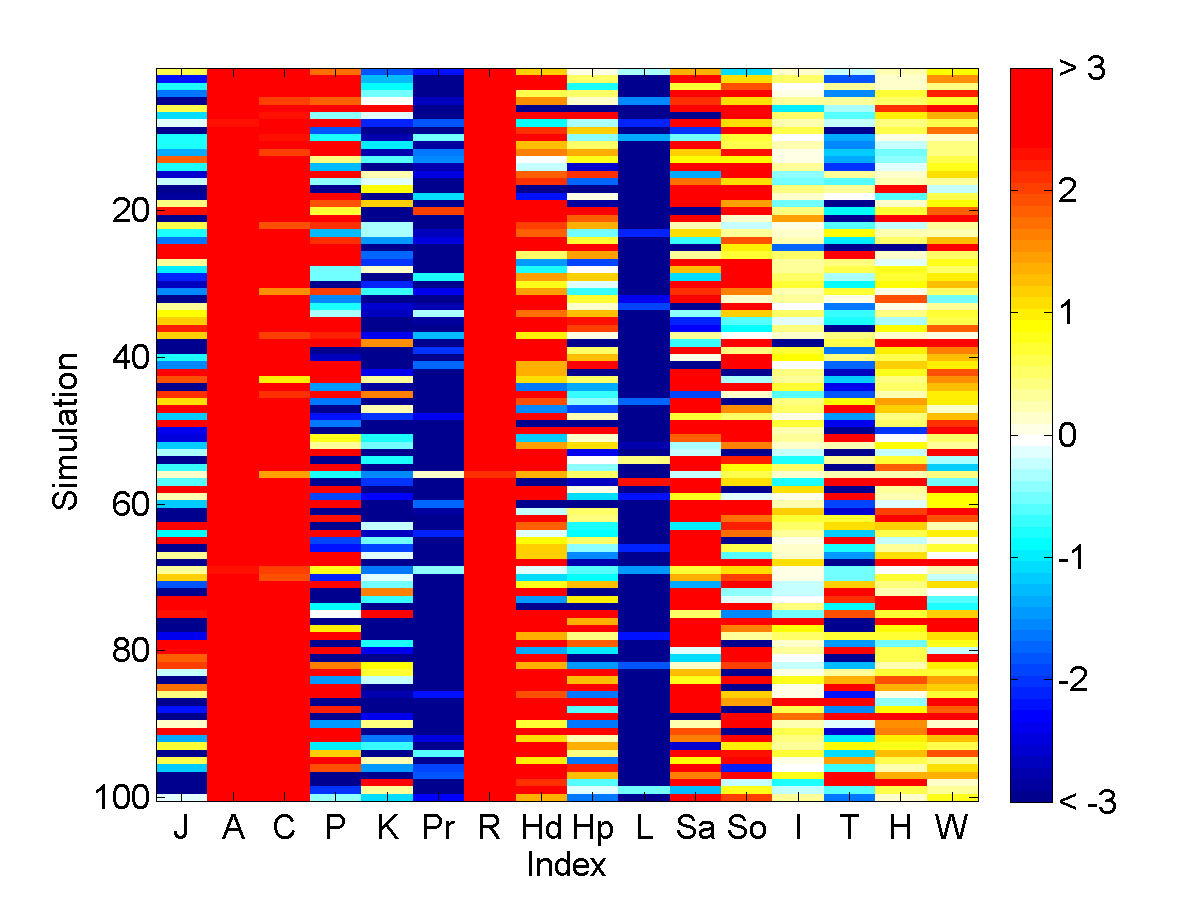}}
\subfigure[Week 9 $\mapsto$ 10, N=20000]{\includegraphics[width=.24\linewidth]{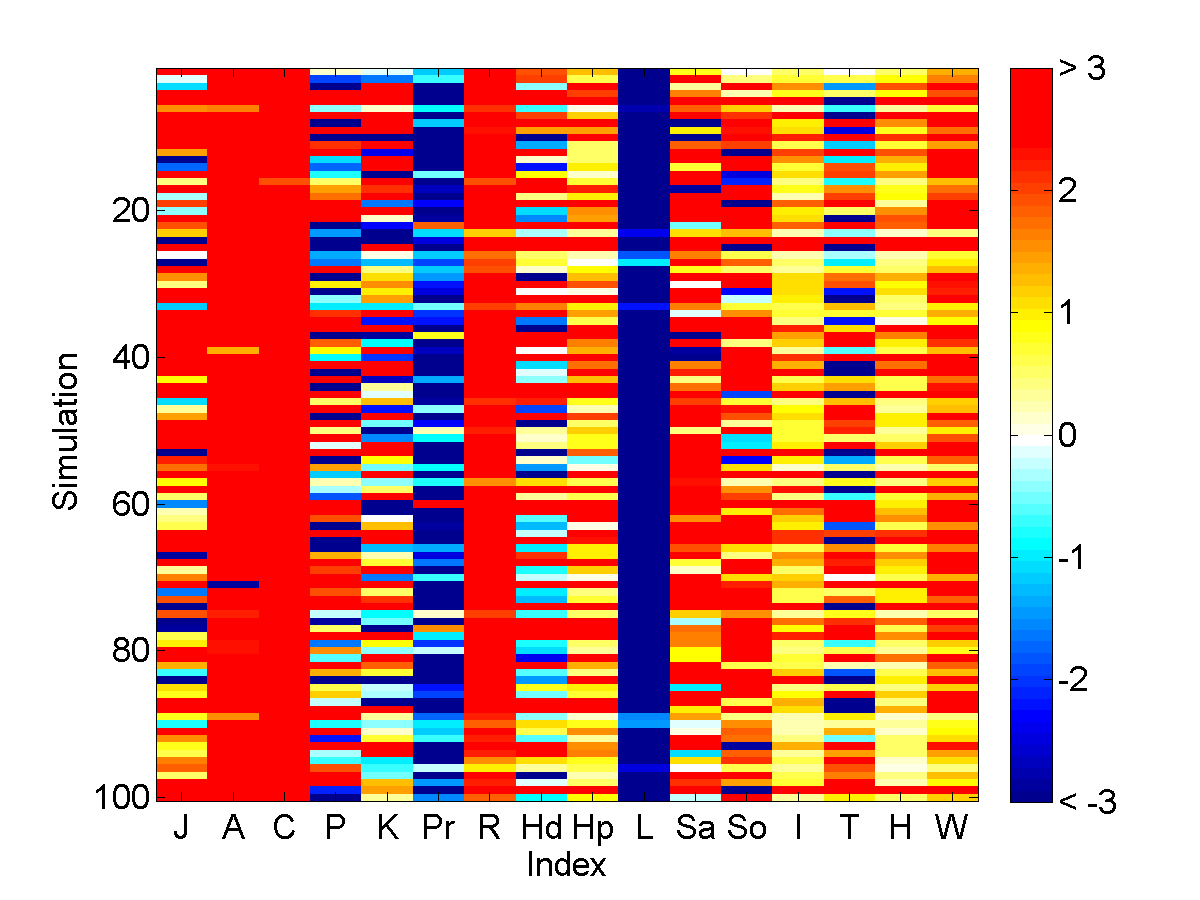}}\\
\subfigure[Week 9 $\mapsto$ 10, N=20]{\includegraphics[width=.24\linewidth]{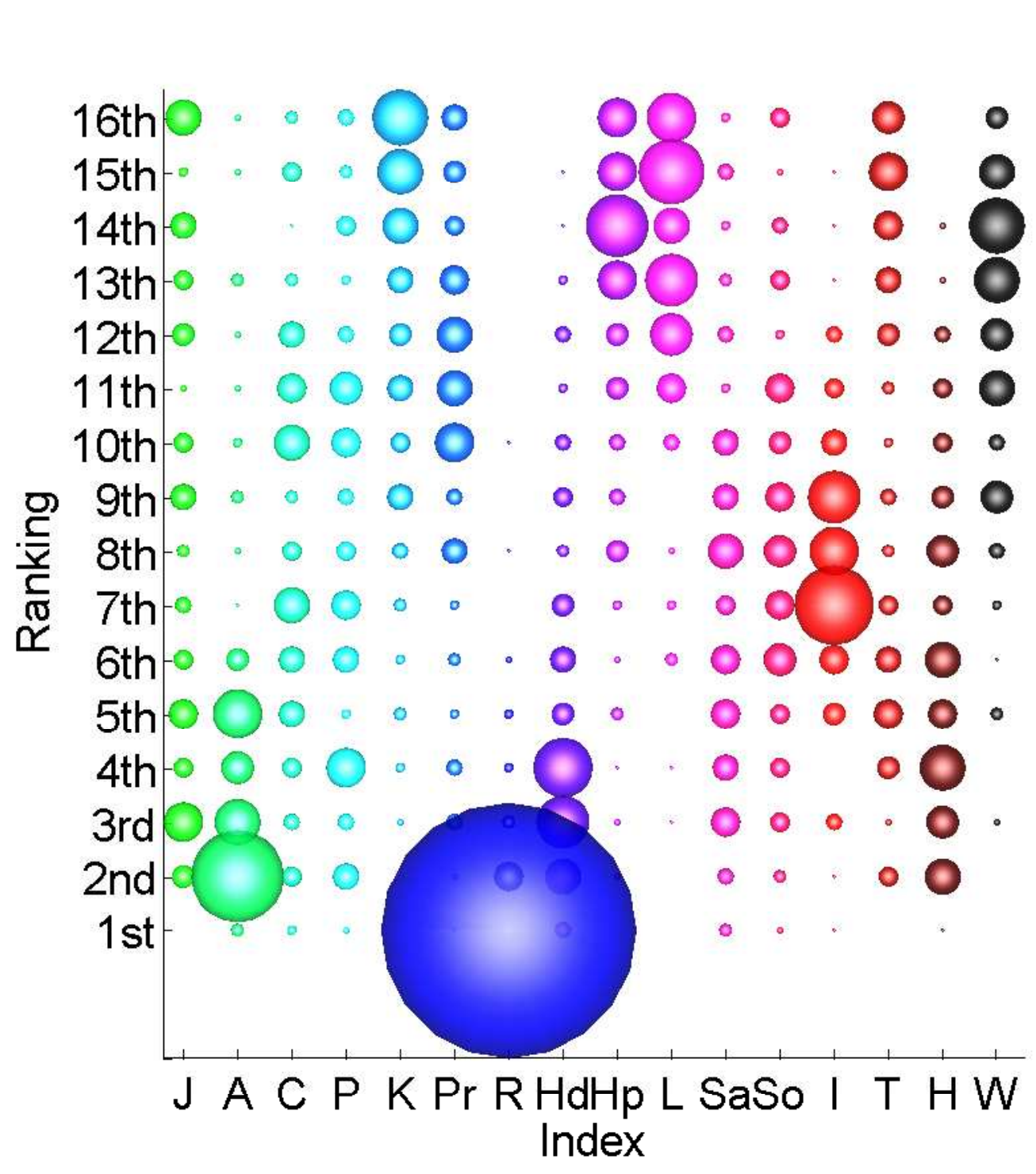}}
\subfigure[Week 9 $\mapsto$ 10, N=200]{\includegraphics[width=.24\linewidth]{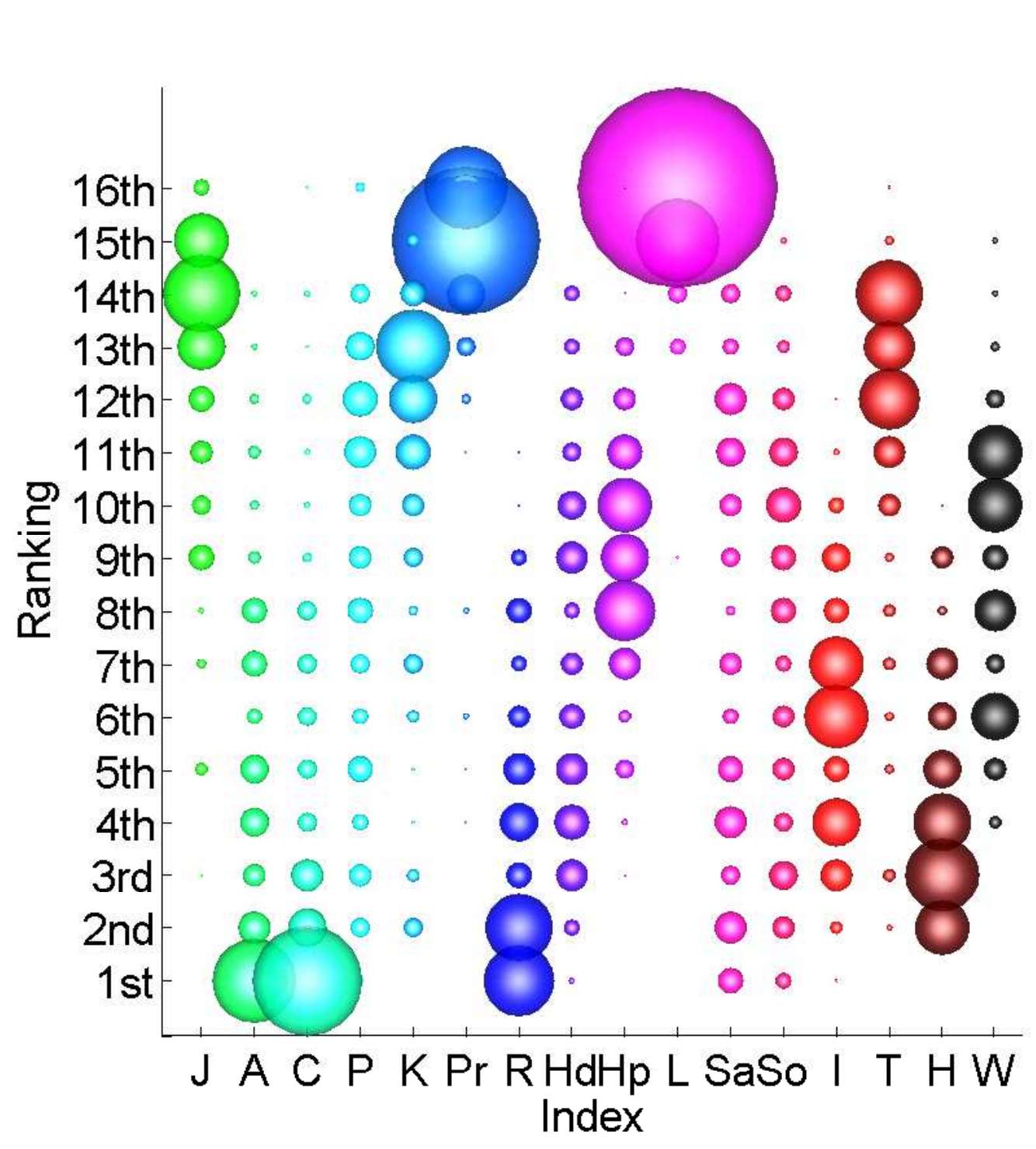}}
\subfigure[Week 9 $\mapsto$ 10, N=2000]{\includegraphics[width=.24\linewidth]{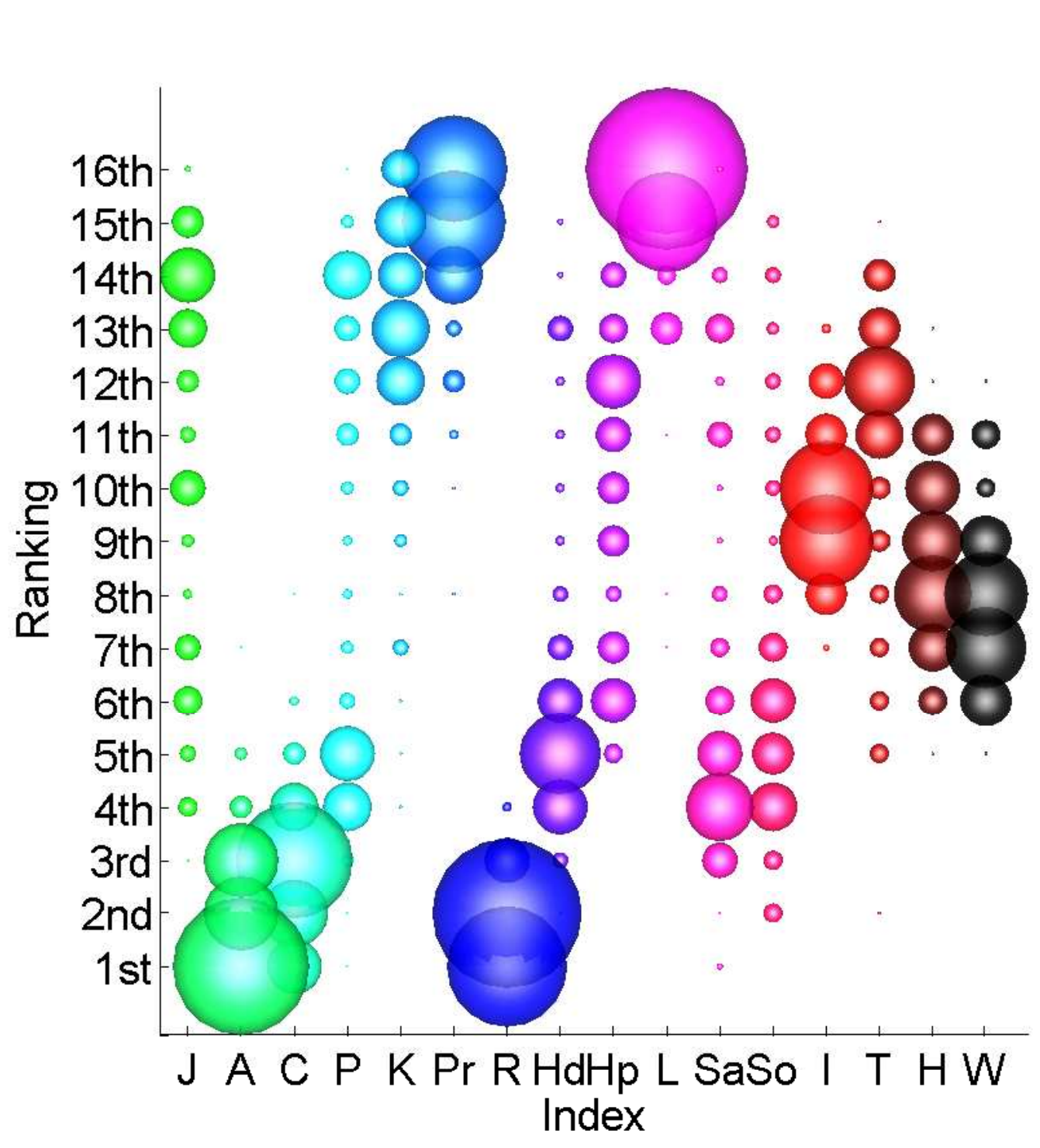}}
\subfigure[Week 9 $\mapsto$ 10, N=20000]{\includegraphics[width=.24\linewidth]{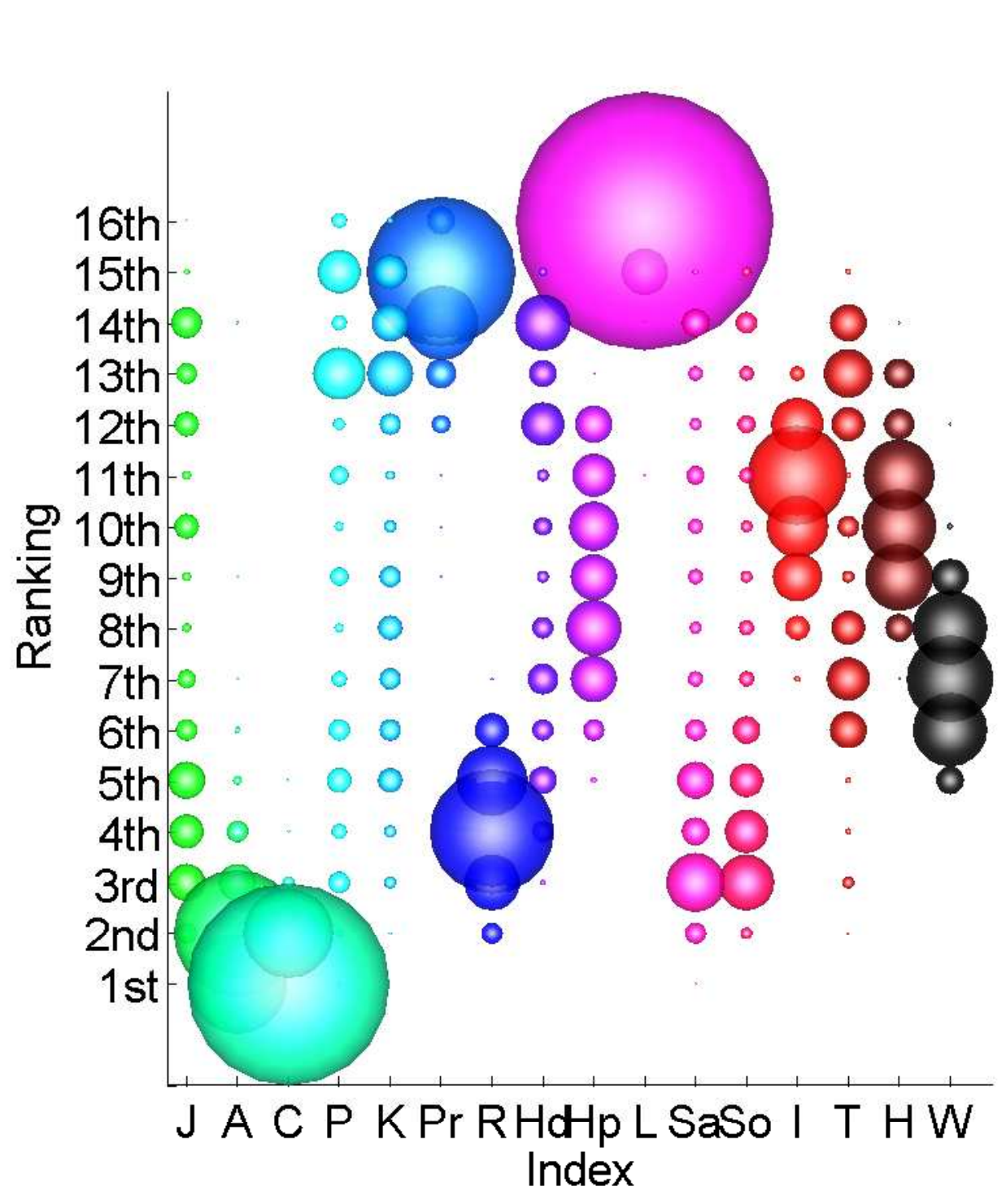}}\\
\caption{Ranking of the value of the evolved coefficients from each of 100 CMA-ES runs when fitness is based on the percent of correctly predicted links from the top $N$ scores. Adamic-Adar is the most frequently chosen top ranking (i.e., heavily weighted) index, followed by common neighbors and resource allocation. The lowest ranking index was LHN. Individual similarity indices, such as happiness, word similarity, Twitter user Id and Tweet count were ranked intermediate. J=Jaccard, A=Adamic-Adar, C=Common neighbors, P=Paths, K=Katz, Pr=Preferential attachment, R=Resource allocation, Hd=Hub depressed, Hp=Hub promoted, L=Leicht-Holme-Newman, Sa=Salton, So=Sorenson, I=Twitter Id similarity, T=Tweet count similarity, H=Happiness similarity, W=word similarity.}
\label{ref:all_solns_weeks79}
\end{figure*}

\end{document}